\documentclass[12pt]{elsarticle_arxiv}

\usepackage{graphicx}
\usepackage{amssymb}
\usepackage{amsmath}
\usepackage{bm}
\usepackage{a4wide}
\usepackage{appendix}
\usepackage{color}
\usepackage{esint}

\graphicspath{{figs/}}

\newcommand\Appx[1]{\ref{#1}}

\def\shp{\varsigma}
\def\eps{\epsilon}
\def\Dlt{\Delta}

\def\Lambf{{\mbox{\boldmath$\Lambda$\unboldmath}}}
\def\Thetabf{{\mbox{\boldmath$\Theta$\unboldmath}}}

\def\ubm{{\bf{u}}}
\def\vbm{{\bf{v}}}
\def\wbm{{\bf{w}}}
\def\pbm{{\bf{p}}}
\def\qbm{{\bf{q}}}
\def\fbm{{\bf{f}}}
\def\hbm{{\bf{h}}}
\def\gbm{{\bf{g}}}
\def\bbm{{\bf{b}}}
\def\dbm{{\bf{d}}}
\def\zbm{{\bf{z}}}
\def\ybm{{\bf{y}}}
\def\xbm{{\bf{x}}}
\def\rbm{{\bf{r}}}
\def\tbm{{\bf{t}}}

\def\Kbm{{\bf{K}}}
\def\Mbm{{\bf{M}}}
\def\Nbm{{\bf{N}}}
\def\Ibm{{\bf{I}}}
\def\Jbm{{\bf{J}}}
\def\Ebm{{\bf{E}}}
\def\Vbm{{\bf{V}}}
\def\Lbm{{\bf{L}}}
\def\Abm{{\bf{A}}}
\def\Bbm{{\bf{B}}}
\def\Cbm{{\bf{C}}}
\def\Sbm{{\bf{S}}}
\def\Wbm{{\bf{W}}}
\def\Dbm{{\bf{D}}}
\def\Hbm{{\bf{H}}}
\def\Gbm{{\bf{G}}}
\def\Fbm{{\bf{F}}}
\def\Rbm{{\bf{R}}}
\def\Qbm{{\bf{Q}}}
\def\Xbm{{\bf{X}}}
\def\Tbm{{\bf{T}}}
\def\Zbm{{\bf{Z}}}

\newcommand\zerobm{\bf{0}}

\newcommand\dlt{\delta}          
\newcommand\vpi{\varpi}          
\newcommand\vsigma{\varsigma}          
\newcommand\lam{\lambda}          

\newcommand\dd{\mathrm{d}}

\newcommand\hh{{\bar{h}}}

\newcommand\xm{{\sf{x}}}
\newcommand\ym{{\sf{y}}}

\newcommand\plper{{\ul{\#}}} 

\newcommand\ipGm[2]{\left \langle{#1},\,{#2}\right\rangle_{\Gamma_0}}

\newcommand\Lpar{[\kern-0.17em{[}}
\newcommand\Rpar{]\kern-0.18em{]}}

\newcommand\Jump[2]{\Lpar {#1} \Rpar_{{#2}}^\pm}

\def\nablad{\hat{\nabla}}
\def\gradpl{\ol{\nabla}}
\def\gradplx{\ol{\nabla}_x}
\def\gradplxS{\ol{\nabla}_x^S}
\def\gradply{\ol{\nabla}_y}
\def\gradplS{\ol{\nabla}^S}
\def\gradplyS{\ol{\nabla}_y^S}
\def\hom{{\rm{H}}}
\def\ext{{\rm{ext}}}
\def\supp{{\rm{supp\,}}}

\newcommand\Hdb{{\bf{H}}^1}
\newcommand\Hpdb{{\bf{H}}_\#^1}
\newcommand\HOdb{\Hbm_{0}^1}

\def\wrt{{w.r.t.{~}}}
\def\rhs{{r.h.s.{~}}}
\def\lhs{{l.h.s.{~}}}
\def\ie{{\it i.e.~}}
\def\eg{{\it e.g.~}}
\def\cf{{\it cf.~}}
\def\etc{{\it etc.}}

\newcommand{\eq}[1]{(\ref{#1})}

\newcommand\diag{{\rm diag}}
\newcommand\diagM[1]{{\sf{~diag}}\left({#1}\right)}

\def\bmi#1{\textbf{\textit{#1}}}

\newcommand\eeb[1]{\eb({#1})}
\newcommand\eeby[1]{\eb_y({#1})}

\newcommand{\wtilde}[1]{\widetilde{#1}}
\newcommand{\ul}[1]{\underline{#1}}
\newcommand{\ol}[1]{\overline{#1}}

\def\pdiff#1#2{\frac{\partial {#1}}{\partial {#2}}}

\def\psibf{{\mbox{\boldmath$\psi$\unboldmath}}}
\def\chibf{{\mbox{\boldmath$\chi$\unboldmath}}}
\def\sigmabf{{\mbox{\boldmath$\sigma$\unboldmath}}}
\def\vthetabf{{\mbox{\boldmath$\vartheta$\unboldmath}}}
\def\thetabf{{\mbox{\boldmath$\theta$\unboldmath}}}
\def\Thetabf{{\mbox{\boldmath$\Theta$\unboldmath}}}
\def\Pibf{{\mbox{\boldmath$\Pi$\unboldmath}}}

\def\R{\hbox{\rm I\kern-0.2em R}}
\def\Z{\hbox{\rm Z\kern-0.3em Z}}
\def\Eop{{{\rm I} \kern-0.2em{\rm E}}}%
\def\Dop{{{\rm I} \kern-0.2em{\rm D}}}%

\def\veps{\varepsilon}
\def\vepsdel{{\varepsilon\delta}}
\def\vkappa{\varkappa}

\def\imu{{\rm{i}}} 

\def\Om{\Omega}
\def\om{\omega}
\def\pd{\partial}

\def\RR{{\mathbb{R}}}
\def\ZZ{{\mathbb{Z}}}
\def\VV{{\mathbb{V}}}

\def\ab{{\bmi{a}}}
\def\bb{{\bmi{b}}}

\def\eb{{\bmi{e}}}
\def\fb{{\bmi{f}}}

\def\mb{{\bmi{m}}}
\def\nb{{\bmi{n}}}
\def\qb{{\bmi{q}}}

\def\ub{{\bmi{u}}}
\def\vb{{\bmi{v}}}
\def\wb{{\bmi{w}}}

\def\Ab{{\bmi{A}}}
\def\Bb{{\bmi{B}}}
\def\Cb{{\bmi{C}}}
\def\Db{{\bmi{D}}}

\def\Hb{{\bmi{H}}}

\def\Lb{{\bmi{L}}}
\def\Mb{{\bmi{M}}}

\def\Sb{{\bmi{S}}}

\def\Acal{\mathcal{A}}
\def\Ccal{\mathcal{C}}
\def\Dcal{\mathcal{D}}
\def\Ecal{\mathcal{E}}
\def\Fcal{\mathcal{F}}
\def\Gcal{\mathcal{G}}
\def\Hcal{\mathcal{H}}
\def\Kcal{\mathcal{K}}
\def\Lcal{\mathcal{L}}
\def\Mcal{\mathcal{M}}
\def\Ncal{\mathcal{N}}
\def\Pcal{\mathcal{P}}

\def\Scal{\mathcal{S}}
\def\Tcal{\mathcal{T}}

\def\Mcalbf{{\mbox{\boldmath$\mathcal{M}$\unboldmath}}}

\def\aYm#1#2{a_m \left ({#1},\,{#2}\right )}
\def\aYc#1#2{a_c \left ({#1},\,{#2}\right )}
\def\bYm#1#2{b_m \left ({#1},\,{#2}\right )}
\def\bhYc#1#2{\hat b_c \left ({#1},\,{#2}\right )}
\def\ipYs#1#2{\intY_{Y^*}#1\cdot#2}
\def\ippYc#1#2{{\left \langle\langle{#1},\,{#2}\right \rangle\rangle_{c}}}
\def\ipXic#1#2{\intY_{\Xi_c} #1\cdot#2}
\def\ipXiS#1#2{\intY_{\Xi_S} #1\cdot#2}
\def\ipGm#1#2{\intY_{\Gamma_0} #1\cdot#2}
\def\TT{\mathcal{T}_{\imu\om}}

\def\Tuf#1{{\mathcal{T}}_\veps{\left ({#1}\right )}} 
\newcommand\Tuftxt{\Tcal_\veps\,}

\def\intY{\fint}

\def\Eop{{{\rm I} \kern-0.2em{\rm E}}}%

\newcommand\FEMapprox{\stackrel{\rm FEM}{\approx}}

\newcounter{theorem}
               {\vspace{7pt} \noindent{\bf Theorem
                   \refstepcounter{theorem}\thetheorem. \protect\label{#1}
                   \hspace{-2mm}}\it}%
               {\vspace*{7pt}}

\newcounter{proposition}
\newenvironment{myproposition}[1]%
               {\vspace{7pt} \noindent{\bf Proposition
                   \refstepcounter{proposition}\theproposition. \protect\label{#1}
                   \hspace{-2mm}}\it}%
               {\vspace*{7pt}}
               
\newcounter{lemma}
\newenvironment{mylemma}[1]%
               {\vspace{7pt} \noindent{\bf Lemma
                   \refstepcounter{lemma}\thelemma. \protect\label{#1}
                   \hspace{-2mm}}\it}%
               {\vspace*{7pt}}

\newcounter{remark}
\newenvironment{myremark}[1]%
               {\vspace{7pt} \noindent{\bf Remark
                   \refstepcounter{remark}\theremark. \protect\label{#1} \hspace{-2mm}}\rm }%
               {\vspace*{-7pt} \flushright $\triangle$\\ } 

\newcounter{definition}
\newenvironment{mydefinition}[1]%
               {\vspace{7pt} \noindent{\bf Definition
                   \refstepcounter{definition}\thedefinition. \protect\label{#1} \hspace{-2mm}}\rm }%
               {\vspace*{-7pt} \flushright $\square$\\ } 

\begin{document}

\begin{frontmatter}

\title{Homogenization of the vibro--acoustic transmission on periodically perforated elastic plates with arrays of resonators}

\author[NTIS]{E.~Rohan\corref{cor1}}
\ead{rohan@kme.zcu.cz}
\cortext[cor1]{Corresponding author}
\author[NTIS]{V.~Luke\v{s}}
\ead{vlukes@kme.zcu.cz}

\address[NTIS]{European Centre of Excellence,
NTIS New Technologies for Information Society,
Faculty of Applied Sciences,\\ University of West Bohemia, \\
Univerzitn\'\i~22, 30614 Plze\v{n}, Czech Republic}

\begin{abstract}
 
Based on our previous work, we propose a homogenized model of acoustic waves
propagating through periodically perforated elastic plates with metamaterial
properties due to embedded arrays of soft elastic inclusions serving for
resonators. Such structures enable to suppress the acoustic transmission for
selected frequency bands. Homogenization of the vibro-acoustic fluid-structure
interaction problem in a 3D complex geometry of the transmission layer leads to
effective transmission conditions prescribed on the acoustic metasurface
associated with the mid-plane of the Reissner-Mindlin plate. Asymptotic
analysis with respect to the layer thickness, proportional to the plate
thickness and to the perforation period, yields an implicit
Dirichlet-to-Neumann operator defined on the homogenized metasurface. An
efficient method is proposed for computing frequency-dependent effective
parameters involved in the homogenized model of the layer. These can change
their signs, thus modifying the acoustic impedance and the effective mass of
the metasurface. The global problem of the acoustic wave propagation in a
waveguide fitted with the plate is solved using the finite element method. The
homogenized interface allows for a significant reduction of the computational
model. Numerical illustrations are presented.

\end{abstract}


\end{frontmatter}

\section{Introduction}\label{sec-intro}

Absorption of acoustic waves using acoustic metasurfaces belongs to challenging
and interesting issues from both the scientific and engineering points of view.
Many structures (engine casing) and devices incorporate perforated plates, or
panels which enable for fluid transport, and simultaneously should reduce the
noise transmission and emission due to the structure vibration. In general,
acoustic absorption is achieved by means of porous and fibrous materials, or
micro-perforated panels. Such structures are relatively very thick to achieve
required absorption. Recently some tunable structures, like space-coiling
structures provide absorbing design solutions with reduced thickness. We omit
approaches based on active absorption which require expensive sophisticated
electrical control. Even passive behaviour of metasurfaces constituted by
elastic composite materials with periodic arrangement of constituents can
exhibit special effective acoustic impedance which admits permeable but also
sound-resistant structure ensuring reduced noise radiation. In many devices, or
constructions, both theses mutually contradictory properties are required to
ensure functionality and environmental feasibility at the same time. While the
former requirement is often related to device functionality, admitting air flow
through the perforations (\eg cooling by ventilation), the latter one is
obvious but nontrivial to achieve -- besides reducing undesired noise
radiation, also suppressing the panel vibrations which may be induced by
incident acoustic wave, or independently by structural vibrations induced the
device (engines, mechanic transmissions \etc).

The aim of the present paper is to present a homogenized interface model of the
vibro-acoustic transmission through a metasurface designed as a perforated
elastic plate with resonators. We derive transmission conditions on the
homogenized interface replacing the problem of the fluid-structure interaction
in a very complex geometry. As the advantage, such a homogenized interface
reduces significantly the related numerical discretized model and, thereby, the
computational complexity when compared to the direct numerical simulations
which consist in solving directly the vibro-acoustic problem with a 3D elastic
structure describing the panel, \cf \cite{Yedeg-CMAME2016-Nitche-Helmholtz}.
Although the homogenization approach is not completely new in the context of
the acoustic transmission, it has not been applied so far to elastic plates
with resonators represented by strong heterogeneities in the elasticity
coefficients.

The homogenization strategy has been applied in situations when thin rigid
perforated plate represented by interface $\Gamma_0$ is characterized by the
thickness $\approx \dlt$ proportional to the size of the perforating holes
$\veps\approx\dlt$. It has been shown \cite{bb2005,Schweizer-JMPA2020} that the
homogenized interface is totally transparent for the acoustic field at the zero
order $o(\veps^0)$ terms of the model which describes the limit behaviour for
$\veps \rightarrow 0$, cf. \cite{DELOURME201228}. To get a nontrivial model
which captures acoustic impedance of the thin interface, a higher order
approximation involving the correctors at order $o(\veps^1)$ must be
considered. A slightly different treatment of the interface homogenization is
based on the so-called inner and outer asymptotic expansions, see \eg
\cite{Clayes-Delourm-AA2013,Marigo-Maurel-JASA2016,Marigo-Maurel-PRSA2016}. In
contrast with \cite{bb2005} dealing with thin perforated interfaces only, in
our previous studies \cite{rohan-lukes-waves07,Rohan-Lukes-AMC2019} we were
concerned with homogenization of a fictitious layer containing the perforated
plate. Nonlocal transmission conditions were obtained as the two-scale
homogenization limit $\veps \rightarrow 0$ of the acoustic field interacting
with the rigid, or elastic plate with an approximation respecting a given
finite scale $\veps_0 > 0$.

Acoustic metasurfaces
\cite{Liang-etal-Nanophotonics2018,Zou-PhysLettA-20201,Liang-PRR2020,Hu-Oskay-CMAME2020}
and their efficient modelling is the second issue of the present paper. The
area of electromagnetic wave propagation in periodic structures equipped with
resonators and the design of photonic crystals provided inspirations how to
manipulate the elastic waves in solids and fluids. However, there are
remarkable differences between the two types of waves due to different physical
phenomena. The so-called Bragg scattering of electromagnetic waves which arises
from the diffraction and refraction of incident waves in a crystalline
structure whose the periodicity is comparable to the wave length, the same
phenomenon in the acoustics would require large structures with the periodicity
in the range of meters, to manipulate waves of audible frequencies, \cf
\cite{Hu-Oskay-CMAME2020} where $10^5$Hz waves in phononic crystals are treated
by virtue of the higher-order homogenization. The second approach based on the
metamaterial design characterized by sub wavelength periodic structures is much
more appropriate to attenuate acoustic waves producing undesired noise.
Nonetheless, some interesting ideas of combined ``hybrid resonances'' in
metasurfaces with the acoustic-electric energy conversion seem to be promising
\cite{Ma-Yang-Xiao-NatMater2014} and challenging for the optimal design of
acoustic bulk metamaterials, as well as metasurfaces,
\cite{Kohei-Noguchi-Yamada-CMAME2013-Acoustic-topopt-Zwickers,Noguchi-CMAME-2018-High-frequency-homog-metamat-topot}.
For that purpose, the use of the two-scale homogenization method is quite
efficient as far as the scale separation between the wave lengths and the
characteristic size of the microstructures holds. Two essential building block
types for the design of acoustic metasurfaces include space-coiling structures
and those based on the resonance effect, constituting arrays of the Helmholtz
resonators. The first type elongate the travelling path of the waves through a
relatively thin panel which is designed as a porous medium with large
tortuosity (helical channels or labyrinth-like cascades), so that the effective
behaviour is comparable with straight waveguides featured by a high refractive
index, or lowered wave speed \cite{Fu-etal-metagrating-NC2019},
\cf~\cite{Liang-PRR2020}.


The second type, \ie periodic structures incorporating the Helmholtz resonators
distributed at a surface, is more suitable for devices operating at a rather
narrow frequency band. It can be designed in a dual way, using the surrounding
fluid itself which operates in resonance chambers -- pores in the panel
\cite{Huang-JASA2019}, or using compliant solid structures -- mass-spring
devices attached to the panel, \cf~\cite{PernasSalomon-2018}. While for the
first group the modelling is based on the poroelasticity theory, for the latter
type of metamaterials and meta surfaces, the effective continuum models were
proposed in \cite{Milton-Willis} due to discrete lattices and the so-called
Willis media with asymmetric terms in the Hooke's law
\cite{Liu-etal-WillisMetamaterialBeam-PhysRevX-2019,Milton-Briane-Willis-NJP2006},
or approaches treating distributed resonators in the framework of the Cosserat
continua, see \eg \cite{Madeo-Neff-etal-2015}. As an alternative,
homogenization-based modelling with the so-called high-contrast material
scaling has been employed to account for anti-resonance effects in elastic
media with periodically distributed soft inclusions
\cite{Auriault-Bonnet,Miara-etal-mms,Smyshlyaev2009}. Although the classical
first-order homogenization method applied to the standard electrodynamics with
periodically oscillating elasticity and density coefficients leads to the
standard effective astronomic model describing a non dispersive medium, it has
been shown that the homogenization method combined with a suitable scaling
ansatz related to the contrast in the periodically oscillating elasticity
coefficients leads to models of highly dispersive media due to effective
anisotropic mass density depending on the frequency of imposed oscillations.
Such a scaling, \ie the dependence of the material coefficients on the
heterogeneity size $\veps$ enables to preserve the information about the finite
period of the medium when passing to the limit $\veps\rightarrow 0$
\cite{Miara-etal-mms}. In effect, the periodic heterogeneity is not smeared out
in the limit completely, so that the frequency band gaps in wave propagation
can be identified by negative mass density of the homogenized model
\cite{rohan-IJES}. The band gap sensitivity and optimization through the
inclusion shape design has been considered in \cite{Vondrejc-IJSS2017}. For
plates established in the frameworks of the Kirchhoff-Love and the
Reissner-Mindlin theories, the soft inclusion metamaterial models were derived
in \cite{rohan-miara-ZAMM2015}. To analyze metamaterial response to an external
harmonic loading by forces with frequencies in range of the band gaps, a
spectral decomposition based method has been proposed in
\cite{Rohan2015bg-plates}.

In this paper, we build on the modelling approach reported in
\cite{Rohan-Lukes-AMC2019}, where the vibro-acoustic transmission conditions
were derived for a simple elastic plate embedded in a waveguide. Here we derive
analogous conditions for ``metamaterial plates'' -- elastic perforated plates
with resonators, which are respected by periodically distributed soft
inclusions. This feature completely modifies the effective model of the
vibro-acoustic transmission. Equations describing a homogenized transmission
layer in which the acoustic fluid-structure interaction is considered involve
frequency dependent coefficients computed using characteristic responses of the
representative cell which consists of the fluid and solid parts. The plate with
resonators is handled using the Reissner-Mindlin kinematics according to
\cite{rohan-miara-ZAMM2015}. When compared to the ``standard'' plate model
\cite{Rohan-Lukes-AMC2019}, the homogenization based two scale modelling of
metamaterial plate leads to complex characteristic problems parametrized by the
incident wave frequency. To resolve characteristic responses of the reference
cell efficiently, a spectral decomposition is employed. The vibro-acoustic
transmission conditions obtained by the homogenization constitute an implicit
Dirichlet-to-Neumann operator which relates the jump of the global acoustic
pressure in the waveguide, as evaluated at the homogenized fictitious layer
faces, with the effective acoustic momenta associated with the two faces. The
proposed modelling approach provides a reduced computational model of the
homogenized interface; the fluid-structure interaction problem imposed in a
complex 3D geometry describing the periodic architecture of the heterogeneous
plate is replaced by a 2D interface model, whereby the geometrical and
mechanical features are retained. In this respect, a Bernoulli plate model
obeying the Biot constitutive theory with time dependent permeability was
treated recently in \cite{Webster-Gurvich2021-weak-poro-elastic-plate} as a
``2.5'' dimensional problem.

The rest of the paper is organized, as follows. In Section~\ref{sec-problem},
the vibro-acoustic problem in a waveguide is introduced and a subproblem
representing the transmission layer response is defined. Its homogenization is
described in Section~\ref{sec-hom}, whereas some technical details are
explained in \Appx{apx-1} and \ref{apx-2}. In Section~\ref{sec-glob}, the
variational formulation of the global acoustic problem is coupled with the
homogenized layer through the Dirichlet-to-Neumann type conditions. Numerical
implementation of the computational homogenization is reported in
Section~\ref{sec-fem}; some details are postponed in \Appx{apx-3}. The
two-scale modelling using the homogenized interface is illustrated in
Section~\ref{sec-numex} where also a comparison with the ``standard'' plate
model \cite{Rohan-Lukes-AMC2019} is shown. Concluding remarks and research
perspectives follow in Section~\ref{sec-concl}.

\paragraph{Notation}
In the paper, the mathematical models are formulated in a Cartesian coordinate
system $\mathcal{R}(\text{O};\eb_1,\eb_2,\eb_3)$ where $O$ is the origin of the
space and $(\eb_1,\eb_2,\eb_3)$ is a orthonormal basis for this space. The
spatial position $x$ in the medium is specified through the coordinates
$(x_1,x_2,x_3)$ with respect to a Cartesian reference frame $\mathcal{R}$. The
boldface notation for vectors, $\ab = (a_i)$, and for tensors, $\bb =
(b_{ij})$, is used. The gradient and divergence operators applied to a vector
$\ab$ are denoted by $\nabla\ab$ and $\nabla \cdot \ab$, respectively. By
$\nabla^S\ub$ we denote the symmetrized gradient $\nabla\ub$, \ie the strain
tensor. When these operators have a subscript which is space variable, it is
for indicating that the operator acts relatively at this space variable, for
instance $\nabla_x = (\pd_i^x)$.
The symbol dot `$\cdot$' denotes the scalar product between two vectors and the
symbol colon `$:$' stands for scalar (inner) product of two second-order
tensors. Throughout the paper, $x$ denotes the global (``macroscopic'')
coordinates, while the ``local'' coordinates $y$ describe positions within the
representative unit cell $Y\subset\RR^3$ where $\RR$ is the set of real
numbers. By latin subscripts $i,j,k,l \in\{1,2,3\}$ we refer to
vectorial/tensorial components in $\RR^3$, whereas subscripts $\alpha,\beta \in
\{1,2\}$ are reserved for the tangential components with respect to the plate
midsurface, \ie coordinates $x_\alpha$ of vector represented by $x' = (x_1,x_2)
= (x_\alpha)$ are associated with directions $(\eb_1,\eb_2)$. Moreover,
$\gradplx = (\pd_\alpha)$ is the ``in-plane'' gradient. The gradient in the
so-called dilated configuration with coordinates $(x',z)$ is denoted by $\hat
\nabla = (\gradpl, \frac{1}{\veps} \pd_z)$. {We also use the jump \wrt the
transversal coordinate, $\Jump{q(\cdot,x_3)}{r} = q(\cdot,r/2) -
q(\cdot,-r/2)$.}

\section{Formulation with the large contrast elasticity plate} \label{sec-problem}

\begin{figure}
  \centering
\includegraphics[width=0.98\linewidth]{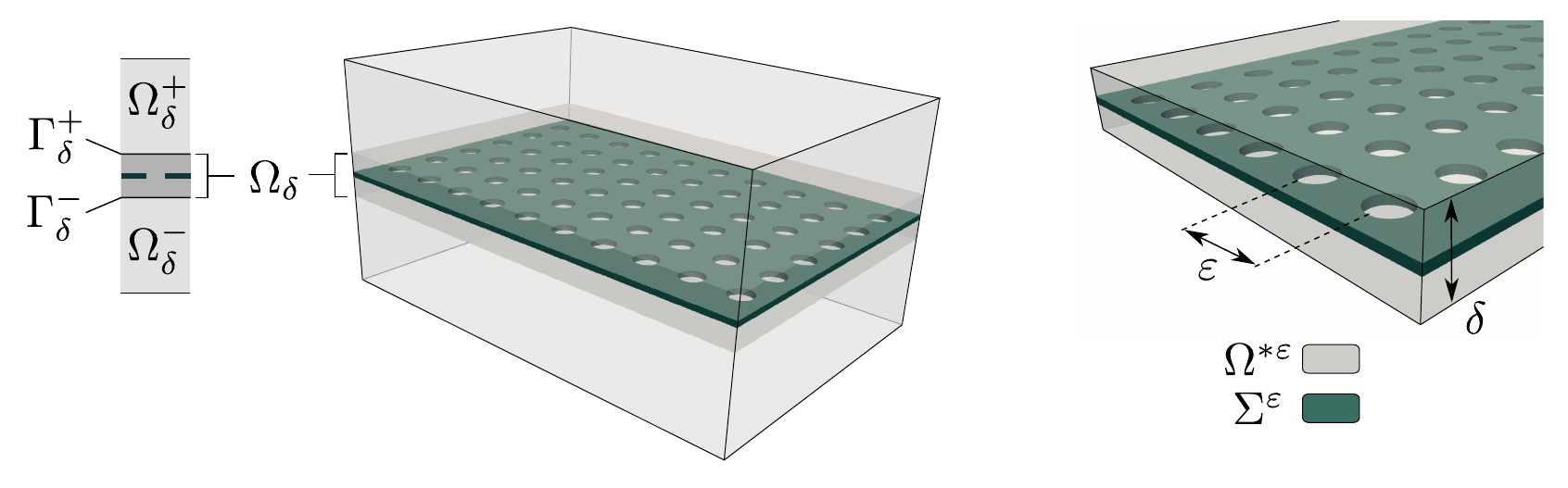}
  
\caption{Transmission layer $\Om_\dlt$ of thickness $\dlt$ embedded in the global domain $\Om^G =\Om_\dlt \cup \Om_\dlt^+ \cup \Om_\dlt^-$. Fluid and solid components occupy the domains $\Om^{*\veps}$ and $\Sigma^\veps$, respectively; note $\dlt = \vkappa\veps$. Interface $\Gamma_0$ represent the homogenized transmission layer, see Fig.~\ref{fig-pd2}.}\label{fig-pd1}
\end{figure}

\begin{figure}
  \centering
\includegraphics[width=0.98\linewidth]{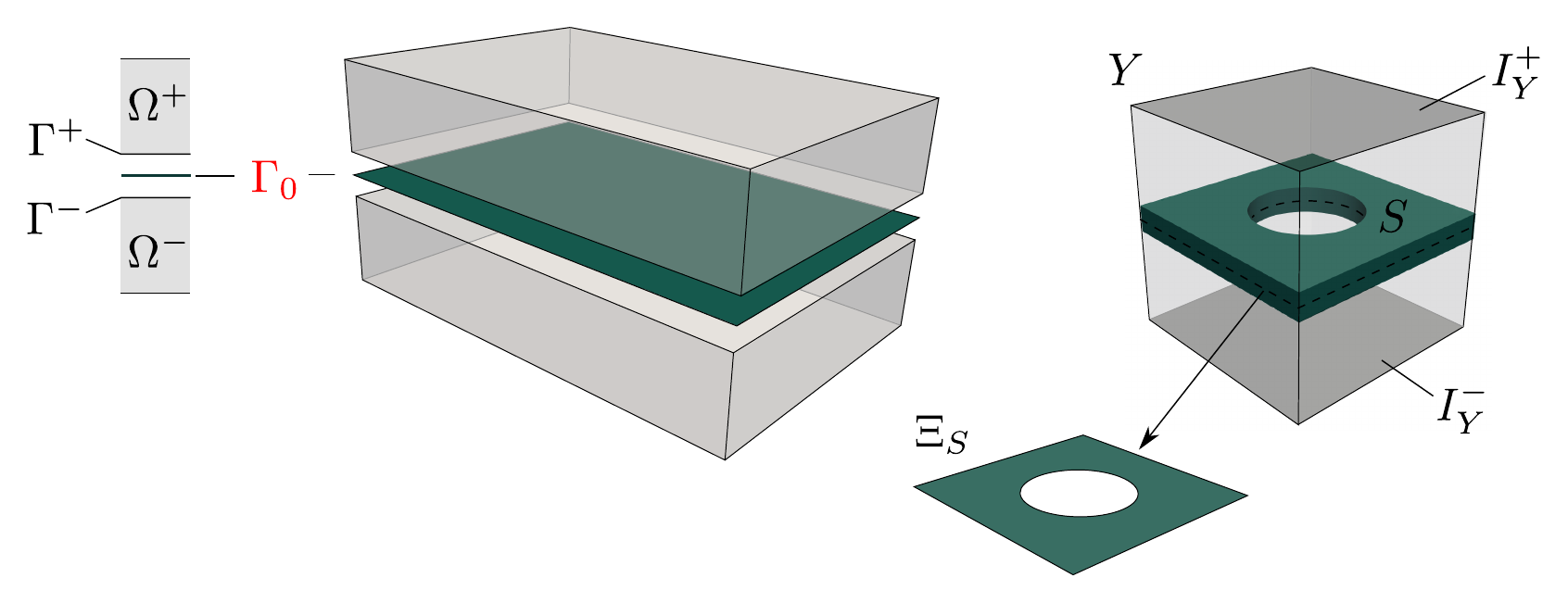}
  
\caption{Left: Transmission layer $\Om_\dlt$ represented by the homogenized interface $\Gamma_0$ at the macro-scale. Right: Representative 3D cell $Y$. The solid part $S$ corresponds to the perforated plate -- its 2D representation $\Xi_S$ within the cell.}\label{fig-pd2}
\end{figure}

In this section we introduce the global problem of the acoustic wave
propagation in a domain $\Om^G\subset \RR^3$ containing a periodically
perforated plate with distributed resonators, otherwise called the metamaterial
plate. We pursue the homogenization procedure proposed in
\cite{rohan-lukes-waves07} dealing with rigid periodically perforated plate,
and further elaborated for ``standard'' compliant plates in
\cite{Rohan-Lukes-AMC2019}. The homogenized vibro-acoustic transmission
conditions were derived using the asymptotic analysis \wrt a scale parameter
$\veps$ which describes the thickness of an elastic plate when considered as a
3D object $\Sigma^\veps \subset \Om^G$, and also the characteristic size of the
microstructure. For the reader's convenience, we recall the main steps of
deriving the transmission conditions for a limit global problem, as explained
in detail in our previous work \cite{Rohan-Lukes-AMC2019}. The global problem
is formulated in a domain $\Om^G\subset \RR^3$ in which the perforated elastic
plate of the Reissner-Mindlin type is embedded, being represented by its
perforated midsurface. Then a transmission layer $\Om_\dlt$ of the thickness
$\dlt$ is introduced via its midsurface $\Gamma_0$, see Fig.~\ref{fig-pd1} and
Fig.~\ref{fig-pd2}. The acoustic fluid occupies domain
$\Om^{*\veps}\subset\Om_\dlt$ and the associated acoustic field in $\Om_\dlt$
is coupled with the one in $\Om^G\setminus \Om_\dlt$ on the fictitious planar
surfaces $\Gamma_\dlt^\pm$. The homogenization procedure is applied to derive
an effective model of the vibro-acoustic interaction in the layer $\Om_\dlt$. In
our study, the layer thickness $\delta$ is proportional to the plate
heterogeneity period $\veps = \dlt / \vkappa$, with a given fixed $\vkappa >
0$, and also related to the plate thickness $h^\veps = \veps\bar{h}$ respected
when describing the fluid-structure interaction on the 3D plate surface. This
double role of $\veps$ is used in the asymptotic analysis the of the
vibro-acoustic problem.

Finally we record the result of \cite{Rohan-Lukes-AMC2019} concerning the limit
global problem for the acoustic waves in the fluid interacting with the
homogenized perforated plate represented by $\Gamma_0$. In this context, a
given plate thickness $h$ is related to a given finite thickness $\dlt_0$ of
the transmission layer, whereby the continuity of the acoustic fields on
interfaces $\Gamma_{\dlt_0}^\pm$ provides the homogenized vibro-acoustic
coupling conditions prescribed on $\Gamma_0$ representing the homogenized
transmission layer.

%
%

Knowing that parameters $\veps$ and $\dlt$ may vary only proportionally as
$\dlt = \vkappa \veps$ with a fixed $\vkappa>0$, in what follows we drop the
subscript $_\dlt$ when referring to geometrical objects depending on on both the
layer thickness and the periodic heterogeneity, thus, $\Sigma_\dlt^\veps \equiv
\Sigma^\veps$.

\subsection{Geometry of the perforated layer}

\begin{figure}
\centerline{\includegraphics[width=0.8\linewidth]{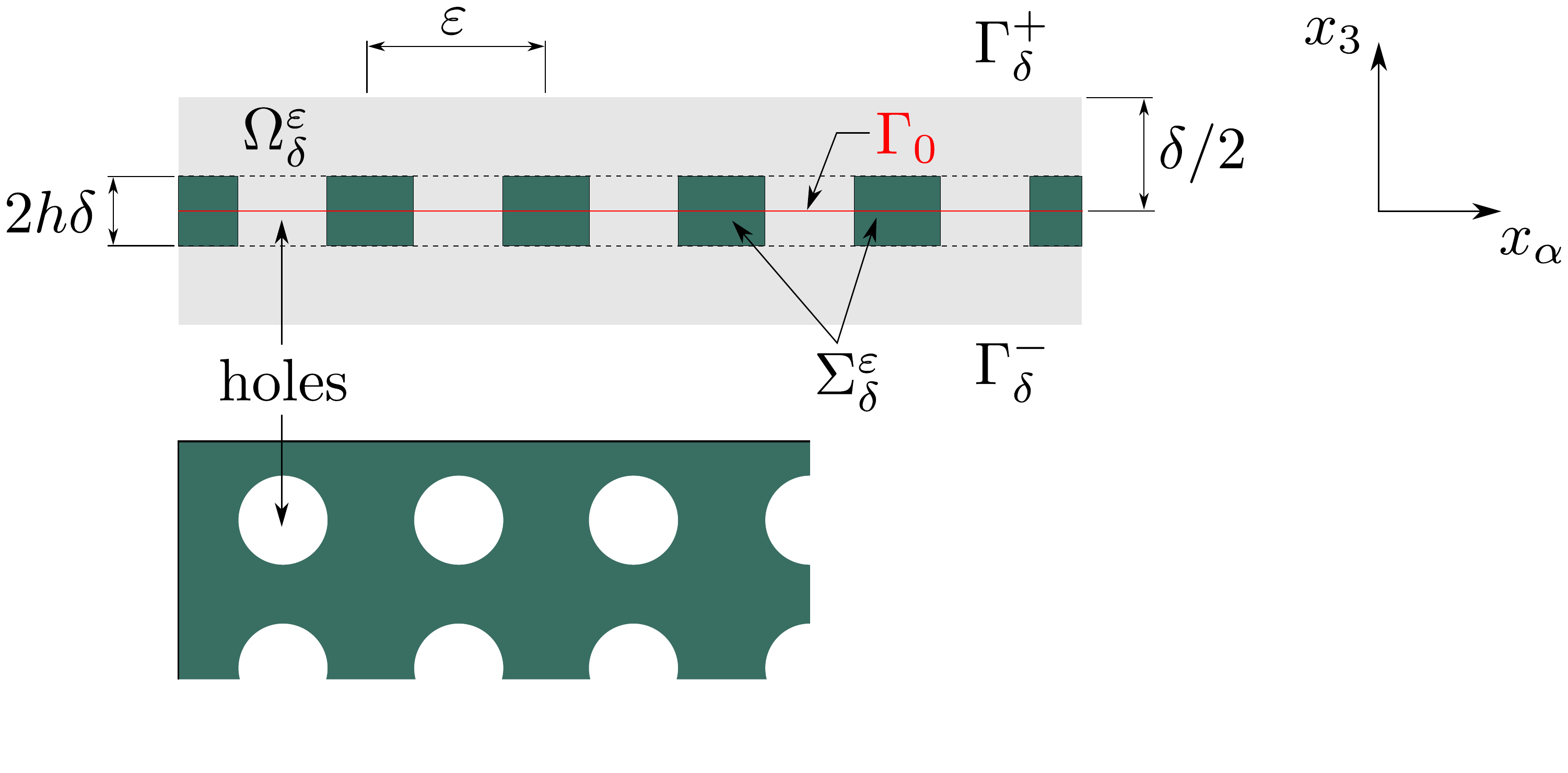}}
\caption{Scheme of the transmission layer $\Om_\dlt$ in which the perforated plate $\Sigma_\dlt^\veps$ (dark gray) is embedded. The complementary domain $\Om_\dlt^\veps$ is occupied by the acoustic fluid (light gray).}\label{fig-scheme-layer}
\end{figure}

Let  $\Gamma_0 \subset \{x \in \Om^G| x_3 = 0\}$ be a bounded 2D planar manifold
representing the plate midsurface.
We define $\Om_\delta = \Gamma_0 \times ]-\delta/2,\delta/2[ \subset \Om^G$, an open bounded domain representing the
transmission layer. This enables to decompose $\Om^G$ into three nonoverlapping
parts, as follows: $\Om^{G} =\Om_\dlt \cup \Om_\dlt^+ \cup \Om_\dlt^-$. Thus,
the transmission layer is bounded by $\pd \Om_\delta$ which splits into three disjoint
parts:
\begin{equation}\label{eq-1}
\begin{split}
\pd \Om_\delta = \Gamma_\delta^+ \cup  \Gamma_\delta^- \cup \pd_\ext \Om_\delta\;,
\quad \Gamma_\delta^\pm = \Gamma_0 \pm {\delta\over{2}} \vec{e_3}\;,
\quad \pd_\ext \Om_\delta = \pd \Gamma_0 \times  ]-\delta/2,\delta/2[\;,
\end{split}
\end{equation}
where $\delta > 0$ is the layer thickness and $\vec{e_3} = (0,0,1)$, see 
Fig.~\ref{fig-scheme-layer}. In the context of the transmission layer definition, we
consider the plate as a 3D domain $\Sigma^{\veps}$ defined in terms of the
perforated midsurface $\Gamma^{\veps}$; the following definitions are employed:
\begin{equation}\label{eq-hh}
\begin{split}
\Sigma^{\veps} & = \Gamma^{\veps}\times \veps \hh ]-1/2, +1/2[\;,\\
\pd\Sigma^{\veps} & = \pd_\circ\Sigma^{\veps} \cup \pd_+ \Sigma^{\veps} \cup  \pd_- \Sigma^{\veps} \cup \pd_\ub\Sigma^{\veps}\;,\\
\mbox{ where } & \\
\pd_\circ\Sigma^{\veps} & = \pd_\circ\Gamma^\veps\times \veps \hh ]-1/2, +1/2[ \;,\\
\pd_\pm \Sigma^{\veps} & = \Gamma^{\veps} \pm \veps \hh/2\;,
\end{split}
\end{equation}
where $\pd_\ub\Sigma^{\veps}$ is the surface where the plate is clamped.

The midsurface $\Gamma^{\veps}$ representing the perforated plate is generated
using a representative cell $\Xi_S \subset \RR^2$, as a periodic lattice. Let
$\Xi = {]0,\ell_1[}\times{]0,\ell_2[}$, where $\ell_1,\ell_2 > 0$ are given
(usually $\ell_1=\ell_2=1$) and consider the hole $\Xi^* \subset \Xi$, whereas
its complement $\Xi_S =\Xi\setminus \ol{\Xi^*}$ defines the solid plate
segment. Then
\begin{equation}\label{100}
\begin{split}
\Gamma^\veps & = \bigcup_{k \in \ZZ^2}  \veps\left(\Xi_K + \sum_{i=1,2}k_i\ell_i\vec{e_i} \right)\cap \Gamma_0 \;.
\end{split}
\end{equation}
Further we introduce the representative periodic cell $Y$ and define its solid part
$S \subset Y$, 
\begin{equation}\label{101}
\begin{split}
Y & = \Xi \times {]-\vkappa/2,+\vkappa/2]}\;,\\
S & = \Xi_S \times{\hh }  ]-1/2,+1/2[\;,
\end{split}
\end{equation}
so that $Y^* = Y\setminus \ol{S}$ is the fluid part. Obviously, in the
transmission layer $\Om_\dlt$, the fluid occupies the part
\begin{equation}\label{102}
\begin{split}
\Om^{*\veps} & = \bigcup_{k \in \ZZ^2}\veps(Y^* + \sum_{i=1,2}k_i\ell_i\vec{e_i})\cap \Om_\dlt\;,
\end{split}
\end{equation}
where $\vec{e_1} = (1,0,0)$ and $\vec{e_2} = (0,1,0)$.

For completeness, by virtue of \eq{eq-hh} we can introduce the decomposition of
boundary $\pd S = \pd_\circ S \cup \pd_\pm S \cup \pd_\# S$. For this we need
the boundary $\pd \Xi_S = \pd_\circ \Xi_S \cup \pd_\#\Xi_S$, where $\pd_\#\Xi_S
\equiv \pd \Xi$, so that the closed curve $\pd_\circ \Xi_S = \pd \Xi^*$
generates the cylindrical boundary $\pd_\circ S$:
\begin{equation}\label{103}
\begin{split}
\pd_\circ S & = \pd_\circ\Xi_S \times{\hh }  ]-1/2,+1/2[\;,\\
\pd_\pm S & = \Xi_S \pm \vec{e_3}\hh/2\;,\\
\pd_\# S & = \pd\Xi \times{\hh }  ]-1/2,+1/2[\;.
\end{split}
\end{equation}
For the sake of simplicity, by $\pd \Xi_S$ we shall refer to $\pd_\circ  \Xi_S$.

\begin{figure}
\centerline{\includegraphics[width=0.75\linewidth]{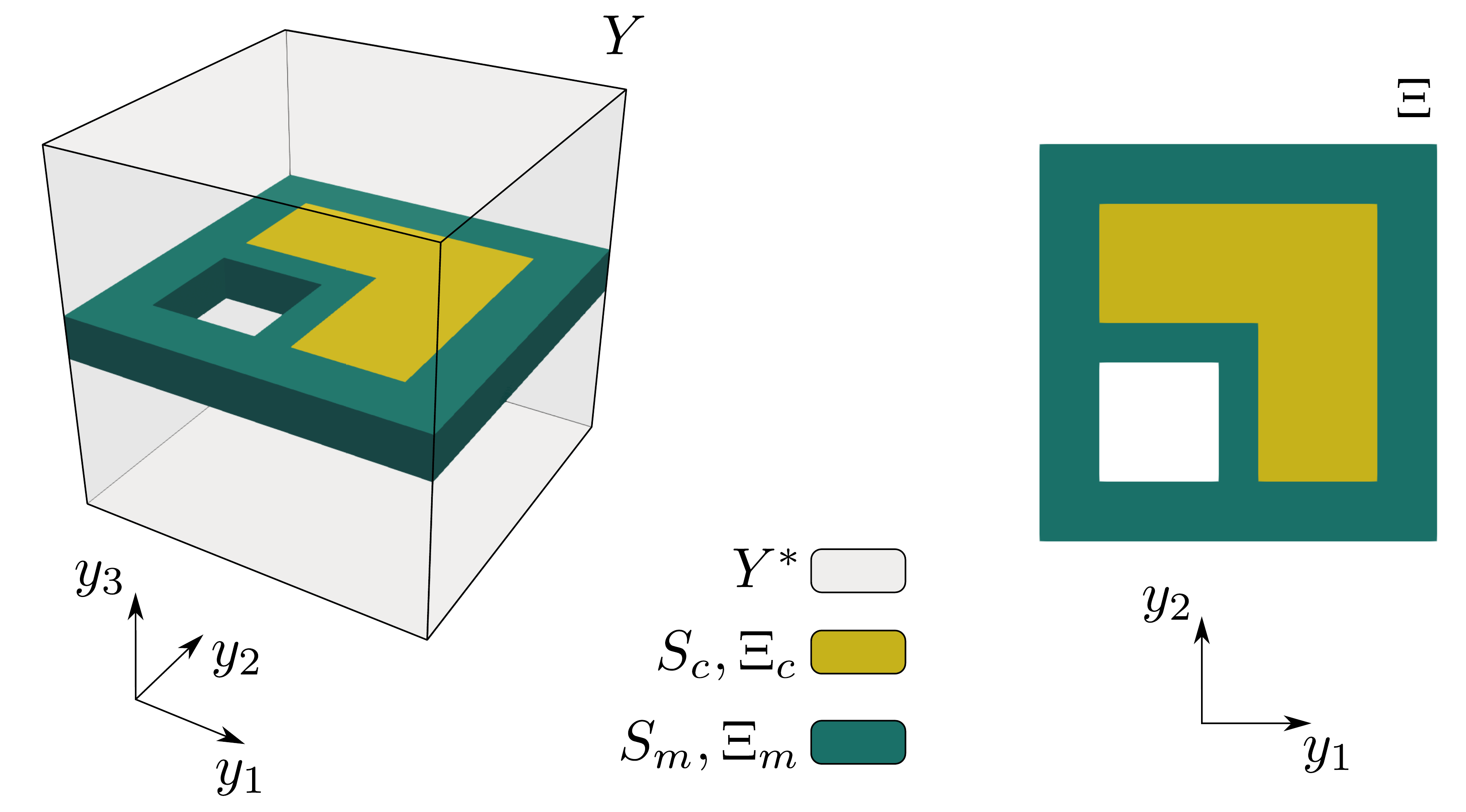}}
\caption{Representative cell $Y$ with the solid part $S$ and its plate representation $\Xi$.}\label{fig-cellY-Xi}
\end{figure}

\subsection{Vibro-acoustic problem in the transmission layer}

The acoustic harmonic wave with the frequency $\omega$ is described by the
acoustic potential $p^\veps:\Om^{*\veps}\ni x \mapsto\RR^3$ in the fluid, the
corresponding wave in the elastic body is described by the displacement field
$\ub^\veps:\Sigma^\veps\ni x \mapsto\RR^3$. The body is fixed to a rigid frame
on the boundary $\pd_\ub\Sigma$ and interacting with the fluid on
$\pd_*\Sigma^\veps = \pd\Sigma^\veps\setminus\pd_\ub\Sigma^\veps$.

Some further notation will be employed: by $c$ we denote the sound speed in the
acoustic fluid, $\rho_0$ is reference fluid density, $\sigmabf^\veps =
\Dop^\veps \eeb{\ub^\veps}$ denotes the stress in the elastic solid
characterized by the elasticity tensor $\Dop^\veps = (D_{ijkl}^\veps)$, and by
$\nb = (n_i)$ we denote the normal vector.

We shall now present the vibro-acoustic problem which will be subject of the
asymptotic analysis $\veps\rightarrow 0$.

Given $g^{\veps\pm}$ on $\Gamma_\dlt^\pm$, find $p^\veps$ in
$\Om^{*\veps}$ and $\ub^\veps$ in $\Sigma^\veps$, such that following equations hold:
wave equations for the fluid and the solid phases:
\begin{equation}\label{eq-G2-02a}
\begin{split}
c^2 \nabla^2 p^\veps + \om^2 p^\veps & = 0 \quad \mbox{ in } \Om^{*\veps} \;, \\
\nabla\cdot \sigmabf^\veps(\ub^\veps) + \om^2 \rho^\veps \ub^\veps & = 0 \quad \mbox{ in } \Sigma^\veps\;,
\end{split}
\end{equation}
fictitious interface conditions:
\begin{equation}\label{eq-G2-02b}
\begin{split}
\pdiff{p^\veps}{n} & = - \imu \om g^{\veps\pm} \quad \mbox{ on } \Gamma_\dlt^{\pm}\;,\\
\end{split}
\end{equation}
acoustic transmission:
\begin{equation}\label{eq-G2-02c}
  \begin{split}
    \left.
\begin{array}{rcl}
\imu \om \nb\cdot\ub^\veps & = & \nb\cdot\nabla p^\veps\\
\nb \cdot \sigmabf^\veps(\ub^\veps) & = & \bb(p^\veps)  = \imu \om \rho_0 p^\veps \nb
\end{array}
\right\}& \quad \mbox{ on } \pd_* \Sigma^\veps\;,
\end{split}
\end{equation}
other boundary conditions for the acoustic fluid:
\begin{equation}\label{eq-G2-02d}
\begin{split}
r \imu\om c p^\veps + c^2 \pdiff{p^\veps}{n} & = s 2\imu\om c\bar p
\quad  \mbox{ on } \pd\Om_\dlt \setminus (\pd_\ub\Sigma^\veps\cup \Gamma_\dlt^\pm)\;, \\
\end{split}
\end{equation}
 clamped elastic structure: 
\begin{equation}\label{eq-G2-02e}
\begin{split}
\ub^\veps & = 0 \quad  \mbox{ on } \pd_\ub\Sigma^\veps\;.
\end{split}
\end{equation}
where $\pd_* \Sigma^\veps = \pd\Sigma^\veps \cap \pd\Om^{*\veps}$ is the
surface of the elastic structure in contact with the fluid, thus,
$\pd\Sigma^\veps = \pd_* \Sigma^\veps \cup \pd_\ub\Sigma^\veps$. The constants
$r,s \in \{0,1\}$ and $\bar p$ are defined to describe incident, reflected, or
absorbed acoustic waves in the fluid, according to a selected part of the
boundary.

On the interfaces $\Gamma_\dlt^\pm$, the acoustic fluid velocity projected into
the normal $\imu\om g^{\veps\pm}$ is provisionally assumed to be prescribed; in
fact, $g^{\veps\pm}$ is an internal variable introduced when decomposing the
global acoustic field into the outer one $P^\dlt$ and the one defined in the
layer, denoted by $p^\veps$. In particular, one requires
\begin{equation}\label{eq-G2-03}
\begin{split}
\imu\om g^{\veps\pm} & = \pdiff{P^\dlt}{n^\pm}  \quad \mbox{ on } \Gamma_\dlt^\pm\;,\\
p^\veps & = \hat p^\veps = P^\dlt \quad \mbox{ on } \Gamma_\dlt^\pm\;,
\end{split}
\end{equation}
where $n^\pm$ refers to normals $\nb^\pm$ outer to domains $\Om_\dlt^\pm$.

\subsection{Plate model}

Following the treatment reported in \cite{Rohan-Lukes-AMC2019}, instead of a 3D
description of the plate, we consider the thin elastic structure being
approximated by the Reissner-Mindlin plate model, thus representing the solid
structure by an extended 2D continuum respecting shear stresses induced by
rotations of the plate cross-sections \wrt the mid-plane.

The plate model can be obtained by the asymptotic analysis of the
corresponding 3D elastic structure while its thickness tends to zero. 
However, the obtained limit plate model is then interpreted in terms of a given
plate thickness  $h^\veps$, being represented by its perforated mean surface $\Gamma^\veps$.
Boundary $\pd_\circ\Gamma^\veps = \pd\Gamma^\veps\setminus\pd_\ext\Gamma^\veps$
describes the perforations distributed periodically with the spatial period $\veps$.
Therefore, all involved variables depend on $\veps$. 
The plate deflections are described by amplitude of the membrane elastic wave
$\ub^\veps= (u_1^\veps,u_2^\veps)$, of the transverse wave $u_3^\veps$ and of
the rotation wave $\thetabf^\veps= (\theta_1^\veps,\theta_2^\veps)$. Two linear
constitutive laws are involved, which depend upon the {second order} tensor
$\Sb^\veps=(S_{ij}^\veps) = \shp \delta_{ij}$, where $\shp>0$ is the shear
coefficient, and the {fourth order} elasticity tensor
$\Eop^\veps=(E_{ijkl}^\veps)$ which is given by the Hooke's law with tensor
$\Dop^\veps$ describing the 3D elasticity model, but here adapted for the plane
stress constraint. We recall that all indices $i,j,k,l = 1,2$.

The forces $\fb = (\ol{\fb},f_3)$ and $\ol{\fb}^\pd$ applied in the plate
volume and on its boundary, as well as the applied moments $\ol{\mb}$ and
$\ol{\mb}^\pd$, depend on the acoustic potential $p^\veps$. The crucial step in
deriving the model of vibro-acoustic transmission consists in describing these
forces and moments in terms of $p^\veps$ imposed on surface $\pd\Sigma^\veps$
in the 3D plate representation.

The acoustic transmission layer and the plate thicknesses are employed in two
contexts. Firstly, the periodically perforated plate model is defined in terms
of the 2D domain $\Gamma^\veps \subset \Gamma_0$ representing the mid-plane,
and the thickness $h = \veps_0 \hh$ with a fixed $\veps_0>0$.
Secondly, the interaction between the 3D elastic structure and the acoustic
fluid is described in terms of the plate surface whose the thickness must be
proportional to $\veps$. 
Therefore, we consider $h^\veps = \veps\hh$ and the elastic body occupying
domain $\Sigma^{\veps}$, see \eq{eq-hh}.

The homogenization of the periodically perforated plate is done by pursuing the
asymptotic analysis $\veps\rightarrow 0$ applied to the plate model which will
be presented below. While $h$ is fixed in the plate equation operator, being
independent of $\veps$, at the \rhs terms we get $1/(\veps\hh)$ which is
coherent with the dilation operation applied when dealing with fluid equation,
see Section~\ref{sec-dil}.

\subsection{Forces and moments acting  on the plate surface $\pd\Sigma^\veps$}

Due to the fluid-structure interaction, the forces and moments involved in the
equations introduced below can be identified using the 3D representation of the
plate surface $\pd_*\Sigma^\veps$ decomposed according to \eq{eq-hh}. The
actual surface traction $b_i^\veps = \imu\om\rho_0 n_i p^\veps$ is given by the
acoustic potential and by the surface normal $\nb = (n_i)$; note that $n_\alpha
= 0$, $\alpha = 1,2$ on $\pd_\pm\Sigma^\veps$, whereas $n_3 = 0$ on
$\pd_\circ\Sigma^\veps$. Hence, loading the plate on its surface by the
acoustic pressure yields the following expressions of forces and moments
coupled with the reduced plate degrees of freedom,
\begin{equation}\label{eq-fb}
\begin{split}
\ol{f}_\alpha^\veps & = 0\;,\quad \ol{m}_\alpha^\veps  = 0\;,\quad f_3^\pd  = 0\;,\\
\ol{f}_3^\veps & = \sum_{s = +,-}b_3(x',s \veps\hh/2) = \imu\om\rho_0 (p^\veps(x',\veps\hh/2)- p^\veps(x',-\veps\hh/2))\;,\\
{f}_\alpha^{\pd,\veps} & =  \int_{-h^\veps/2}^{h^\veps/2}b_\alpha(x',x_3)\dd x_3
 = \imu\om\rho_0 \int_{-h^\veps/2}^{h^\veps/2} n_\alpha(x') p^\veps(x',x_3)\dd x_3
\;,\\
{m}_\alpha^{\pd,\veps} & = -\int_{-h^\veps/2}^{h^\veps/2}x_3 b_\alpha(x',x_3)\dd x_3= -\imu\om\rho_0\int_{-h^\veps/2}^{h^\veps/2}x_3 n_\alpha(x') p^\veps(x',x_3)\dd x_3 \;.
\end{split}
\end{equation}
Note that $\ol{f}_\alpha^\veps$ and $\ol{m}_\alpha^\veps$ express the volume
forces which are disregarded, whereas $f_3^\pd = 0$ is the consequence of the
inviscid acoustic fluid.

\subsection{Fluid structure interaction on the plate surface $\pd\Sigma^\veps$}

The plate displacements $\wb^\veps$ defined on the surface
$\pd\Sigma^{\veps} $ are expressed using the mid-plane kinematic fields. It holds that
\begin{equation}\label{eq-wf4}
\begin{split}
\wb^\veps(x',x_3) &  = ( w_k^\veps(x',x_3))\;,\\
w_\alpha^\veps(x',x_3) & = u_\alpha^\veps(x') - x_3 \theta_\alpha^\veps(x')\;,\quad\alpha = 1,2 \;,\\
w_3^\veps(x',x_3) & = u_3^\veps(x')\;,
\end{split}
\end{equation}
where $x' \in \Gamma^\veps$, $x_3 \in \veps \hh ]-1/2,1/2[$. In analogy, the
test displacements $\tilde w_k^\veps(x',x_3)$ $k=1,2,3$ can be introduced in terms
of the test functions $(\vb^\veps,\psibf^\veps)$ involved in the weak formulation of the vibro-acoustic problem. It incorporates
the virtual power 
\begin{equation}\label{eq-fba}
\begin{split}
\int_{\pd_*\Sigma^\veps} \bb^\veps\cdot \tilde \wb^\veps
\end{split}
\end{equation}
of the external forces $\bb^\veps = \imu\om\rho_0 \nb p^\veps$ acting on the plate


In the equation associated with the acoustics in the fluid, the kinematic
condition in \eq{eq-G2-02c}$_1$ prescribed on $\pd\Sigma^\veps$, involves the
displacement field $\wb$ which must be expressed in terms of the mid-plane
displacements and rotations $\ub^\veps$ and $\thetabf^\veps$, as introduced in
\eq{eq-wf4}. This yields (recall the jump $\Jump{q(\cdot,x_3)}{\veps \hh} =
q(\cdot,\veps \hh/2) - q(\cdot,-\veps \hh/2)$)
\begin{equation}\label{eq-wf5}
\begin{split}
\int_{\pd\Sigma^\veps} \nb\cdot\wb^\veps q^\veps = 
\int_{\Gamma^\veps} u_3^\veps \Jump{q^\veps(\cdot,x_3)}{\veps \hh} + \int_{\pd\Gamma^\veps}
\veps\hh \int_{-1/2}^{1/2} \bar \nb \cdot (\ub^\veps - \veps\hh \zeta \thetabf^\veps) q^\veps(\cdot,\veps\hh \zeta) \dd \zeta\;.
\end{split}
\end{equation}

\subsection{Variational formulation of the vibro-acoustic problem in the layer}

In order to derive the homogenized model of the transmission layer, we shall
need the variational formulation of problem \eq{eq-G2-02a}-\eq{eq-G2-02d} with
the fluid-structure interaction terms \eq{eq-fba}-\eq{eq-wf5} involving forces
and moments \eq{eq-fb}.

Find $p^\veps \in H^1(\Om^{*\veps})$ and $(\ub^\veps,\thetabf^\veps) \in
(H^1_0(\Om))^5$ such that
\begin{equation}\label{eq-va6}
\begin{split}
c^2 \int_{\Om^{*\veps}} \nabla p^\veps \cdot \nabla q^\veps
- \om^2 \int_{\Om^{*\veps}} p^\veps q & =
-\imu \om c^2\left ( \int_{\Gamma^{\pm\veps}} g^{\veps\pm} q^\veps\,d\Gamma
+ \int_{\pd \Sigma^{\veps}} \nb\cdot\wb^\veps q^\veps\,d\Gamma \right ),
\end{split}
\end{equation}
for all $q \in H^1(\Om^{*\veps})$, where $\nb$ is outward normal to
domain $\Sigma^\veps$, and 
%
\begin{equation}\label{eq-va7}
\begin{split}
& \om^2 h \int_{\Gamma^\veps} \rho \ub^\veps \cdot \vb^\veps 
  \om^2 \frac{h^3}{12} \int_{\Gamma^\veps} \rho\thetabf^\veps\cdot \psibf^\veps
  -h \int_{\Gamma^\veps} [\Eop^\veps \gradplS \ol{\ub}^\veps ]: \gradplS \ol{\vb}^\veps\\
&-h \int_{\Gamma^\veps} [\Sb^\veps (\gradpl u_3^\veps - \thetabf^\veps)]\cdot
(\gradpl v_3^\veps - \psibf^\veps) - \frac{h^3}{12}
\int_{\Gamma^\veps} [\Eop^\veps \gradplS  \thetabf^\veps]: \gradplS \psibf^\veps\\
 = &
 \int_{\Gamma^\veps}   \fb^\veps(p^\veps)\cdot\vb^\veps  +
\int_{\Gamma^\veps} \ol{\mb}^\veps(p^\veps)\cdot \psibf^\veps
+\int_{\pd_\circ\Gamma^\veps}   \ol{\fb}^{\pd,\veps}(p^\veps)\cdot\ol{\vb}^\veps 
+\int_{\pd_\circ\Gamma^\veps}   \ol{\mb}^{\pd,\veps}(p^\veps)\cdot\psibf^\veps
\;,
\end{split}
\end{equation}
for all test functions $(\vb^\veps,\psibf^\veps)\in (H^1_0(\Om))^5$. In
\eq{eq-va6}, the displacements $\wb^\veps$ defined on the surface
$\pd\Sigma^{\veps} $ are expressed using \eq{eq-wf5}.

\subsection{Dilated formulation}\label{sec-dil}

We can now state the vibro-acoustic problem in the dilated layer $\hat\Om =
\Gamma_0 \times ]-\vkappa/2, +\vkappa/2[$, where the fluid occupies domain
$\hat\Om^* = \{(x',\veps^{-1}x_3) \in \RR^3|x\in\Om^{*\veps}\}$, see \eq{102}.
Upon introducing coordinates $(x',z)$ in the dilated configuration, whereby $z
= \veps^{-1}x_3$ and $x' = (x_\alpha)$, the gradients are $\nablad =
(\pd_\alpha,\veps^{-1}\pd_z)$.

Due to the transformations consisting of the dilation and the periodic
unfolding, the vibro-acoustic problem can be reformulated in the domain which
does not change with $\veps$. Consequently, the standard means of convergence
can be used to obtain the limit model. In the dilated configuration, we keep
the same notation for all functions depending on $x_3$ and, thereby, on $z$, to
simplify the notation. Thus, $\nabla p(x) = \nablad p(x_\alpha,z)$.

Equation \eq{eq-va6} with the substitution \eq{eq-wf5} can now be transformed
by the dilatation which yields
\begin{equation}\label{eq-wf6}
\begin{split}
\int_{\hat\Om^\veps} \hat\nabla p^\veps \cdot \hat\nabla q^\veps - \frac{\om^2}{c^2} \int_{\hat\Om^\veps} p^\veps q^\veps 
 = -\frac{\imu \om}{ \veps} 
\int_{\Gamma^\pm} \hat g^{\veps\pm} q^\veps \\
 -\frac{\imu \om}{ \veps} \left[\int_{\Gamma^\veps} u_3^{\veps} \Jump{q^\veps(\cdot,x_3)}{\veps \hh}
+ \int_{\pd_\circ\Gamma^\veps}
\veps\hh \int_{-1/2}^{1/2} \bar \nb \cdot (\ol{\ub}^{\veps} - \veps\hh \zeta \thetabf^{\veps}) q^\veps(\cdot, \veps\hh \zeta) \dd \zeta
\right]\;.
\end{split}
\end{equation}
Further we employ \eq{eq-fb} to rewrite \eq{eq-va7} which is divided by
$h^\veps$; the plate thickness is given for a fixed size of the
heterogeneities, \ie $h = \veps_0\hh$. However, when dealing with the \rhs
interaction terms, $h := h^\veps = \veps\hh$ in accordance with the dilation
transformation. Thus, we get the plate equation in the following form
%
\begin{equation}\label{eq-wp7} 
\begin{split}
& \om^2 \int_{\Gamma^\veps} {\rho} \ub^\veps \cdot \vb^\veps +
\om^2 \frac{h^2}{12} \int_{\Gamma^\veps} {\rho}\thetabf^\veps\cdot \psibf^\veps\\
& - \int_{\Gamma^\veps} [\Eop^\veps \gradplS \ol{\ub}^\veps ]: \gradplS \ol{\vb}^\veps
-\int_{\Gamma^\veps} [\Sb^\veps (\gradpl u_3^\veps - \thetabf^\veps)]\cdot
(\gradpl v_3^\veps - \psibf^\veps) - \frac{h^2}{12}
\int_{\Gamma^\veps} [\Eop^\veps \gradplS  \thetabf^\veps]: \gradplS \psibf^\veps\\
 = &
\frac{\imu \om\rho_0}{\veps\hh} \left[
\int_{\Gamma^\veps} v_3^\veps  \Jump{p^\veps(\cdot,x_3)}{\veps \hh}
+ \veps\hh \int_{\pd_\circ\Gamma^\veps}\int_{-1/2}^{1/2}p^\veps(\cdot, \veps\hh \zeta) \bar\nb \cdot (\ol{\vb}^\veps - \veps\hh\zeta \psibf^\veps)\dd \zeta
\right]
\;.
\end{split}
\end{equation}
It is worth noting that, in \eq{eq-wf6} and \eq{eq-wp7}, the \rhs integrals
provide a symmetry of the following formulation.
\paragraph{The vibro-acoustic problem formulation}
The acoustic response in the dilated layer $\hat\Om_\dlt$ is described by
$(p^\veps,\ub^\veps,\thetabf^\veps) \in H^1(\hat\Om^{*\veps})\times
(H_0^1(\Gamma^\veps))^5$ which satisfy equations \eq{eq-wf6}-\eq{eq-wp7} for
any test fields $(q^\veps,\vb^\veps,\psibf^\veps) \in H^1(\hat\Om^{*\veps})\times (H_0^1(\Gamma^\veps))^5$. 
The momentum fluxes depending on $\veps$ are assumed to be given in the following form

\begin{equation}\label{eq-aes2}
\begin{split}
\hat g^{\veps+}(x') & = g^{0}(x')  + \veps g^{1+}(x',\frac{x'}{\veps}) \;,\\
\hat g^{\veps-}(x')  & = -g^{0}(x')  - \veps g^{1-}(x',\frac{x'}{\veps})\;,
\end{split}
\end{equation}
where $g^{0} \in L^2(\Gamma_0)$ and $g^{1\pm}(x',y') \in L^2(\Gamma_0 \times
\RR^2)$, whereby $g^{1\pm}(x',\cdot)$ being $\Xi$-periodic in the second
variable.

%
For any $\veps > 0$ and $\hat g^{\veps\pm}$ defined according to \eq{eq-aes2},
the vibro-acoustic interaction problem constituted by equations
\eq{eq-wf6}-\eq{eq-wp7} possesses a unique solution
$(p^\veps,\ub^\veps,\thetabf^\veps)$. To prove its existence and uniqueness,
the \emph{a~priori} estimates must be derived in analogy with the treatment
reported in \cite{Rohan-Lukes-AMC2019}, Appendix A, whereby the results of
\cite{rohan-miara-ZAMM2015} dealing with the strongly heterogeneous
Reissner-Mindlin plates must be employed, to adapt to the strong heterogeneity
of the elastic properties introduced below in \eq{eq-elast1}.

\section{Homogenization of the transmission layer}\label{sec-hom}

In this section we present the two-scale limit of the vibro-acoustic problem
\eq{eq-wf6}-\eq{eq-wp7}. For this, in terms of the reference cell $\Xi$, we
first introduce a strongly heterogeneous elasticity of the perforated plate.
Accordingly, the asymptotic expansions are defined which enable to pass to the
limit $\veps\rightarrow 0$ in all integrals of the variational formulation
\eq{eq-wf6}-\eq{eq-wp7} processed by the unfolding transformation, see \eg
\cite{Cioranescu2008-Neumann-sieve}. We employ the unfolding operator
$\Tuftxt{}: L^2(\Om_\delta;\RR) \rightarrow L^2(\Gamma_0 \times Y;\RR)$ which
transforms a function $f(x)$ defined in $\Om_\delta$ into a function of two
variables, $x' \in \Gamma_0$ and $y \in Y$. For any $f \in L^1(Y)$, the cell
average involved in all unfolding integration formulae will be abbreviated by
\begin{equation}\label{eq-uf4}
\begin{split}
\frac{1}{|\Xi|}\int_\Xi f= \intY_\Xi f\;\quad \frac{1}{|\Xi|}\int_{D} f =:  \intY_D f\;,
\end{split}
\end{equation}
whatever the domain $D \subset \ol{Y}$ of the the integral is (\ie volume, or surface).

\subsection{Heterogeneous perforated plate in the transmission layer}

The plate with a periodic structure is represented by the cell $\Xi \in \RR^2$
consisting of its solid part $\Xi_S$ and the hole $\Xi^*$, see
Fig.~\ref{fig-cellY-Xi},
\begin{equation}\label{eq-geom1}
\begin{split}
  \Xi_S & = \Xi_m \cup \Xi_c \cup \pd \Xi_c\;,\quad \ol{\Xi_c} \cap \ol{\Xi^*} = \emptyset\;,\\
  S & = \Xi_S \times I_\hh\;,\\
  S_d & = \Xi_d  \times I_\hh\;,\; d = m,c\;,
\end{split}
\end{equation}
where $I_\hh = {\hh }  ]-1/2,+1/2[$. Note that $\pd\Xi_c \cap \pd\Xi^* = \emptyset$.

According to this split of $\Xi_S$ the material stiffnesses and the densities of the plate are defined,
\begin{equation}\label{eq-elast1}
\begin{split}
  \Eop^\veps(x) & = \chi_m^\veps(x) \Eop_m + \veps^2 \chi_c^\veps(x) \Eop_c\;,\\
  \Sb^\veps(x) & = \chi_m^\veps(x) \Sb_m + \veps^2 \chi_c^\veps(x) \Sb_c\;,\\
   \rho^\veps(x) & = \chi_m^\veps(x) \rho_m + \chi_c^\veps(x) \rho_c\;.
\end{split}
\end{equation}
Thus, the contrast in the elasticity is assumed, whereas the density can vary
only moderately. Let us recall that this scaling ansatz has been used when
analysing stop bands of the wave propagation in 3D elastic structures
\cite{Miara-etal-mms,rohan-IJES} and plates
\cite{rohan-miara-ZAMM2015,Rohan2015bg-plates}.

\subsection{Limit two-scale equations of the transmission layer}

Although the rigorous convergence results can be obtained, in this paper, we
derive the limit model of the transmission layer using the formal asymptotic
expansion method. The following truncated expansions
$(p^\veps,\ub^\veps,\thetabf^\veps)$ defined in terms of unfolded fields are
consistent with the convergence results supported by the a~priori estimates,
\begin{equation}\label{eq-va13a}
\begin{split}
\Tuf{p^\veps} & = p^0(x') + \veps p^1(x',y)\;, \\
\Tuf{\ol{\ub}^\veps} & =
\ol{\ub}^0(x') + \veps\ol{\ub}^1(x',y')+ \chi_c \hat{\ol{\ub}}(x',y')\;,\\
\Tuf{u_3^\veps} & =
u_3^0(x') + \veps u_3^1(x',y') + \chi_c \hat u_3(x',y')\;,\\
\Tuf{\thetabf^\veps} & = \thetabf^0(x') + \veps \thetabf^1(x',y') + \chi_c \hat\thetabf(x',y')\;,
\end{split}
\end{equation}
where $x' \in \Gamma_0$, $ y' \in \Xi$. In~\eq{eq-va13a}, all the two-scale
functions are $\Xi$-periodic in the second variable $y'$, or $y$ in the case of
$p$; recall that the couple $y = (y',z) \in Y$ determines positions in the
reference cell $Y$.


We now present the limit coupled system of the plate and the acoustic fluid
which governs the acoustic pressure and the plate deflections and rotations,
see \Appx{apx-1} for details. The response to the transverse acoustic momentum
represented by $g^0$ and $\Delta g^{1}:=g^{1+}-g^{1-}$ involves the macroscopic
fields $(p^0,\ub^0,\thetabf^0)$ and the two-scale functions
$(p^1,\ub^1,\hat\ub,\thetabf^1,\hat\thetabf)$. For the assumed boundary
conditions describing a clamped plate, the following admissibility sets are
introduced,
\begin{equation}\label{eq-p1a}
  \begin{split}
    \VV^0(\Gamma_0) & = \{(\vb,\thetabf)| \bar\vb,\thetabf \in \HOdb(\Gamma_0),\;v_3 \in H_0^1(\Gamma_0)\} = [H_0^1(\Gamma_0)]^5\;,\\
    \VV^1(\Gamma_0,\Xi_S) & = \{\qb = (\vb,\thetabf) \in L_2(\Gamma_0)\times [H_\#^1(\Xi_S)]^5,\; \int_{\Xi_S} q_i = 0,\; i = 1,\dots,5\}\;\\
    \hat\VV(\Gamma_0,\Xi_S) & = \{\hat\qb = (\hat\vb,\hat\thetabf) \in L_2(\Gamma_0)\times [H^1(\Xi_S)]^5,\; \supp \hat\qb = \Xi_c\}\;.
\end{split}
\end{equation}

The plate equation holds for all test fields $(\vb^0,\psibf^0) \in
\VV^0(\Gamma_0)$, $(\vb^1,\psibf^1) \in \VV^1(\Gamma_0,\Xi_S)$ and
$(\hat\vb,\hat\psibf) \in \hat\VV(\Gamma_0,\Xi_S)$,
\begin{equation}\label{eq-p1}
\begin{split}
& -\om^2 \int_{\Gamma_0}\intY_{\Xi_S}\intY_{\Xi_S} \rho_S \left(( \ub^0 + \chi_c \hat u_3)\cdot(\vb^0 + \chi_c \hat v_3) + 
  \frac{h^2}{12} (\thetabf^0  + \chi_c\hat\thetabf)\cdot(\psibf^0 + \chi_c\hat\psibf)\right) \\
  &  \int_{\Gamma_0}\left(\Pcal_m((\ub^0,\ub^1,\thetabf^0,\thetabf^1),(\vb^0,\vb^1,\psibf^0,\psibf^1)) + \Pcal_c((\hat\ub,\hat\thetabf),(\hat\vb,\hat\psibf))\right) \\
& = 
\frac{\imu\om\rho_0}{\hh}\int_{\Gamma_0}\left(v_3^0\intY_{\Xi_S}\Jump{p^1}{\hh}
  +\intY_{\Xi_c}\hat v_3 \Jump{p^1}{\hh}\right) \\
&  + \imu\om\rho_0 \int_{\Gamma_0}\left(\ol{\vb}^0 \cdot \int_{-1/2}^{1/2}\intY_{\pd\Xi_S}\ol{\nb} p^1 + p^0 \intY_{\pd\Xi_S}\ol{\vb}^1\cdot \ol{\nb}\right),
\end{split}
\end{equation}
 where, at the \lhs, the abstract notation $\Pcal_m$ and $\Pcal_c$ introduced in \eq{eq-wp7-L2} represent operators associated with the plate elasticity.
The fluid is governed by the following equation to hold for all $q^0 \in H^1(\Gamma_0)$ and $q^1 \in H_\#^1(Y^*)$,
\begin{equation}\label{eq-f6}
\begin{split}
& c^2 \int_{\Gamma_0} \intY_{Y^*} (\gradplx p^0 + \gradply p^1)\cdot
(\gradplx q^0 + \gradply q^1) + c^2 \int_{\Gamma_0} \intY_{Y^*}\pd_z p^1 \pd_z q^1 - \om^2 \int_{\Gamma_0} \intY_{Y^*} p^0 q^0\\
& = 
-\imu\om c^2 \int_{\Gamma_0} \left[
q^0 \intY_\Xi \Delta g^{1} + g^0\left(\intY_{I_y^+}q^1 -  \intY_{I_y^-}q^1\right)\right.\\
& \left. + u_3^0\intY_{\Xi_S} \Jump{q^1}{\hh}
+ \ol{\ub}^0\cdot\intY_{\pd\Xi_S}\ol{\nb}\hh\int_{-1/2}^{1/2}q^1 \dd\zeta + 
q^0 \hh \intY_{\pd\Xi_S}\ol{\nb}\cdot\ol{\ub}^1+ \intY_{\Xi_c}  \hat u_3 \Jump{q^1}{\hh}\right]\;,
\end{split}
\end{equation}
recalling the abbreviation $\Delta g^{1}=g^{1+}-g^{1-}$.

\subsection{Local problems}
We consider a fix position $x' \in \Gamma_0$. The local problems are identified
in the limit equations \eq{eq-p1}-\eq{eq-f6} for vanishing macroscopic test
fields, \ie upon substituting there $\vb^0 \equiv \zerobm, \psibf^0 \equiv
\zerobm$ and $q^0 \equiv 0$. In doing so, the system is decomposed into two
subsystems which can be solved independently for a given macroscopic responses.
Furthermore, due to the linearity of the obtained equations for unknowns
comprising macroscopic fields and two-scale fields, the latter fields can be
decomposed using a multiplicative split into the macroscopic variables and the
characteristic responses.

\subsubsection{Characteristic responses of the fluid and soft inclusions in the plate }

These characteristic responses describe dynamic properties of the two-phase
heterogeneous structure where the soft inclusions constitute a kind of
resonators. The first group of autonomous local problems imposed in $Y^*\times
\Xi_c$ for $(p^1(x',\cdot),u_3(x',\cdot)) \in W_\plper$, where $W_\plper =
H_0^1(\Xi_c)\times H_\plper^1(Y^*)$, is extracted from \eq{eq-p1}-\eq{eq-f6}
while vanishing macroscopic test functions and all two-scale test function
except of $q^1$ and $\hat v_3$,
\begin{equation}\label{eq-Lp1}
\begin{split}
  & \intY_{Y^*}\nablad_y p^1 \cdot \nablad_y q^1 + \imu\om\intY_{\Xi_c}  \hat u_3 \Jump{q^1}{\hh}
  = - \intY_{Y^*}\nablad_y q^1 \cdot \nablad_y y_\alpha \pd_\alpha^x p^0\\
 &  - \imu\om \left[ g^0\left(\intY_{I_y^+}q^1 -  \intY_{I_y^-}q^1\right) +  u_3^0\intY_{\Xi_S} \Jump{q^1}{\hh}+ \ol{\ub}^0\cdot\intY_{\pd\Xi_S}\ol{\nb}\hh\int_{-1/2}^{1/2}q^1 \dd\zeta\right]\;,\\
 & \intY_{\Xi_c}\left( [\Sb_c \gradpl_y \hat u_3] \cdot\gradpl_y \hat v_3 - \om^2 \rho_c \hat u_3 \hat v_3 \right) - \frac{\imu\om\rho_0}{\hh}\intY_{\Xi_c}\hat v_3\Jump{p^1}{\hh}
  = \om^2 u_3^0\intY_{\Xi_c}\rho_c \hat v_3\;,
\end{split}
\end{equation}
for all $(q^1,\hat v_3) \in W_\plper$. 
By virtue of the linearity of \eq{eq-Lp1}, $(p^1,\hat u_3)$ can be expressed using the linear combinations of the macroscopic variables $(\pd_\beta^x p^0,g^0,\ub^0)$ and the characteristic response functions $\pi^\beta,\xi$ and $\eta^k$, $\beta = 1,2$, $k = 1,2,3$, 
\begin{equation}\label{eq-f7}
\begin{split}
p^1(x',y) &= \pi^\beta(y)\pd_\beta^x p^0(x') + \imu\om \xi(y) g^0(x') +  \imu\om \eta^k(y) u_k^0(x')\;,\\
\hat u_3(x',y) &= \hat w^\beta(y)\pd_\beta^x p^0(x') + \imu\om \hat\vsigma(y) g^0(x') +  \imu\om \hat \vpi^k(y) u_k^0(x')\;.
\end{split}
\end{equation}
Using the following inner products and bilinear forms,
\begin{equation}\label{eq-lp1b}
\begin{split}
\bhYc{\hat{w}}{\hat{z}} & = \intY_{\Xi_c} [\Sb_c \nabla_y \hat  w] \cdot\nabla_y \hat z\;,\\
\ipXic{\ub}{\vb} & = \intY_{\Xi_c} \ub\cdot \vb\;,\\
\ipYs{\ub}{\vb} & = \intY_{Y^*} \ub\cdot \vb\;,\\
\end{split}
\end{equation}
we define operator $\TT$, which is employed in the problems for the characteristic responses,
\begin{equation}\label{eq-lp2}
\begin{split}
  \ippYc{\TT(\hat w,\hat p)}{(\hat v,\hat q)} & = \frac{\hh}{\rho_0} \left[ \bhYc{\hat w}{\hat v} - \om^2 \ipXic{\rho \hat w}{\hat v}\right]\\
  & \quad - \imu\om \ipXic{\hat v}{ \Jump{\hat p}{\hh}} + \imu\om \ipXic{\hat w}{ \Jump{\hat q}{\hh}} + \ipYs{\nabla_y \hat p}{\nabla_y \hat q}\;.
\end{split}
\end{equation}
We can now introduce the local problems related to the fluid and fluid-solid
interaction which enable to compute the characteristic responses involved in
\eq{eq-f7}.


\begin{enumerate}
\item Find $(\hat w^\beta,\pi^\beta) \in W_\plper$, such that
\begin{equation}\label{eq-lp3}
  \begin{split}
    \ippYc{\TT(\hat w^\beta,\pi^\beta)}{(\hat v,q)} & = - \ipYs{\gradply y_\beta}{\gradply q}\;,\quad \forall (\hat v,q) \in  W_\plper\;.
\end{split}
\end{equation}
\item Find $(\hat \vsigma,\xi) \in W_\plper$, such that
\begin{equation}\label{eq-lp4}
  \begin{split}
    \ippYc{\TT(\hat \vsigma,\xi)}{(\hat v,q)} & = -\left(\intY_{I_y^+} q
- \intY_{I_y^-} q\right) \;,\quad \forall (\hat v,q) \in  W_\plper\;.
\end{split}
\end{equation}
\item Find $(\hat \vpi^\alpha,\eta^\alpha) \in W_\plper$, such that
\begin{equation}\label{eq-lp5}
  \begin{split}
    \ippYc{\TT(\hat \vpi^\alpha,\eta^\alpha)}{(\hat v,q)} & = -\hh \intY_{\pd \Xi_S} n_\alpha \int_{-1/2}^{1/2} q(\cdot,\zeta)\dd\zeta \;,\quad \forall (\hat v,q) \in  W_\plper\;.    
\end{split}
\end{equation}
\item Find $(\hat \vpi^3,\eta^3) \in W_\plper$, such that
\begin{equation}\label{eq-lp6}
  \begin{split}
    \ippYc{\TT(\hat \vpi^3,\eta^3)}{(\hat v,q)} & = -\imu\om\frac{\hh}{\rho_0} \intY_{\Xi_c} \rho \hat v - \ipXiS{\Jump{q}{\hh}}{1}\;,\quad \forall (\hat v,q) \in  W_\plper\;.    
\end{split}
\end{equation}  

\end{enumerate}

For the particular type of structures considered in this study, some of the
above problems can be simplified due to the vanishing part of the solution.

\begin{mydefinition}{def-1}
 Cell $Y$ decomposed in the fluid part $Y^*$ and the solid part $S$ is called $z$-symmetric, if 
\begin{equation}\label{eq-lp-zsym1}
  \begin{split}
    (y',z) \in Y^* \Leftrightarrow (y',-z) \in Y^*\;, \mbox{ and }
    (y',z) \in \pd Y^* \Leftrightarrow (y',-z) \in \pd Y^*\;.
\end{split}
\end{equation} 

\end{mydefinition}
Although the $z$-symmetry property is related directly to the cell
decomposition rather than to the cell $Y$ itself, the notion of $z$-symmetry
applies in the context of $Y$ and its decomposition.

\begin{mydefinition}{def-2}
 Let $Y$ is $z$-symmetric.
 Function $q$ defined in $Y_d \subset Y$, with traces defined on $\pd Y_d$ is called $z$-symmetric, if 
\begin{equation}\label{eq-lp-zsym2}
  \begin{split}
    q(y',z) = q(y',-z) \mbox{ for any } (y',z) \in \ol{Y_d}\;.
\end{split}
\end{equation}

\end{mydefinition}

\begin{mylemma}{thm-0}
For any $z$-symmetric function $q$, $\Jump{q}{\hh} = 0$ so that $\ipXiS{v}{\Jump{q}{\hh}} = 0$ for  any $v \in H^1(\Xi_c)$.
\end{mylemma}
As the result, the characteristic deflections in problems \eq{eq-lp3} and \eq{eq-lp5} vanish.

\begin{myproposition}{thm-1}
  Let $Y$ be $z$-symmetric. Then problems \eq{eq-lp3} and \eq{eq-lp5} can be reformulated in terms of the characteristic pressures $\pi^\beta$ and $\eta^\alpha$ only, which satisfy
\begin{equation}\label{eq-lp3a}
  \begin{split}
  \ipYs{\nabla_y (\pi^\alpha + y_\alpha)}{\nabla_y \hat q} = 0\quad \forall q \in H_\plper^1(Y^*)\;,
  \end{split}
\end{equation}
\begin{equation}\label{eq-lp5a}
  \begin{split}
    \ipYs{\nabla_y \eta^\alpha}{\nabla_y \hat q} = -\hh \intY_{\pd \Xi_S} n_\alpha \int_{-1/2}^{1/2} q(\cdot,\zeta)\dd\zeta
    \quad \forall q \in H_\plper^1(Y^*)\;,
  \end{split}
\end{equation}
 whereby $\hat w^\beta = \hat\varpi^\alpha \equiv 0$.

\end{myproposition}

The proof is based on Lemma~\ref{thm-0}. It can be seen, that solutions
$\pi^\alpha$ and $\eta^\alpha$ are $z$-symmetric, so that equations in
\eq{eq-lp3} and \eq{eq-lp5} are satisfied with vanishing $\hat w^\beta$ and
$\hat\varpi^\alpha$.

In this paper, we consider a $z$-symmetric cell $Y$, so that
Proposition~\ref{thm-1} applies. However, for the sake of generality, we shall
keep the formulations \eq{eq-lp3} and \eq{eq-lp5}. This will induce a general
set of homogenized effective tensors involved in the macromodel. Due to the
$z$-symmetry, some components of these coefficients vanish. In future studies,
we intend to consider plates with more complex perforations, such that the
$z$-symmetry will not apply.

The second group of autonomous local problems is imposed in the soft plate
inclusions $\Xi_c$ and governs in-plane displacements $\hat{\ol{\ub}}$ and
rotations $\hat\thetabf$, whereby the only nonvanishing test functions are
$\hat{\ol{\vb}}$ and $\hat\psibf$.
The following identities related to the dynamic properties are to be satisfied
by $\hat\thetabf(x,\cdot) \in \HOdb(\Xi_c)$, $\hat u_\alpha \in H_0^1(\Xi_c)$,
\begin{equation}\label{eq-lp7}
\begin{split}
  \aYc{\hat\thetabf}{\hat\psibf} - \om^2 \ipXic{\rho (\hat\thetabf + \thetabf^0)}{\hat\psibf} & = 0\;,\quad \forall \hat\psibf \in \HOdb(\Xi_c)\;, \\
 \aYc{\hat{\ol{\ub}}}{\hat\psibf} - \om^2 \ipXic{\rho (\hat{\ol{\ub}} + \ol{\ub}^0)}{\hat\psibf} & = 0\;,\quad \forall \hat\psibf \in \HOdb(\Xi_c)\;,  
\end{split}
\end{equation}
where 
\begin{equation}\label{eq-lp1}
\begin{split}
\aYc{\ol{\ub}}{\ol{\vb}} & = \intY_{\Xi_c} [\Eop_c \eeby{\ol{\ub}}]:\eeby{\ol{\vb}}\;.
\end{split}
\end{equation}
Both problems in \eq{eq-lp7} yield the following eigenvalue problem: Find $(\Thetabf^r, \lam^r) \in \HOdb \times \RR$ for $r = 1,2,\dots$, such that
\begin{equation}\label{eq-lp8}
\begin{split}
  \aYc{\Thetabf^r}{\hat\psibf} & =  \lam^r \ipXic{\rho \Thetabf^r}{\hat\psibf}\;,\quad \forall \hat\psibf \in \HOdb(\Xi_c)\;, \\
\end{split}
\end{equation}
so that the local responses are expressed in the bases constituted by  the eigenfunctions 
\begin{equation}\label{eq-lp9}
\begin{split}
  \hat\thetabf & = \sum_{r\geq 1} c^r \Thetabf^r\;,\quad \hat{\ol{\ub}} = \sum_{r\geq 1} \bar c^r  \Thetabf^r\;,
\end{split}
\end{equation}
where the coefficients $c^r$ and  $\bar c^r$ are given by 
\begin{equation}\label{eq-M3a}
\begin{split}
  c^k & = \frac{\om^2}{\lam^k-\om^2 } \intY_{\Xi_c}\rho \Thetabf^k \cdot \thetabf^0\;,\quad
  \bar c^k  = \frac{\om^2}{\lam^k-\om^2 } \intY_{\Xi_c}\rho \Thetabf^k \cdot \ol{\ub}^0\;.
\end{split}
\end{equation}

\subsubsection{Characteristic responses associated with the plate stiffness}

Local problems characterizing the static elasticity of the plate associated
with the normal and shear strains arise from \eq{eq-p1} for nonvanishing test
functions $\vb^1$ and $\psibf^1$. These problems can be expressed using the
bilinear forms
\begin{equation}\label{eq-lp1c}
\begin{split}
\aYm{\ol{\ub}}{\ol{\vb}} & = \intY_{\Xi_m} [\Eop_m \gradplyS{\ol{\ub}}]:\gradplyS{\ol{\vb}}\;,\\
\bYm{w}{z} & =  \intY_{\Xi_m} [\Sb_m \nabla_y w]\cdot\nabla_y z\;,
\end{split}
\end{equation}
and in terms of coordinate combinations $\Pibf^{\alpha\beta} = (\Pi_\nu^{\alpha\beta})$, $\Pi_\nu^{\alpha\beta} = y_\beta\delta_{\alpha \nu}$ with $\nu,\alpha,\beta = 1,2$, so that
$\gradplxS \ol{\ub}^0 =  \gradplyS \Pibf^{\alpha\beta} \pd_\alpha^x \ol{u}_\beta^0$.
For a fixed $x' \in \Gamma_0$, functions $\thetabf^1(x',\cdot), \ol{\ub}^1(x',\cdot) \in [H_\#^1(\Xi_m)]^2$ and $ u_3^1(x',\cdot) \in H_\#^1(\Xi_m)$ satisfy
\begin{equation}\label{eq-va30}
\begin{split}
\aYm{\thetabf^1 + \Pibf^{\alpha\beta}\pd_\alpha^x \theta_\beta^0}{\psibf^1} = 0\;,\quad \forall\psibf^1  \in [H_\#^1(\Xi_m)]^2\;,\\
 \aYm{\ol{\ub}^1 + \Pibf^{\alpha\beta}\pd_\alpha^x \ol{u}_\beta^0}{\ol{\vb}^1} = \imu \om \rho_0\int_{\Gamma_0}p^0\intY_{\pd\Xi_S} \ol{\nb}\cdot\ol{\vb}^1
\;,\quad \forall \ol{\vb}^1 \in [H_\#^1(\Xi_m)]^2\;,\\
 \bYm{ u_3^1 + y_\alpha (\pd_\alpha^x u_3^0 - \theta_\alpha^0)}{v_3^1} = 0\;,\quad \forall v_3^1 \in H_\#^1(\Xi_m);.
\end{split}
\end{equation}
Again using the linearity of \eq{eq-va30}, for the two-scale depending on the
macroscopic responses $\gradplS \ol{\ub}^0$, $\gradplS \thetabf^0$, $\gradpl
u_3$, and $p^0$, the following multiplicative decompositions can be introduced
\begin{eqnarray}
\ol{\ub}^1 & =& \ol{\chibf}^{\alpha\beta} (\gradplS \ol{\ub}^0)_{\alpha\beta} + \ol{\chibf}^* \imu\om\rho_0 p^0\;,\label{eq-ms1}\\
u_3^1 & =& \chi^k \left( (\gradpl u_3)_k - \theta_k\right) 
\;,\label{eq-ms2}\\
\thetabf^1 & =& \ol{\chibf}^{\alpha\beta} (\gradplS \thetabf^0)_{\alpha\beta} 
\;,\label{eq-ms3}
\end{eqnarray}
involving the local characteristic responses
$\ol{\chibf}^{\alpha\beta},\ol{\chibf}^*\in\Hpdb(\Xi_m)$, and $\chi^k \in
H_\#^1(\Xi_m)$, usually called the corrector functions. It is worth noting that
the same functions $\ol{\chibf}^{\alpha\beta}$ are involved in both
$\ol{\ub}^1$ and $\thetabf^1$ due to the similar structure of \eq{eq-va30}$_1$
and \eq{eq-va30}$_2$.
The following three local autonomous problems have to be solved,
\begin{itemize}
\item Find  $\ol{\chibf}^{\alpha\beta}\in \Hpdb(\Xi_m)/\RR^2$ such that
\begin{equation}\label{eq-mic3}
\begin{split}
\aYm{\ol{\chibf}^{\alpha\beta} + \Pibf^{\alpha\beta}}{\ol{\vb}}  & = 0 \quad \forall \ol{\vb} \in \Hpdb(\Xi_m)\;.
\end{split}
\end{equation}

\item Find $\chi^\alpha \in H_\#^1(\Xi_m/\RR$ such that
\begin{equation}\label{eq-mic4}
\begin{split}
\bYm{\chi^\alpha + y_\alpha}{ \tilde z} & = 0
\quad \forall \tilde z \in H_\#^1(\Xi_m)\;, \quad \alpha = 1,2\;.
\end{split}
\end{equation}
\item Find $\ol{\chibf}^* \in \Hpdb(\Xi_m)/\RR^2$ such that
\begin{equation}\label{eq-mic5}
\begin{split}
\aYm{\ol{\chibf}^*}{\ol{\vb}} = \intY_{\pd \Xi_S} \ol{\nb}\cdot\ol{\vb}
\;,\quad \forall \ol{\vb} \in \Hpdb(\Xi_m)\;.
\end{split}
\end{equation}
\end{itemize}


\subsection{Macroscopic model equations}

In the limit equations \eq{eq-p1}-\eq{eq-f6}, we now consider nonvanishing the
macroscopic test functions $\vb^0, \psibf^0$ and $q^0$ only, whereas all other
test functions vanish. Thus, the following equations which describe behaviour of
the plate and the fluid in the fictitious layer, are obtained, such that
\begin{equation}\label{eq-M1p}
\begin{split}
& -\om^2 \int_{\Gamma_0}\intY_{\Xi_S} \rho_S \left(( \ub^0 + \chi_c \hat \ub)\cdot\vb^0 + 
\frac{h^2}{12} (\thetabf^0  + \chi_c\hat\thetabf)\cdot\psibf^0 \right) \\
& +  \int_{\Gamma_0}\left(\intY_{\Xi_m}[\Eop_m(\gradplxS\ol{\ub}^0 + \gradplyS\ol{\ub}^1)]:\gradplxS\ol{\vb}^0 + \intY_{\Xi_m}[\Sb_m(\gradpl_x u_3^0 + \gradpl_y u_3^1 - \thetabf^0)]\cdot \psibf^0\right)\\
& + \frac{h^2}{12}\int_{\Gamma_0} \left(\intY_{\Xi_m}[\Eop_m(\gradplxS \thetabf^0 + \gradplyS \thetabf^1)]:\gradplxS \psibf^0 \right)\\
& = 
\frac{\imu\om\rho_0}{\hh}\int_{\Gamma_0}v_3^0\intY_{\Xi_S}\Jump{p^1}{\hh}
  + \imu\om\rho_0 \int_{\Gamma_0}\ol{\vb}^0 \cdot \int_{-1/2}^{1/2}\intY_{\pd\Xi_S}\ol{\nb} p^1\;,
\end{split}
\end{equation}
holds for all couples $(\vb^0,\psibf^0) \in \VV^0$, and 
\begin{equation}\label{eq-M1f}
\begin{split}
& c^2 \int_{\Gamma_0} \intY_{Y^*} (\gradplx p^0 + \gradply p^1)\cdot
\gradplx q^0   - \om^2 \int_{\Gamma_0} \intY_{Y^*} p^0 q^0\\
& = 
-\imu\om c^2 \int_{\Gamma_0} q_0 \left(
\intY_\Xi \Delta g^{1} + \hh \intY_{\pd\Xi_S}\ol{\nb}\cdot\ol{\ub}^1\right)\;,
\end{split}
\end{equation}
holds for all $q^0 \in L^2(\Gamma_0)$.

\subsubsection{Homogenized coefficients}\label{sec-HC1}

We shall first identify homogenized coefficients in the fluid equation
\eq{eq-M1f}. Upon substituting there the multiplicative splits \eq{eq-f7} and
\eq{eq-ms1}, it yields
\begin{equation}\label{eq-f11}
\begin{split}
  A_{\alpha\beta} & = \intY_{Y^*}{\nabla_y (\pi^\beta + y_\beta)}\cdot{\nabla_yy_\alpha}
  = \ipYs{\nabla_y (\pi^\beta + y_\beta)}{\nabla_y (\pi^\alpha + y_\alpha)}\;,\\
B_\alpha & = \intY_{Y^*} \pd_\alpha^y \xi\;,\\
D_{\alpha k}^* & =  \intY_{Y^*} \pd_\alpha^y \eta^k \;,\\
H_{\alpha\beta}&  = \intY_{\pd \Xi_S}\ol{\nb}\cdot\ol{\chibf}^{\alpha\beta}\;,\\
K & = \intY_{\pd \Xi_S}\ol{\nb}\cdot\ol{\chibf}^{*}\;.
\end{split}
\end{equation}
The homogenized coefficients $\Ab = (A_{\alpha\beta})$, $\Bb = (B_\alpha)$, and
$\Db^* = (D_k^*)$, are associated with the integral involving $p^1$, whereas
$\Hb = (H_{\alpha\beta})$ and $K$ are identified in the last \rhs integral
involving $\ol{\ub}^1$.

Now the homogenized fluid equation \eq{eq-M1f} can be rewritten in terms of the
coefficients \eq{eq-f11}. It is satisfied by macroscopic functions
$(p^0,g^0,\ub^0)\in H^1(\Gamma_0)\times L^2(\Gamma_0) \times \Hdb(\Gamma_0)$,
\begin{equation}\label{eq-f12}
\begin{split}
& c^2\int_{\Gamma_0} ( \Ab \gradpl_x p^0 )\cdot \gradpl_x q^0
- \zeta^* \om^2 \int_{\Gamma_0} p^0 q^0 + \imu \om c^2\int_{\Gamma_0} g^0 \Bb\cdot\gradpl_x q^0 + \imu \om c^2\int_{\Gamma_0}\ \gradpl_x q^0 \cdot \Db^*\ub^0\\
= & -\imu \om c^2\hh \int_{\Gamma_0} q^0\left (\frac{\Dlt G^1}{\hh} + \imu \om\rho_0
 K p^0  +  \Hb:\gradplS_x\ol{\ub}^0
\right )\;,
\end{split}
\end{equation}
for all $q^0 \in H^1(\Gamma_0)$, where $\zeta^* = |Y^*|/|\Xi|$ and $\Dlt G^1 =
{|\Xi|}^{-1}\int_\Xi \Delta g^{1}$. Recall that the transversal momentum flux
difference $\Delta g^{1}$ was introduced in \eq{eq-f6}. In
Section~\ref{sec-glob}, we shall establish $g^0$ and $\Delta g^{1}$ using
averaged momentum fluxes $\hat G_0^\pm$ associated with the interfaces
$\Gamma^\pm$, see Fig.~\ref{fig-pd2}.


Further we consider the plate equation \eq{eq-M1p} and denote $\gamma:=
\frac{\hh}{\rho_0}$ which will be involved in some of the following expressions
for the homogenized coefficients. The inertia terms associated with the plate
in-plane velocities and cross-sectional rotations can be expressed using the
homogenized mass coefficients $\tilde\Mcalbf = (\Mcal_{\alpha\beta})$,
\begin{equation}\label{eq-M2}
\begin{split}
    \intY_{\Xi_S} \rho_S ( {\ol{\ub}}^0 + \chi_c \hat{\ol{\ub}})\cdot{\ol{\vb}}^0 & = v_\alpha^0 \tilde\Mcal_{\alpha\beta} u_\beta^0\;,\\
    \intY_{\Xi_S} \rho_S ( \thetabf^0 + \chi_c \hat \thetabf)\cdot\psibf^0 & = \psi_\alpha^0 \tilde\Mcal_{\alpha\beta} \theta_\beta^0 \;,
\end{split}
\end{equation}
where $\tilde\Mcalbf$ is identified upon  substituting \eq{eq-lp9}-\eq{eq-M3a} in \eq{eq-M2},
\begin{equation}\label{eq-M3}
\begin{split}
  \gamma^{-1}\tilde\Mcal_{\alpha\beta} = \intY_{\Xi_S}\rho_S \delta_{\alpha\beta} -\sum_{r\geq 1} \frac{\om^2}{\om^2 - \lam^r} \intY_{\Xi_c}\rho \Theta_\alpha^r \intY_{\Xi_c}\rho \Theta_\beta^r \;.
\end{split}
\end{equation}

To proceed, in \eq{eq-M1p}, we consider the integrals related to the
elasticity, as represented by tensors $\Eop_m$ and $\Sb_m$, which can be
expressed using the effective elasticity tensors $\Eop^\hom =
(E_{\alpha\beta\mu\nu}^\hom)$ and $\Sb^\hom = (S_{\alpha\beta}^\hom)$, and
using the pressure-strain coupling tensor $\Hb^* = (H_{\alpha\beta})$ defined,
as follows
\begin{equation}\label{eq-M0}
\begin{split}
E_{\alpha\beta\mu\nu}^\hom & =  \intY_{\Xi_m} \Eop_m \gradplS_y({\ol{\chibf}^{\mu\nu} + \Pibf^{\mu\nu}}):\gradplS_y({\ol{\chibf}^{\alpha\beta} + \Pibf^{\alpha\beta}})\;,\\
S_{\alpha\beta}^\hom & = \intY_{\Xi_m} \left[\Sb_m\nabla_y (\chi^\alpha + y_\alpha)\right]\cdot
\nabla_y (\chi^\beta + y_\beta)\;,\\
H_{\alpha\beta}^* & =  -\intY_{\Xi_m} \Eop_m\gradplS_y\ol{\chibf}^* : \gradplS_y\ol{\chibf}^{\alpha\beta}\;.
\end{split}
\end{equation}
The above expressions can be obtained using the split forms \eq{eq-ms1}-\eq{eq-ms3}.

Now we shall consider the other terms in \eq{eq-M1p}. These are related to the
transverse inertia due to the plate deflections, and to the \rhs terms
describing the fluid-structure interaction. The following expressions are
obtained upon substituting there the split form of two-scale functions, see
\eq{eq-f7} and \eq{eq-ms1}-\eq{eq-ms3},
\begin{equation}\label{eq-M4}
\begin{split} 
  &   -\om^2\intY_{\Xi_S} \rho_S \left(\hat u_3^0 + \chi_c[\hat w^\beta \pd_\beta p^0 + \imu\om \hat\vsigma g^0 + \imu\om \hat\vpi^k u_k^0]\right) v_3^0 \\
  & -\frac{\imu\om\rho_0}{\hh}v_3^0\intY_{\Xi_S}\Jump{\pi^\beta\pd_\beta^x p^0 + \imu\om \xi g^0 +  \imu\om \eta^k u_k^0}{\hh}\\
  &   -\imu\om \rho_0 v_\alpha^0 \intY_{\pd\Xi_S} n_\alpha  \int_{-1/2}^{1/2}(\pi^\beta\pd_\beta^x p^0 + \imu\om \xi g^0 +  \imu\om \eta^k u_k^0)\\
  & = \om^2  v_3^0 u_3^0\underbrace{\left[\frac{\rho_0}{\hh}\intY_{\Xi_S}\Jump{\eta^3}{\hh} - \intY_{\Xi_S} \rho_S (1 +\imu\om\chi_c\hat\vpi^3 )\right]}_{-\gamma^{-1}M_{33}} \\
  & + \om^2 v_3^0 u_\alpha^0 \underbrace{\left[\frac{\rho_0}{\hh}\intY_{\Xi_S}\Jump{\eta^\alpha}{\hh} - \imu\om\intY_{\Xi_c} \rho_c \hat\vpi^\alpha\right]}_{-\gamma^{-1}M_{3\alpha}} \\
  & + \om^2  v_\alpha^0 u_k^0\underbrace{\left[\rho^0 \intY_{\pd\Xi_S} n_\alpha  \int_{-1/2}^{1/2}\eta^k\right]}_{- \gamma^{-1}M_{\alpha k}}\\
  & - \imu\om v_3^0 \pd_\beta p^0\underbrace{\frac{\rho_0}{\hh}\left[\intY_{\Xi_S}\Jump{\pi^\beta}{\hh} - \imu\om\frac{\hh}{\rho_0}\intY_{\Xi_c} \rho_c \hat w^\beta\right]}_{\gamma^{-1}D_{3 \beta}} \\ 
  & - \imu\om v_\alpha^0 \pd_\beta p^0\underbrace{\frac{\rho_0}{\hh} \left[\hh\intY_{\pd\Xi_S} n_\alpha  \int_{-1/2}^{1/2}\pi^\beta\right] }_{\gamma^{-1}D_{\alpha \beta}} \\
  & + \om^2 v_3^0 g^0 \underbrace{\frac{\rho_0}{\hh}  \left[\intY_{\Xi_S} \Jump{\xi}{\hh} - \imu\om \frac{\hh}{\rho_0} \intY_{\Xi_c} \rho_c \hat \vsigma\right]}_{\gamma^{-1}C_3}
  + \om^2  v_\alpha^0 g^0 \underbrace{\rho_0 \intY_{\pd\Xi_S} n_\alpha  \int_{-1/2}^{1/2}\xi}_{\gamma^{-1}C_\alpha}
\end{split}
\end{equation}
Note that $\rho_0 = \gamma^{-1}\hh$, which can be employed in the definition of $C_\alpha$ and $M_{\alpha k}$.





It is now possible to define the homogenized mass tensor,
\begin{equation}\label{eq-M6}
\begin{split}
  \Lb & = \left[
\begin{array}{ll}
  (\tilde\Mcal_{\alpha\beta} +  M_{\alpha\beta} ) \;, &  (M_{\alpha 3}) \\
  (M_{3 \beta}) \;, & ( M_{33})
\end{array}  \right] \;,
\end{split}
\end{equation}
so that the plate inertia associated with the displacements is expressed by $\om^2 \Lb \cdot \ub^0$. 

Using the homogenized coefficients, the macroscopic (homogenized) plate
equation \eq{eq-M1p} satisfied by macroscopic functions $(p^0,g^0,\ub^0,\thetabf^0)$ can be rewritten, as follows 
\begin{equation}\label{eq-M8}
\begin{split}
& -\om^2 \int_{\Gamma_0}\gamma^{-1}\left( (\Lb \ub^0)\cdot \vb + \frac{h^2}{12}
(\tilde \Mcalbf\thetabf^0)\cdot\vthetabf \right) +
 \int_{\Gamma_0} (\Sb^\hom(\gradpl_x u_3^0 - \thetabf^0))\cdot
(\gradpl_x  v_3 - \vthetabf)\\
& + \frac{h^2}{12} \int_{\Gamma_0}(\Eop^\hom \gradplS_x\thetabf^0):\gradplS_x\vthetabf + \int_{\Gamma_0}(\Eop^\hom \gradplS_x\ol{\ub}^0):\gradplS_x{\ol{\vb}} \\
&- \imu\om\rho_0 \int_{\Gamma_0} p^0\Hb:\gradplS_x{\ol{\vb}}
- \imu\om \gamma^{-1}\int_{\Gamma_0} \vb\cdot \left(\Db\gradpl p^0 +  \imu\om  \Cb g^0 \right) = 0\;,
\end{split}
\end{equation}
recalling $\gamma^{-1}=\frac{\rho_0}{\hh}$.

\subsubsection{Coupling conditions on $\Gamma_0$}

This condition is necessary to close the system of equations describing the
acoustic field $p^\veps$ in the transmission layer $\Om_\dlt$ with the one
describing the global acoustic field $P^\dlt$ in $\Om_\dlt^+$ and $\Om_\dlt^-$.
The following identity is a weak formulation of condition \eq{eq-G2-03}$_2$,
\begin{equation}\label{eq-cc1}
\begin{split}
 \int_{\Gamma_\delta^+} \psi P^\dlt  -  \int_{\Gamma_\delta^-}\psi P^\dlt   =  \int_{\Gamma_0} \psi \int_{-\delta/2}^{\delta/2} \pd_{x_3} \tilde p^\vepsdel \quad \forall \psi \in L^2(\Gamma_0)\;,
\end{split}
\end{equation}
where we assume $\psi = \psi(x')$, $x' \in \Gamma_0$; by $\tilde{}$ we
denote an extension of $p^\vepsdel$ to the whole $\Om_\delta$. Further we proceed as
in \cite{Rohan-Lukes-AMC2019} by introducing an approximation for a finite layer thickness
$\delta_0 = \vkappa\veps_0 >0$, such that \eq{eq-cc1} yields the following limit condition
\begin{equation}\label{eq-cc3}
\begin{split}
\frac{1}{\veps_0} \int_{\Gamma_0} \psi (\hat P^{+} - \hat P^{t-} ) = \int_{\Gamma_0} \psi \left(\intY_{I_y^+} p^1 - \intY_{I_y^-} p^1\right)
 \quad \forall \psi \in L^2(\Gamma_0)\;,
\end{split}
\end{equation}
where $\hat P^{+/-}$ are the limit traces of $P^\dlt$ on $\Gamma_\dlt^{+/-}$ for $\dlt \rightarrow 0$.
Upon substituting there $p^1$ by the split form \eq{eq-f7}, we get
\begin{equation}\label{eq-M9}
\begin{split}
\int_{\Gamma_0} \psi \left(
\Bb'\cdot\gradpl_x p^0 - \imu \om F g^0 + \imu \om \Cb'\cdot \ub^0
\right) & = \frac{1}{\veps_0} \int_{\Gamma_0} \psi (\Delta{\hat P})\quad \forall \psi \in L^2(\Gamma_0)\;.
\end{split}
\end{equation}
where $\Delta{\hat P} = \hat P^+ - \hat P^+$ and  
\begin{equation}\label{eq-M10}
\begin{split}
F & = - \intY_{I_y^+} \xi + \intY_{I_y^-} \xi\;, 
\\
{B'}_\alpha & = \intY_{I_y^+} \pi^\alpha - \intY_{I_y^-}  \pi^\alpha = \intY_{Y_\vkappa^*} \pd_\alpha^y \xi = B_\alpha\;,\quad \alpha = 1,2\;,\\
{C'}_k  &  = \intY_{I_y^+} \eta^k - \intY_{I_y^-}\eta^k = \intY_{\pd S} n_k \xi = C_k\;,\quad k = 1,2,3\;.
\end{split}
\end{equation}

In the next section, we shall formulate the global problem whose the solution
describe the acoustic field in $\Om^G = \hat\Om^+\cup\hat\Om^- \cup \Gamma_0$.
Acoustic waves in the homogenized fictitious layer represented by $\Gamma_0$ 
with embedded perforated plate are governed by the system of equations
\eq{eq-f12}, \eq{eq-M8}, and \eq{eq-M9} expressing the transmission conditions
involving the homogenized coefficients. This system is featured by some
symmetries due to the following theorem.

\begin{myproposition}{thm-3}
\begin{list}{}{}  
\item (i) Coefficients involved in the coupled transmission conditions \eq{eq-f12}, \eq{eq-M8}, and \eq{eq-M9} satisfy the following relationships (recall $\alpha, \beta = 1,2$, $k = 1,2,3$)
\begin{equation}\label{eq-thm-3-1}
\begin{split}
  M_{\alpha\beta} & = M_{\beta\alpha}\;,\quad  M_{3\beta} = M_{\beta 3}\;,\\
  D_{\alpha\beta} & = D_{\beta\alpha}^*\;,\quad  D_{3\beta} = D_{\beta 3}^*\;,\\
  H_{\alpha\beta}^* & = H_{\beta\alpha} = H_{\alpha\beta}\;,\\
  A_{\alpha\beta} & = A_{\beta\alpha}\;,\\
  B_\alpha & = B_\alpha'\;,\\
  C_k & = C_k'\;.
\end{split}
\end{equation}
\item (ii) Coefficients $\tilde\Mcalbf,\Mb,\Db,\Ab,\Cb,\Bb$ and $F$ depend on $\om$, in general.
\item (iii) For $z$-symmetric cells $Y$, see Definition~\ref{def-1},
  \begin{equation}\label{eq-thm-3-2}
\begin{split}
C_\alpha = 0\;,\quad B_\alpha = 0\;, \quad D_{3\alpha} = 0\mbox{ and also } M_{3\alpha} = 0\;,\quad \alpha, \beta = 1,2\;.
\end{split}
  \end{equation}
\item (iv) Moreover, for $z$-symmetric cells $Y$, coefficients $A_{\alpha\beta}$ and $D_{\alpha k}$ do not depend on $\om$.
\end{list} 
\end{myproposition}

The proof of Proposition~\ref{thm-3}, assertions (i) and (ii), is given in
\Appx{apx-2}. Although the frequency is involved in the $\TT$ operator and,
thereby, the characteristic responses of problems \eq{eq-lp3}-\eq{eq-lp6}
should depend on $\om$, it may not be clear if the homogenized coefficients do
depend on $\om$ as well. Assertion (iii) is the direct consequence of
Proposition~\eq{thm-1}; in the case of $z$-symmetric cells $Y$, $\hat w^\beta =
\hat\varpi^\alpha \equiv 0$ and the ``reduced'' local problems \eq{eq-lp3a} and
\eq{eq-lp5a} do not involve $\om$, which yields assertion (iv).




\section{Global problem with homogenized transmission layer} \label{sec-glob}

The acoustic wave propagation in the bulk $\Om^G$ and in the homogenized layer
$\Gamma_0$ is described in terms of the global acoustic pressure $\hat P$
defined in subdomains $\hat \Om^+$ and $\hat \Om^-$, and in terms of
$(p^0,\ub^0,\theta^0)$ involved in equations \eq{eq-f12}, \eq{eq-M8} and
\eq{eq-M9} governing the vibro-acoustic interaction in the layer embedding the
perforated plate. These fields are coupled trough the limit conditions arising
form \eq{eq-G2-03}.

\subsection{Coupling condition and problem formulation}

We summarize the equations describing fields $(p^0,\ub^0,\theta^0)$ defined in
the homogenized layer represented by $\Gamma_0$. Due to symmetries stated in
Proposition~\ref{thm-3}, after some straightforward manipulations, the system
of equations \eq{eq-f12}, \eq{eq-M8} and \eq{eq-M9} attains a symmetric form.
From \eq{eq-f12} we get
\begin{equation}\label{eq-MN1}
\begin{split}
& \int_{\Gamma_0} ( \Ab \gradpl_x p^0 )\cdot \gradpl_x q^0
-\om^2 \int_{\Gamma_0} ( \frac{\zeta^*}{c^2} + \hh\rho_0 K) p^0 q^0 \\
& + \imu\om\int_{\Gamma_0}\left(q^0\hh\Hb:\gradplS_x\ol{\ub}^0 + \gradpl_x q^0 \cdot \Db\ub^0\right)
+  \imu \om \int_{\Gamma_0}\left( g^0 \Bb\cdot\gradpl_x q^0 + q^0 \Dlt G^1\right) = 0\;,
\end{split}
\end{equation}
for all $q \in H^1(\Gamma_0)$. Upon multiplication by $\gamma = \hh/\rho_0$ in \eq{eq-M8}, it becomes 
\begin{equation}\label{eq-MN2}
\begin{split}
& -\om^2 \int_{\Gamma_0}\left( (\Lb \ub^0)\cdot \vb + \frac{h^2}{12}
(\tilde \Mcalbf\thetabf^0)\cdot\vthetabf \right) +
 \int_{\Gamma_0} \gamma(\Sb^\hom(\gradpl_x u_3^0 - \thetabf^0))\cdot
(\gradpl_x  v_3 - \vthetabf)\\
& + \frac{\gamma h^2}{12} \int_{\Gamma_0}(\Eop^\hom \gradplS_x\thetabf^0):\gradplS_x\vthetabf + \int_{\Gamma_0}\gamma(\Eop^\hom \gradplS_x\ol{\ub}^0):\gradplS_x{\ol{\vb}} \\
&- \imu\om \int_{\Gamma_0} \left (p^0\hh\Hb:\gradplS_x{\ol{\vb}}
+\vb\cdot \Db\gradpl p^0 \right) +  \om^2\int_{\Gamma_0} \vb\cdot \Cb g^0  = 0\;,
\end{split}
\end{equation}
for all $(\ub^0,\thetabf^0) \in [H_0^1(\Gamma_0)]^5$. Finally, \eq{eq-M9} multiplied by $-\imu\om$ yields
\begin{equation}\label{eq-MN3}
\begin{split}
\frac{\imu\om}{\veps_0} \int_{\Gamma_0} \psi (\Delta{\hat P}) -\int_{\Gamma_0} \psi \left(
\imu\om\Bb'\cdot\gradpl_x p^0 + \om^2 F g^0  - \om^2 \Cb'\cdot \ub^0
\right) = 0\;,\quad \forall \psi \in L^2(\Gamma_0)\;.
\end{split}
\end{equation}

In \cite{Rohan-Lukes-AMC2019}, we derived a Dirichlet-to-Neumann (DtN) operator
which relates $\Dlt P$ and $p^0$, satisfying
\begin{equation}\label{eq-G3-03}
\begin{split}
\Delta P & = \frac{1}{\veps_0}\left(\hat P^+ - \hat P^-\right)\;,\quad\mbox{ and }
p^0  = \frac{1}{2}\left(\hat P^+ + \hat P^-\right)\;,
\end{split}
\end{equation}
with the averaged momentum fluxes $\hat G_0^\pm$ defined,  as follows
\begin{equation}\label{eq-G3-00}
\begin{split}
\hat G_0^\pm(x') &  = \frac{1}{|\Xi|}\int_\Xi \left(g^0(x') + \veps_0 g^{1,\pm}(x',y')\right)\dd y' \\
& = g^0(x') + \veps_0 G^{1\pm}(x')\;,\quad\mbox{ where }  G^{1\pm} = \frac{1}{|\Xi|}\int_\Xi g^{1\pm}(x',y')\dd y'\;.
\end{split}
\end{equation}
Now, using $\hat G_0^\pm$, we can express $g^0$ and $\Dlt G^1$ involved in \eq{eq-f12}.
It holds that
\begin{equation}\label{eq-G3-04}
\begin{split}
g^0 & \approx \frac{1}{2}\left(\hat G_0^+ + \hat G_0^-\right) = g^0 + \frac{\veps_0}{2}(G^{1+} + G^{1-})\;,\\
\Delta G^1 & = \frac{1}{|\Xi|}\int_\Xi \Dlt g^1(x',y')\dd y' = G^{1+} - G^{1-} = \frac{1}{\veps_0}\left(\hat G_0^+ - \hat G_0^-\right)\;.
\end{split}
\end{equation}

We may now introduce the global problem of acoustic field in a waveguide
equipped with the perforated plate. By $\Gcal$ we refer the DtN operator
defined in an implicit way, using the equations representing the homogenized
layer \eq{eq-MN1}-\eq{eq-MN3} which together with \eq{eq-G3-03} and
\eq{eq-G3-04} establish the mapping $\Gcal:\hat P \mapsto \hat G_0^\pm$. In
response to an incident wave with its amplitude $\bar p$, we find $\hat P$ in
$\Om^G$ satisfying
\begin{equation}\label{eq-G3-01}
\begin{split}
  c^2 \nabla^2 \hat P + \om^2 \hat P & = 0 \quad \mbox{ in } \hat\Om^+\cup\hat\Om^-\;,\\
  r \imu\om c \hat P + c^2 \pdiff{\hat P}{n} & = s 2\imu\om c\bar p
\quad  \mbox{ on } \pd_\ext \Om^G\;,\\
\mbox{ \textbf{ interface condition:} } & \\
 \pdiff{\hat P^s}{n^s}|_{\Gamma_{\veps_0}^s} & =  s\imu\om \hat G^s\quad \mbox{ on } \Gamma_0\;,\quad s = +,-\;,\\
 \Gcal(\hat P,\hat G_0^\pm) & = 0
\quad  \mbox{ on }  \Gamma_0\;,
\end{split}
\end{equation}
where $\Gamma_{\veps_0}^s$, $s = +,-$ is the layer surface for a given $\veps_0
> 0$, and $n^\pm$ are the unit normal vector outward to
$\hat\Om_{\veps_0}^\pm$. Furthermore, the constants $r,s \in \{0,1\}$ and $\bar
p$ are defined to describe incident, reflected, or absorbed acoustic waves in
the fluid, according to a selected part of the boundary.


\subsection{Weak formulation  of the global problem}

For numerical simulations reported in the following section, we need the weak
formulation of problem \eq{eq-G3-01} with the DtN mapping introduced above, so
that a computational algorithm based on the finite element method can be used.
In what follows, we consider $z$-symmetric microstructures which arise
naturally due to the transverse isotropy of the plate. By the consequence, see
Proposition~\ref{thm-3}, (iii), some homogenized coefficients vanish. In this
respect, we denote $\ol{\Db} = (D_{\alpha\beta})$ and $\ol{\Mb} =
(M_{\alpha\beta})$. We employ the following bilinear forms involving the
homogenized coefficients introduced in the preceding sections,
\begin{equation}\label{eq-bilin}
\begin{split}
\Acal(p,q) & = \ipGm{\Ab \gradpl p}{\gradpl q}\;,\\
\Scal((w,\thetabf),(v,\vthetabf)) & = \frac{\hh}{\rho_0}\ipGm{\Sb^\hom (\gradpl w - \thetabf)}{\gradpl v - \vthetabf}\;,\\
\Ecal(\thetabf,\vthetabf) & = \frac{\hh}{\rho_0}\ipGm{\Eop^H\gradplS \thetabf}{\gradplS \vthetabf}\;,\\
\Fcal(g,\psi) & = \ipGm{Fg}{\psi}\;,\\
\Hcal(\ol{\vb},p) & =\hh\ipGm{\Hb:\gradplS\ol{\vb}}{p} \;,\\
\Ccal(u,\psi) & = \ipGm{C_3 u}{\psi}\;,\\
\Kcal(p,q) & = \ipGm{(\frac{\zeta^*}{c^2}+\rho_0 \hh K)p}{q}\;,\\
\Lcal(\ol{\ub},\ol{\vb}) & = \ipGm{(\ol{\Mb} + \tilde\Mcalbf)\ol{\ub}}{\ol{\vb}}\;,\\
\Mcal(\thetabf,\vthetabf) & = \ipGm{\tilde\Mcalbf\thetabf}{\vthetabf}\;,\\
\Ncal(u_3,v_3) & = \ipGm{M_{33} u_3}{v_3}\;,\\
\Dcal(\ol{\vb},p) & = \ipGm{\ol{\Db}\ol{\vb}}{\gradpl p}\;.
\end{split}
\end{equation}
%


We obtain the weak formulation of \eq{eq-G3-01} governing the global acoustic
field $\hat P \in H^1(\hat \Om^+\cup\hat \Om^+)$ which satisfies
\begin{equation}\label{eq-DtN-0}
\begin{split}
c^2\int_{\hat \Om^+\cup\hat \Om^-}\nabla \hat P \cdot \nabla Q - \om^2  \int_{\hat \Om^+\cup\hat \Om^-}\hat P Q 
+ \int_{\pd_\ext \Om^G} r \imu\om c \hat P Q & \\
- \imu\om c^2 \left(\ipGm{\hat G_0^+}{Q^+} - \ipGm{\hat G_0^-}{Q^-}\right) 
& = \int_{\pd_\ext \Om^G} s\imu\om c \bar{p} Q \;,
\end{split}
\end{equation}
for all $Q \in H^1(\hat\Om^+\cup\hat\Om^-)$, whereby $Q^\pm$ denotes the trace
of $Q$ on $\pd \hat\Om^\pm$.

The DtN operator involving functions $(\hat P^\pm,p^0,\ol{\ub}) \in
[L^2(\Gamma_0)]^2 \times H^1(\Gamma_0)\times [H_0^1(\Gamma_0)]^2$ and $(\hat
G_0^\pm,u_3,\thetabf) \in [L^2(\Gamma_0)]^2\times H_0^1(\Gamma_0)\times
[H_0^1(\Gamma_0)]^2$ is represented by the following equalities arizing from
\eq{eq-MN1}-\eq{eq-MN3}, where we employ \eq{eq-G3-03} and
\eq{eq-G3-04}, so that we have
 \begin{equation}\label{eq-DtN-1}
\begin{split}
  \Acal(p^0,q) - \om^2\Kcal(p^0,q) + \imu\om\left(\Dcal(\ol{\ub},q) + \Hcal(\ol{\ub},q)\right)
+  \frac{\imu\om}{\veps_0}\ipGm{[\hat G_0^+ - \hat G_0^-]}{q}  & = 0\;,\\
-\imu\om\left(\Dcal(\ol{\vb},p^0) + \Hcal(\ol{\vb},p^0)\right)
+ \Ecal(\ol{\ub},\ol{\vb}) - \om^2\Lcal(\ol{\ub},\ol{\vb}) & = 0\;,
\end{split}
\end{equation}
%
for all $(q,\ol{\vb}) \in H^1(\Gamma_0)\times [H_0^1(\Gamma_0)]^2$, and 
\begin{equation}\label{eq-DtN-2}
\begin{split}
\frac{\om^2}{2}\Fcal([\hat G_0^+ + \hat G_0^-],\psi) - \om^2\Ccal(u_3,\psi) {-} \frac{\imu\om}{\veps_0}\ipGm{[\hat P^+ - \hat P^-]}{\psi} & = 0\;,\\
\frac{\om^2}{2}\Ccal([\hat G_0^+ + \hat G_0^-],v_3) - \om^2\left(\Ncal(u_3,v_3)
 +\frac{h^2}{12}\Mcal(\thetabf,\vthetabf) \right) & \\
+\Scal((u_3,\thetabf),(v_3,\vthetabf)) + \frac{h^2}{12}\Ecal(\thetabf,\vthetabf)  & = 0\;,
\end{split}
\end{equation}
for all $(\psi,v,\vthetabf) \in L^2(\Gamma_0)\times H_0^1(\Gamma_0)\times
[H_0^1(\Gamma_0)]^2$. Recall the coupling equation \eq{eq-G3-03},
\begin{equation}\label{eq-DtN-3}
\begin{split}
\ipGm{2 p^0 - [\hat P^+ + \hat P^-]}{q} & = 0\quad \forall q \in L^2(\Gamma_0)\;.
\end{split}
\end{equation}

Now we can state the main result of this section. 

\paragraph{Global acoustic problem with the homogenized perforated plate}
Given the incident acoustic wave represented by $\bar p$ on $\Gamma_{\rm in}$,
find the acoustic potential $\hat P$ defined in $\hat\Om^G =
\hat\Om^+\cup\hat\Om^-$ and other functions $(p^0,\hat G_0^\pm,\ub,\thetabf)$
defined on $\Gamma_0$ such that the variational equalities
\eq{eq-DtN-0}-\eq{eq-DtN-3} hold.

\section{Numerical implementation of the transmission conditions}\label{sec-fem}

In this section, we explain a finite-element based implementation of the
homogenized layer model \eq{eq-DtN-1}-\eq{eq-DtN-3}. Its coupling with the
discretized problem of the acoustic waveguide is postponed in
Section~\ref{sec-numex}.

\subsection{Discretized micro-problems}

We shall describe a procedure of computing the homogenized coefficients
involved in the model of the homogenized transmission layer. For this, both
subparts of the computational cell $Y$ are dicretized by finite elements. Due
to the specific geometry of the perforation, $Y$ is $z$-symmetric with a
cylindrical hole $Y^* \cap S$, see \eq{101}, positions of the discretization
nodes in $\Xi_S$ representing the plate are shifted positions of those on
either of the surfaces $\pd_S^\pm Y = \{y = (y',z) \in Y|\;z = \pm\hh/2\}$.
Moreover, by virtue of the elliptic differential operators involved in $\TT{}$,
see \eq{eq-lp2}. We shall describe the discretized form of the micro-problems
related to the acoustic fluid-structure interaction coupling the response in
the solid plate inclusion $\Xi_c$ with the fluid response in $Y^*$. The
effective plate stiffness coefficients \eq{eq-M0} are given by the
characteristic stationary responses \eq{eq-mic3}-\eq{eq-mic5} of the plate
matrix $\Xi_m$ which are computed using a standard finite element
discretization of the elliptic problems.

\subsubsection{Problems related to the acoustic fluid-structure interaction}\label{sec-FE-FSI}

We introduce a generic form of the characteristic problems
\eq{eq-lp3}-\eq{eq-lp6}. We use the matrix notations, such that $\ubm$ and
$\pbm$ represent the displacement field $\vpi$ and pressure field $\eta$
involved in problem \eq{eq-lp5}, recalling $\gamma = \hh/\rho^0$. For the other
problems \eq{eq-lp3}-\eq{eq-lp5}, the correspondence with the characteristic
response functions (the correctors) is obvious, see Tab.~\ref{tab-hc1}. The
following self-explaining matrix notation is employed to introduce the
discretized characteristic problems,
\begin{equation}\label{eq-nm1a}
\begin{split}
  \bhYc{\hat w}{\hat v} & \FEMapprox \vbm^T \Abm \wbm\;,\\
  \ipXic{\rho \hat w}{\hat v} & \FEMapprox \vbm^T \Mbm\wbm\;,\\
  \ipXic{\hat w}{ \Jump{\hat q}{\hh}} & \FEMapprox \qbm^T \Bbm \wbm\;,\\
  \ipYs{\nabla_y \hat p}{\nabla_y \hat q} & \FEMapprox \qbm^T \Cbm \pbm\;,\\
\end{split}
\end{equation}
where $\wbm$ and $\vbm$ contain the displacement DOFs, whereas $\qbm$ and
$\pbm$ represent the pressure DOFs. Further, the \rhs in problems
\eq{eq-lp3}-\eq{eq-lp6} are approximated using vectors $\bbm$ and $\rbm$, so
that

\begin{equation}\label{eq-nm1b}
\begin{split}
  \ipYs{\gradply y_\alpha}{\gradply q} & \FEMapprox \qbm^T\ol{\bbm}_\alpha\;,\quad \ol{\rbm}_\alpha = \zerobm\;, \\
  \intY_{I_y^+} q - \intY_{I_y^-} q  & \FEMapprox \qbm^T\bbm_\xi\;,\quad {\rbm}_\xi = \zerobm\;, \\
  \hh\rho_0\intY_{\pd\Xi_S} n_\alpha  \int_{-1/2}^{1/2}\xi & \FEMapprox \qbm^T\bbm_\alpha\;,\quad \rbm_\alpha = \zerobm\;, \\
  \ipXiS{\Jump{q}{\hh}}{1} & \FEMapprox \qbm^T\bbm_3\;,\quad \intY_{\Xi_c} \rho \hat v \FEMapprox \hat\vbm^T\rbm_3\;,
\end{split}
\end{equation}
Note that above the subscripts $_\alpha,_\xi,_3$ are the labels and do not mean
the vector component.

All problems  \eq{eq-lp3}-\eq{eq-lp6} attain  the following discretized form,
\begin{equation}\label{eq-nm1}
  \begin{split}
    \underbrace{\left[
  \begin{array}{ll}
\gamma(\Abm - \om^2\Mbm)\;,& - \imu\om \Bbm^T \\
       \imu\om \Bbm\;, & \Cbm
    \end{array}\right]}_{\Kbm}
\underbrace{\left[
  \begin{array}{c}
    \ubm \\ \pbm
\end{array}\right]}_{\qbm} & = - 
\underbrace{\left[
  \begin{array}{c}
    \imu\om \gamma \rbm \\ \bbm
\end{array}\right] }_{\fbm} \;, 
\end{split}
\end{equation}
thus, $\Kbm\qbm = - \fbm$. Obviously, matrix $\Kbm$ depending on $\om$ can
become singular for certain $\om$ approaching resonances.

\subsubsection{Solution procedure}\label{sec-FE-solve}

We shall provide a computationally efficient method for robust computing the
characteristic responses of problems \eq{eq-lp3}-\eq{eq-lp6} within any
frequency range.

Let $\vbm = \imu\om\ubm$, then \eq{eq-nm1} can be written as
\begin{equation}\label{eq-nm5}
\begin{split}\left[
\begin{array}{ll}
\gamma\Dbm_\om\;,&  \Bbm^T \\
        \Bbm\;, & \Cbm
\end{array}\right]\left[
\begin{array}{c}
    \vbm \\ \pbm
\end{array}\right] = 
\left[
  \begin{array}{c}
    \gamma \rbm \\ -\bbm
\end{array}\right]\quad
  \mbox{ with }\Dbm_\om = \om^{-2} \Abm - \Mbm\;.
\end{split}
\end{equation}

It shows that both $\vbm$ and $\pbm$ are real column matrices, hence $\ubm$ is
purely imaginary. Further we consider the eigenvalues $\{\lam_j\}$ and
eigenvectors $\{\wbm^i\}$,
\begin{equation}\label{eq-nm6}
\begin{split}
\Abm\Wbm = \Mbm\Wbm\Lambf\;, \quad \Wbm^T\Mbm\Wbm = \Ibm\;,\quad \Lambf = \diagM{\lam_j}\;,\quad \Wbm = [\wbm^1,\dots,\wbm^n]\;,
\end{split}
\end{equation}
so that $\Wbm$ is the modal matrix composed of eigenvectors $\wbm^j$ associated
with eigenvalues $\lam^j \in \RR$, $j = 1,\dots,n$, where $n$ is the size and
also the rank of matrix $\Abm$.

Using the modal transformation, $\vbm = \sum_{i=1}^n \wbm^i \xm_i = \Wbm\xbm$,
where $\xbm = (\xm_i)$, due to \eq{eq-nm6}, problem \eq{eq-nm5} becomes
\begin{equation}\label{eq-nm7}
\begin{split}\left[
\begin{array}{ll}
\gamma(\om^{-2}\Lambf - \Ibm)\;,&  \Rbm^T \\
        \Rbm\;, & \Cbm
\end{array}\right]\left[
\begin{array}{c}
    \xbm \\ \pbm
\end{array}\right] = 
\left[
  \begin{array}{c}
    \gamma \Wbm^T\rbm \\ -\bbm
\end{array}\right]\;,\quad \mbox{ with } \Rbm = \Bbm\Wbm\;.
\end{split}
\end{equation}
Thus, system matrix in \eq{eq-nm7} becomes singular near resonances $\om^2 = \lam_j$ for some $j = 1,\dots,n$.

As the next step, we express $\pbm  = -\Cbm^{-1}(\Rbm\xbm + \bbm)$ from \eq{eq-nm7}$_2$, which yields
\begin{equation}\label{eq-nm8}
\begin{split}
  \Gbm \xbm = -\hbm\;,\quad \mbox{ with }
  \Gbm & = \Hbm - \om^{-2}\Lambf\;,\\
  \Hbm & = \Ibm + \gamma^{-1}\Rbm^T\Cbm^{-1}\Rbm \;,\\
  \hbm & = \Wbm^T\rbm +\gamma^{-1}\Rbm^T\Cbm^{-1}\bbm\;.
\end{split}
\end{equation}
While $\Hbm$ is a regular matrix, $\Gbm$ can become singular because of the resonances.
The eigenvalues $\{\eps_j\}$ and eigenvectors $\{\zbm^j\}$ of matrix $\Hbm$ satisfy
\begin{equation}\label{eq-nm9}
\begin{split}
\Hbm\Zbm = \Lambf\Zbm\Ebm\;, \quad \Zbm^T\Lambf\Zbm = \Ibm\;,\quad \Ebm = \diagM{\eps_j}\;,\quad \Zbm = [\zbm^1,\dots,\zbm^n]\;,
\end{split}
\end{equation}
The modal matrix $\Zbm$ is composed of eigenvectors $\zbm^j$ associated with eigenvalues $\eta^j \in \RR$, $j = 1,\dots,n$, where $n$ is the size of matrix $\Hbm$.
Using the modal transformation, $\xbm = \sum_{i=1}^n \zbm^i \ym_i = \Zbm\ybm$, where $\ybm = (\ym_i)$, problem \eq{eq-nm9} multiplied by $\Zbm^T$ from the left becomes
\begin{equation}\label{eq-nm10}
\begin{split}
  \Zbm^T(\Hbm - \om^{-2}\Lambf)\Zbm\ybm & = -\Zbm^T\hbm\;,\\
  \mbox{ hence }\quad \ybm & = -(\Ebm - \om^{-2}\Ibm)^{-1} \Zbm^T\hbm\\
  & = \diagM{\frac{\om^2}{1 - \om^2\eps_j}}\Zbm^T\hbm\;.
\end{split}
\end{equation}
Using all the backward substitutions, the characteristic responses of \eq{eq-nm1} can be expressed, as follows
\begin{equation}\label{eq-nm11}
\begin{split}
  \imu\om\ubm = \vbm = \Wbm\xbm = \Wbm\Zbm\ybm\;,\\
  \pbm = -\Cbm^{-1}(\Rbm\Zbm\ybm + \bbm)\;.
\end{split}
\end{equation}

\begin{myremark}{rem-nm1}
  Both the eigenvalue problems \eq{eq-nm6} and  \eq{eq-nm9} are defined in terms of real symmetric positive definite matrices, so that eigenvalues $\lam_j$ and $\eps_j$ are real positive.
  The resonance effect $\om^2 \rightarrow 1/\eps_j$ for any $j=1,\dots,n$ depends on the spectral properties of the matrices $\Abm$ and $\Cbm$ by virtue of eigenvalues $\{\lam_j\}$ and $\{\eps_j\}$, respectively; in the latter case, the projected inverse of $\Cbm$ is involved in the definition of matrix $\Hbm$ through the term $\Rbm^T\Cbm^{-1}\Rbm$ which is of the same dimension, as $\Abm$.

  To avoid computing of the inverse matrix $\Cbm^{-1}$, the following  problems are solved for matrix $\Sbm$ and vectors $\dbm_s$,
\begin{equation}\label{eq-nm11a}
  \begin{split}
  \Cbm\Sbm  & = \Rbm\;,\quad \Rightarrow \Rbm^T\Cbm^{-1}\Rbm = \Rbm^T\Sbm\;,\\
  \Cbm \dbm_s & = \bbm_s\;,\quad \mbox{ where } \bbm_s \mbox{ is substituted by }\bbm_\alpha,\bar\bbm_\alpha,\bbm_\xi,\bbm_3\;.
\end{split}
\end{equation}  
\end{myremark}

\subsubsection{Homogenized coefficients}\label{sec-FE-HC}

We shall now present formulae for computing the homogenized coefficients depending on the characteristic responses of microproblems \eq{eq-lp3}-\eq{eq-lp6} which all can be represented by the generic form of the microproblem \eq{eq-nm1} discretized by finite elements (FE).
The homogenized coefficients are evaluated using matrix objects introduced in Tab.~\ref{tab-hc1} associated with the formulae presented in Section~\ref{sec-HC1} by virtue of the FE discretization. Vectors $\bbm_s$ and $\rbm_s$, for index $s$ understood in the context of Tab.~\ref{tab-hc1}, also  constitute the \rhs vectors in the generic microproblem \eq{eq-nm1}. Furthermore, according to \eq{eq-nm8} and using \eq{eq-nm11a}, we compute column vectors 
\begin{equation}\label{eq-hc1}
  \begin{split}
  \hbm_s = \Wbm^T\rbm_s + \gamma^{-1}\Rbm^T\Cbm^{-1}\bbm_s = \Wbm^T\rbm_s + \gamma^{-1}\Sbm^T\bbm_s = \Wbm^T\rbm_s + \gamma^{-1}\Rbm^T\dbm_s\;, 
\end{split}
\end{equation}
which are involved in the expressions for $\ub_s$. 

\begin{table}
  \begin{tabular}{l|lll|l}
    micro- & char.  & FE-discretized & assoc. r.h.s.  & for $z$-symm.  \\
    problem & resp. & solutions & vectors &  cells $Y$ \\
    \hline
\eq{eq-lp3}  &  $(\hat w^\beta,\pi^\beta)$ & $(\ubm_\beta,\pbm_\beta)$ & $\zerobm$,\quad $\ol{\bbm}_\beta$ & $\ol{\ubm}_\beta \equiv  \zerobm$ \\
\eq{eq-lp4}  &     $(\hat \vsigma,\xi)$ & $(\ubm_\xi\pbm_\xi)$ & $\zerobm$,\quad  $\bbm_\xi$ & \\
\eq{eq-lp5}  &     $(\hat \vpi^\alpha,\eta^\alpha)$ & $(\ubm_\alpha,\pbm_\alpha)$ & $\zerobm$,\quad  $\bbm_\alpha$  & $\ubm_\alpha \equiv  \zerobm$ \\
\eq{eq-lp6}  &     $(\hat \vpi^3,\eta^3)$ & $(\ubm_3,\pbm_3)$ & $\zerobm$,\quad  $\bbm_3$ & \\
    \hline
  \end{tabular}
  \caption{Notation associated with microproblems \eq{eq-lp3}-\eq{eq-lp6}. In the last column, the cancellations follow due to  Proposition~\ref{thm-3}. }\label{tab-hc1}
  \end{table}

Below we introduce in detail only the matrix expression of coefficient
$M_{33}$, for the others the expressions are stated under the $z$-symmetry
assumption which applies in the case of the plate perforations considered in
the present study. For more general plate perforations, expressions of the
homogenized coefficients are derived in \Appx{apx-3} using the spectral matrix
objects $\Zbm\diag({\om^2}/{1-\om^2\eps_j})\Zbm^T$ and vectors $\bbm$ and
$\hbm$ introduced in \eq{eq-nm1b} and \eq{eq-nm8}.

\paragraph{ Coefficient $M_{33}$} This coefficient depends on the mean plate density, $\ol{\rho_S}$, in particular
$M_{33} = \ol{\rho_S} + \wtilde{M}_{33}(\om)$, where $\wtilde{M}_{33}$ is expressed by $(\hat \vpi^3,\eta^3)$ being approximated by the couple $(\ubm_3,\pbm_3)$. Using the discretized expressions \eq{eq-nm1b}$_4$ 
we get 
\begin{equation}\label{eq-nm12}
\begin{split}
 \wtilde{M}_{33}(\om) & = \imu\om\gamma\intY_{\Xi_c} \rho\hat\vpi^3  - \intY_{\Xi_S}\Jump{\eta^3}{\hh} \\
&   \FEMapprox \imu\om \gamma\rbm_3^T \ubm_3 - \bbm_3^T\pbm_3 = \gamma\rbm_3^T \Wbm\xbm + \bbm_3^T\Cbm^{-1}(\Rbm\xbm + \bbm_3)\\
  & = \gamma\hbm_3^T\Zbm\ybm + \bbm_3^T\Cbm^{-1}\bbm_3\\
  & = \gamma\hbm_3^T\Zbm\diagM{\frac{\om^2}{1-\om^2\eps_j}}\Zbm^T\hbm_3 + \bbm_3^T\Cbm^{-1}\bbm_3\;,\\
  \mbox{ hence }\quad M_{33} &= \gamma\ol{\rho_S} + \bbm_3^T\Cbm^{-1}\bbm_3 + \gamma\hbm_3^T\Zbm\diagM{\frac{\om^2}{1-\om^2\eps_j}}\Zbm^T\hbm_3\;,
\end{split}
\end{equation}
where $\xbm$ and $\ybm$ are associated to $\ubm$ by virtue of \eq{eq-nm11}.

For all other homogenized coefficients depending on characteristic responses \eq{eq-lp3}-\eq{eq-lp6} the expressions are derived in analogy with \eq{eq-nm12} using the $\bbm_s$ and $\hbm_s$ defined in \eq{eq-nm1b} and \eq{eq-nm8}, taking into account the $z$-symmetry property of cell $Y$. Hence, as the consequence of Lemma~\ref{thm-0}, 
\begin{equation}\label{eq-hc2}
\begin{split}
    \bbm_3^T\pbm_\beta = \zerobm\;,\quad  \bbm_3^T\ol{\pbm}_\beta =  \zerobm\;,\\
    \bbm_\xi^T\pbm_\beta = \zerobm\;,\quad  \bbm_\xi^T\ol{\pbm}_\beta =  \zerobm\;.
\end{split}
\end{equation}
Recalling Proposition~\ref{thm-3}, $\ol{\ubm}_\alpha$ and $\ubm_\alpha$ vanish, which is indicated consistently by vanishing projections
\begin{equation}\label{eq-hc3}
\begin{split}
 \Zbm^T \bbm_\alpha = \zerobm\;,\quad \mbox{ and } \quad \Zbm^T \ol{\bbm}_\alpha= \zerobm\;.
\end{split}
\end{equation}
By the consequence, expressions of some of the homogenized coefficients simplify. With reference to Proposition~\ref{thm-3} (iii), we first present the nonvanishing coefficients,
\begin{equation}\label{eq-hc4}
\begin{split}
 {M}_{\alpha\beta}(\om) & = -\hh\intY_{\pd \Xi_S} n_\alpha \int_{-1/2}^{1/2}\eta^\beta\dd\zeta \FEMapprox    - \bbm_\alpha^T\pbm_\beta = \bbm_\alpha^T\Cbm^{-1}\ol{\bbm}_\beta\;,\\
   D_{\alpha\beta} & = \hh\intY_{\pd\Xi_S} n_\alpha  \int_{-1/2}^{1/2}\pi^\beta \FEMapprox -\bbm_\alpha^T\Cbm^{-1}\bar\bbm_\beta\;,\\
   C_3 & =  \intY_{\Xi_S} \Jump{\xi}{\hh} - \imu\om \gamma \intY_{\Xi_c} \rho_c \hat \vsigma  \FEMapprox -\bbm_3^T\Cbm^{-1}\bbm_\xi - \gamma \hbm_3^T\Zbm\diagM{\frac{\om^2}{1-\om^2\eps_j}}\Zbm^T\hbm_\xi\;,\\
   F & = -\left(\intY_{I_y^+} \xi - \intY_{I_y^-}  \xi \right) \FEMapprox \bbm_\xi^T\Cbm^{-1}\bbm_\xi + \gamma\hbm_\xi\Zbm\diagM{\frac{\om^2}{1-\om^2\eps_j}}\Zbm^T\hbm_\xi \;.
\end{split}
\end{equation}
The acoustic propagation tensor  $A_{\alpha\beta}$ is computed using the solution of problem \eq{eq-lp3a}. The generic discretized representation \eq{eq-nm1} involves $(\bar\ubm_\beta,\bar\pbm_\beta)$ as the approximations of $(\hat \vpi^\beta,\pi^\beta)$. Using the notation reported in \eq{eq-nm1b} and $\hbm_\alpha$ evaluated according to \eq{eq-hc1}, the following expressions are obtained, 
\begin{equation}\label{eq-hc5}
\begin{split}
  A_{\alpha\beta} & =\ipYs{\nabla_y (\pi^\beta + y_\beta)}{\nabla_yy_\alpha} =
  \frac{|Y^*|}{|\Xi|}\delta_{\alpha\beta} + \wtilde A_{\alpha\beta}\;,\\
 \mbox{ where }\quad \wtilde A_{\alpha\beta} & = \ipYs{\gradply y_\alpha}{\gradply \pi^\beta} \FEMapprox \bar\bbm_\alpha^T\bar\pbm_\beta\\
  & = -\bar\bbm_\alpha^T\Cbm^{-1}\bar\bbm_\beta - \gamma\bar\hbm_\alpha^T\Zbm\diagM{\frac{\om^2}{1-\om^2\eps_j}}\Zbm^T\bar\hbm_\beta\;,
\end{split}
\end{equation}
hence the symmetry $A_{\alpha\beta} = A_{\beta\alpha}$ is confirmed. 

Finally, the  coefficients vanishing due to \eq{eq-hc2} and \eq{eq-hc3}  are listed
\begin{equation}\label{eq-hc6}
\begin{split}
  {M}_{3\beta}(\om) & = \imu\om\gamma\intY_{\Xi_c} \rho\hat\vpi^\beta  - \intY_{\Xi_S}\Jump{\eta^\beta}{\hh} =  \FEMapprox \imu\om  \gamma\rbm_3^T \ubm_\beta - \bbm_3^T\pbm_\beta = 0\;,\\
   D_{3 \beta} & =   \intY_{\Xi_S}\Jump{\pi^\beta}{\hh} - \imu\om \gamma\intY_{\Xi_c} \rho_c \hat w^\beta
   \FEMapprox \bbm_3^T\bar\pbm_\beta - \imu\om  \gamma\rbm_3^T\bar\ubm_\beta = 0\;,\\
   C_\alpha & = \gamma\rho_0\intY_{\pd\Xi_S} n_\alpha  \int_{-1/2}^{1/2}\xi \FEMapprox \bbm_\alpha^T \pbm_\xi = 0\;,\\
   B_{\alpha} & = \ipYs{\gradply y_\alpha}{\gradply \xi} \FEMapprox \bar\bbm_\alpha^T\pbm_\xi = 0\;.
\end{split}
\end{equation}

\subsection{Macroscopic problem --- matrix formulation}


We consider a finite thickness $\dlt_0 = \veps_0\vkappa$ of the transmission layer with $\veps_0 > 0$ reflecting the size of the perforations. By the consequence, the global domain $\Om_G^{\dlt_0}$ splits into two parts $\Om_+^{\dlt_0}$ and $\Om_-^{\dlt_0}$ separated by the transmission layer $\Gamma_{\dlt_0}$ which is represented by the homogenized transmission conditions \eq{eq-DtN-1}-\eq{eq-DtN-3} prescribed on $\Gamma_0$. The acoustic field $\hat P^\pm$ satisfying \eq{eq-DtN-0} is aproximated by the 3D conforming P2 Lagrangian tetrahedral elements in  $\Om_G^{\dlt_0}$. Accordingly, identic discretizations of interfaces $\Gamma_+^{\dlt_0}$, $\Gamma_-^{\dlt_0}$ and $\Gamma_0$ by triangular P2 elements are considered, so that  condition \eq{eq-DtN-3} can be satisfied in the sense of the collocation at corresponding nodes. The fluxes $g^0$ and $\hat G_0^\pm$ are approximated by P1 elements on the matching triangular mesh.
As for the plated displacements $\ub$ and rotations $\thetabf$ on $\Gamma_0$, these are approximated by P2 elements and P1 elements, respectively. The vectors containing the nodal DOFs associated with discretised fields are designated in the following table:  

\begin{center}
  \begin{tabular}{llll}
    vector / DOFs & & approximated field & FE type \\
    \hline 
$\pbm^\pm$ &  & $\hat P^\pm$, pressure traces on $\Gamma_0^\pm$ & P2 \\
$\gbm^\pm$   &  & $\hat G^\pm$, acoustic momentum fluxes  on $\Gamma_0^\pm$ & P1  \\
$\pbm^0$   &  &  $p^0$, mean acoustic pressure on $\Gamma_0$ & P2 \\
$\gbm^0$   &  & $g^0$, mean acoustic momentum flux on $\Gamma_0$ & P1 \\
$\bar\ubm$ &  & $u_3$, deflections on $\Gamma_0$ & P2 \\
$\wbm$ &  & $\thetabf$, rotations on $\Gamma_0$ & P1 \\
    \hline
\end{tabular}
\end{center}

\subsubsection{FE discretized formulation of the transmission conditions}
The FE-discretization of 
\eq{eq-DtN-1}-\eq{eq-DtN-3} leads to a matrix formulation involving the global vector of unknowns.  Two alternatives  can be used 
\begin{equation}\label{eq-MN4}
  \begin{split}
\bar\qbm & = [\pbm^\pm; \gbm^\pm; \ubm; \wbm]\;, \quad \mbox{ with } \ubm = [\bar\ubm;\hat\ubm]\;,  \\  
\mbox{ or } \qbm & = [\pbm^0;  \gbm^0; \ubm; \wbm]\;. 
\end{split}
\end{equation}
where the Matlab notation ``;'' means the ``line break'', thus, $\qbm$ is the column vector.

To express $p^0$ and $g^0$, nodal (collocation at FE nodes) operations are introduced,
\begin{equation}\label{eq-MN5}
  \begin{split}
\gbm^0 = \Xbm \gbm^\pm\;, \quad \pbm^0 = \Xbm \pbm^\pm\;,
\end{split}
\end{equation}
whereas the jumps are expressed reciprocally (collocation at FE nodes),
\begin{equation}\label{eq-MN6}
  \begin{split}
\Dlt  G^1 \FEMapprox \Zbm_\veps \gbm^\pm\;, \quad \Dlt \hat P^\pm = \Zbm_\veps \pbm^\pm\;,
\end{split}
\end{equation}
where $\Zbm_\veps$ involves $\veps^{-1}$, the (finite) scale, see \eq{eq-G3-03} and \eq{eq-G3-04}.
When approximating the DtN operator given in its implicit form by \eq{eq-DtN-1}-\eq{eq-DtN-3}, the following matrix expressions of the discretized bilinear forms hold,

\begin{equation}\label{eq-bilin-FE1}
\begin{split}
  \Acal(p^0,q^0) \FEMapprox (\pbm^0)^T \Abm \qbm^0\;, & \quad \Kcal(p^0,q^0) \FEMapprox  (\pbm^0)^T \Qbm \qbm^0 \;,\\
  \Ecal(\thetabf,\vthetabf) \FEMapprox \tbm^T \Ebm \wbm \;, & \quad \Mcal(\thetabf,\vthetabf)   \FEMapprox \tbm^T \Tbm \wbm\;,\\
  \Hcal(\ol{\ub},p^0) +  \Dcal(\ol{\ub},p^0)  \FEMapprox (\pbm^0)^T \Dbm \bar\ubm\;, \quad
  & \ipGm{g^0}{p^0}  \FEMapprox (\pbm^0)^T \Jbm \gbm^0\\
  \Ccal(u_3,g^0) \FEMapprox (\gbm^0)^T \Cbm \hat\ubm \;, \quad  & \Fcal(g^0,\tilde g) \FEMapprox \tilde \gbm^T \Fbm \gbm^0\;,\\
  \Lcal(\ol{\ub},\ol{\vb}) \FEMapprox \bar\vbm^T \Lbm \bar\ubm \;, \quad  &
\Ncal(u_3,v_3) \FEMapprox \hat\Vbm^T \Nbm \hat\ubm\;.
\end{split}
\end{equation}
Matrix $\Jbm$ represents the scalar pruducts (products of P2 and P1 functions of the FE partitioning, in our case) 
\begin{equation}\label{eq-MN9}
  \begin{split}
    \int_{\Gamma_0} \psi (\Delta{\hat P}) \;,\quad \mbox{ and }  \int_{\Gamma_0}q^0 \Dlt G^1\;.
  \end{split}
\end{equation}
The shear-rotation coupling in the bilinear form $\Scal$ yields the following discretized representation,
\begin{equation}\label{eq-bilin-FE2}
  \begin{split}
    \Scal((u_3,\thetabf),(v_3,\vthetabf)) & \FEMapprox  [\bar\vbm;\hat\vbm;\tbm]^T
    \left[\begin{array}{lll}
 \zerobm &   \zerobm & \zerobm   \\  
\zerobm & \Sbm & \Rbm \\ \zerobm & \Rbm^T & \zerobm
      \end{array}\right] \left[\begin{array}{l}
\bar\ubm \\ \hat\ubm \\ \wbm
      \end{array}\right]
    \;.\\   
  \end{split}
\end{equation}

Further we employ the stiffness and the mass matrices, denoted by $\Kbm$ and $\Mbm$, respectively,
\begin{equation}\label{eq-MN7}
  \begin{split}
& \Kbm = \left[
\begin{array}{ll}
  \Ebm & \zerobm \\
  \zerobm & \Sbm
\end{array}  \right] \;,\quad
\Mbm = \left[
\begin{array}{ll}
  \Lbm & \zerobm \\
  \zerobm & \Nbm
\end{array}  \right] \;,
  \end{split}
\end{equation}
which enable to express the strain  and kinetic energies of the plate denoted by  $E_S$ and $E_K$, respectively, being associated with the harmonic wave propagation,
\begin{equation}\label{eq-MN7a}
  \begin{split}
     E_S = \frac{h}{2}[\ubm;\wbm]^T
\left[
\begin{array}{ll}
  \Kbm & -\Rbm \\
  -\Rbm^T &  \frac{h^2}{12}\Ebm
\end{array}  \right]
[\ubm;\wbm]\;, \\ 
E_K = -\om^2\frac{h}{2}
[\ubm;\wbm]^T
\left[
\begin{array}{ll}
  \Mbm & \zerobm\\
  \zerobm & \frac{h^2}{12}\Tbm
\end{array}  \right]
[\ubm;\wbm]\;.
  \end{split}
\end{equation}
Thus, matrix $\Kbm$ expresses the stiffness \wrt the displacements, \ie involving the Young and the shear moduli, whereas matrix $\Mbm$ describes the frequency-dependent mass associated with plate displacements. Note that the elastic and kinetic energies depend also on the rotation.

Using a self-explaining notation, the system \eq{eq-MN1}-\eq{eq-MN3} is represented by the following matrix equation,
\begin{equation}\label{eq-MN8}
  \begin{split}
 & \bar \Hbm \bar\qbm  = \zerobm\;,
\end{split}
\end{equation}
where
\begin{equation*}\label{eq-MN8a}
\begin{split}
& \bar \Hbm = \left[
  \begin{array}{llll}
    \Xbm^T(\Abm -\om^2 \Qbm)\Xbm & -\imu\om\Xbm^T\Jbm\Zbm_\veps &  \imu\om \Xbm^T\Dbm & \zerobm \\
    \imu\om \Zbm_\veps^T\Jbm\Xbm & -\om^2 \Xbm^T\Fbm\Xbm & \om^2 \Xbm^T\Cbm & \zerobm \\
    -\imu\om \Dbm^T\Xbm &  \om^2 \Cbm^T\Xbm  & \Kbm - \om^2\Mbm & -\Rbm \\
    \zerobm & \zerobm & -\Rbm^T & \frac{h^2}{12}( \Ebm - \om^2 \Tbm )
  \end{array}  \right]\;.  
\end{split}
\end{equation*}
is Hermitean matrix. Recalling Proposition~\ref{thm-3}, matrices $\Mbm, \Tbm, \Cbm$ and $\Fbm$ depend on the frequency $\om$ and  are affected by ``micro-level'' resonances through the homogenized coefficients;; in the case of $\Mbm$, these resonances are associated with \eq{eq-nm10} through coefficients $M_{kl}$, see \eq{eq-nm12} and \eq{eq-hc4}$_1$,  and also with the resonances featuring coefficients $\tilde\Mcal_{\alpha\beta}$, see \eq{eq-M3}.


\section{Numerical examples}\label{sec-numex}

This section with numerical examples is divided into two parts. The aim of the
first part is to compare the results calculated by the newly derived model and
by the previously published vibroacoustic model, see
\cite{Rohan-Lukes-AMC2019}. Our new model of the vibroacoustic transmission
involves frequency-dependent homogenized coefficients and will be further
referred to as the {\em metamaterial model}. The model published and validated
in \cite{Rohan-Lukes-AMC2019} will be denoted as a {\em standard model}. In the
second part of this section, we demonstrate the microstructure with a modified
inclusion, with which we can effectively change the distribution of resonant
frequencies.

The presented numerical simulations are base on the discretization by means of
the finite element method and have been implemented in {\em SfePy} -- Simple
Finite Elements in Python \cite{sfepy-multiscale-2019}.

\subsection{Wave propagation in a waveguide}

\begin{figure}[ht]
    \centering
    \includegraphics[width=0.8\linewidth]{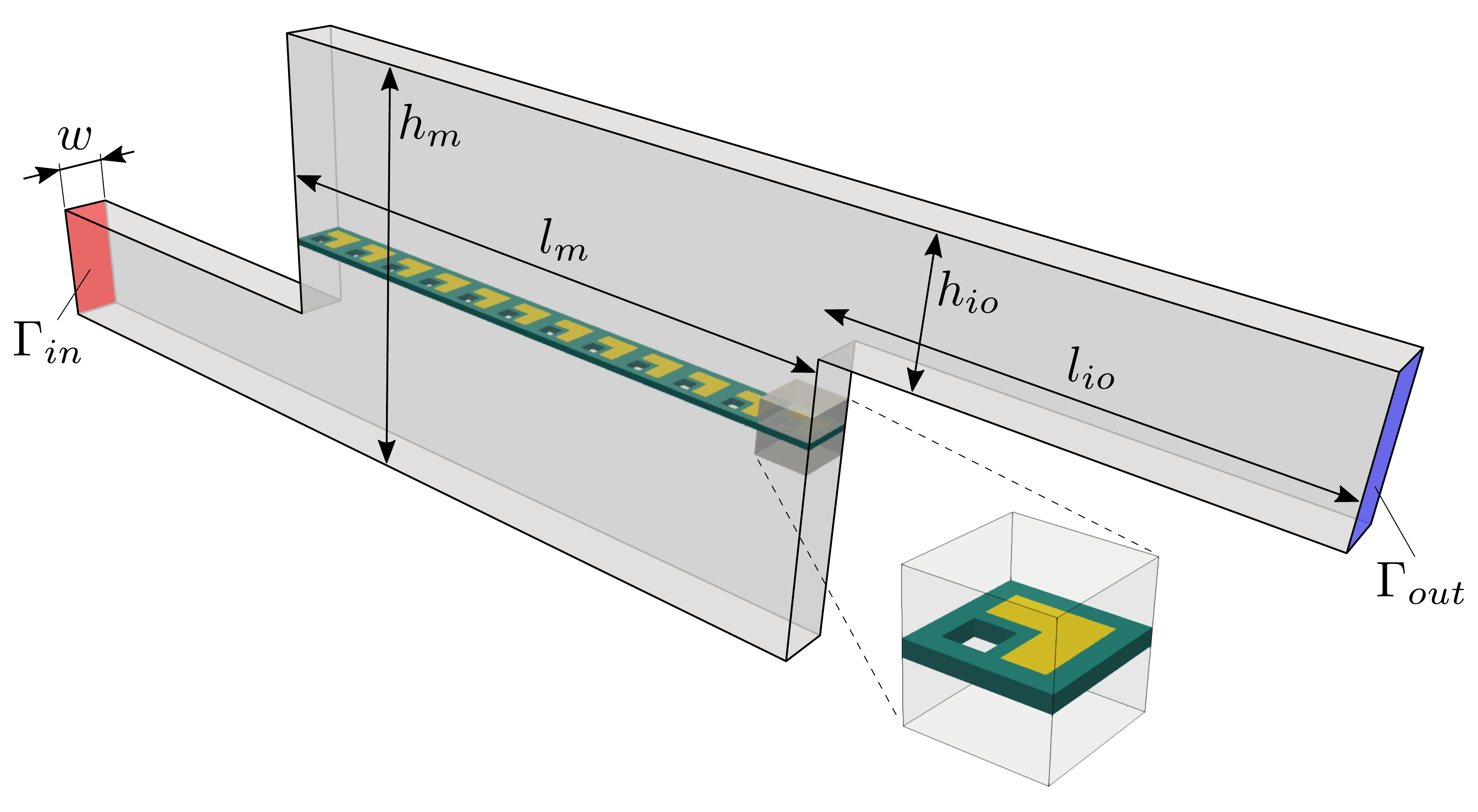}
    \caption{The acoustic domain with the embedded perforated elastic plate
             and the detailed view of the perforation.}
    \label{fig:num-wg}
\end{figure}

We consider the waveguide which domain is divided by the perforated plate into
two parts of the same shape and size as illustrated in Fig.~\ref{fig:num-wg}.
The waveguide dimensions are the following: $l_m = 0.3$\,m, $h_m = l_{io} =
0.2$\,m, $h_{io} = 0.0625$\,m and $w = 0.01$\,m.

\begin{figure}[ht]
    \centering
    \includegraphics[width=0.98\linewidth]{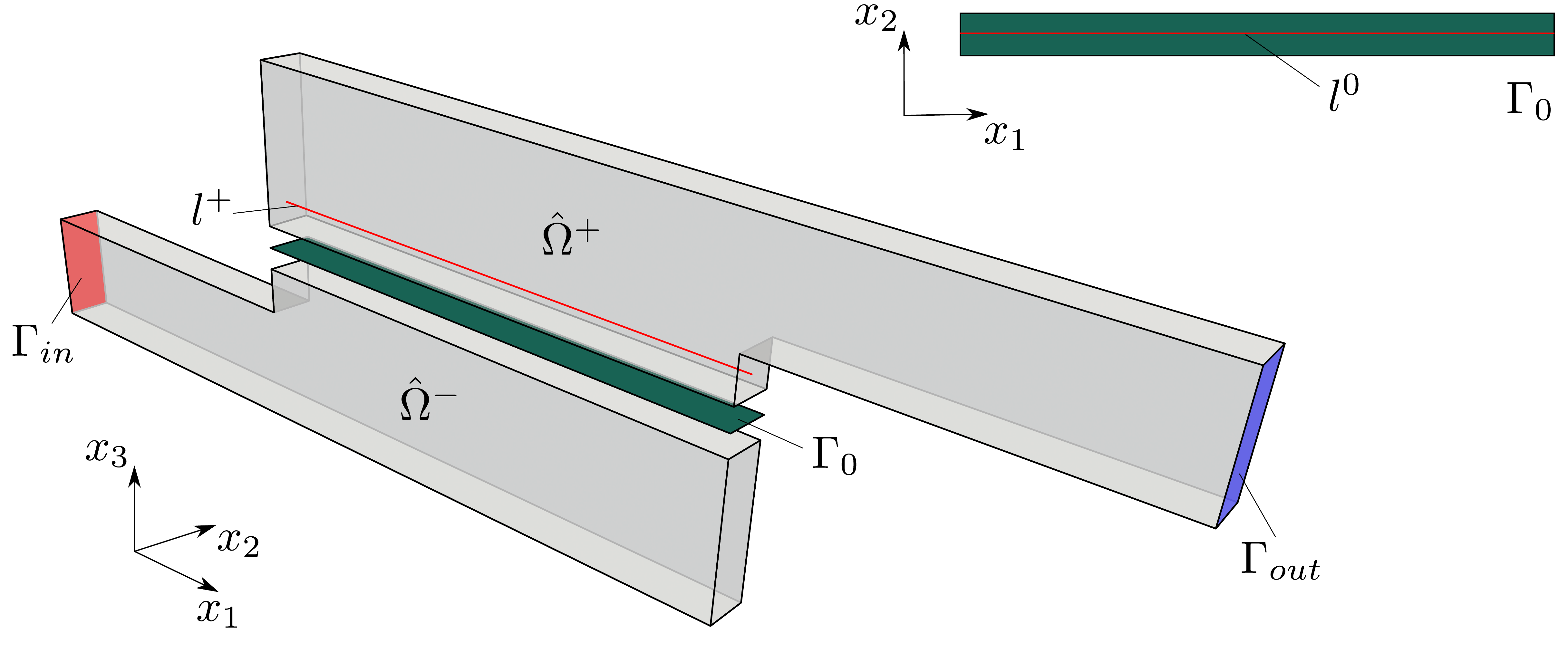}
    \caption{The computational domains $\hat\Omega^+$, $\hat\Omega^-$, $\Gamma_0$
             employed in the numerical simulations of the acoustic transmission
             at the global (macroscopic) level.}
    \label{fig:num-domain-mac}
\end{figure}

The computational domains $\hat\Omega^+$, $\hat\Omega^-$ and $\Gamma_0$,
employed for the numerical simulations of the acoustic transmission, are
depicted in Fig.~\ref{fig:num-domain-mac}. The waveguide input is labelled as
$\Gamma_{in}$ and an incident wave is imposed on this part of the boundary.
This condition is respected in \eq{eq-G2-02d} by values $r = s = 1$ and the
amplitude of the incident wave, appearing in \eq{eq-G2-02d}, is assumed to be
$\bar p = 30$\,Pa. At the waveguide output $\Gamma_{out}$, we consider the
anechoic condition achieved by choosing $r = 1$, $s = 0$ in \eq{eq-G2-02d} on
surface $\Gamma_{out}$. The elastic plate is fixed at the edges parallel to the
$x_2$-axis, see Fig.~\ref{fig:num-domain-mac}, such that $\bar{\bm{u}} =
\bm{0}$, $w_3 = 0$ and $\bm\theta = \bm{0}$. The periodic boundary conditions
are applied on the faces orthogonal to the $x_2$-axis. This requirement leads
to the homogeneous distribution of the macroscopic fields in the $x_2$-direction.

\begin{figure}[ht]
    \centering
    \includegraphics[width=0.7\linewidth]{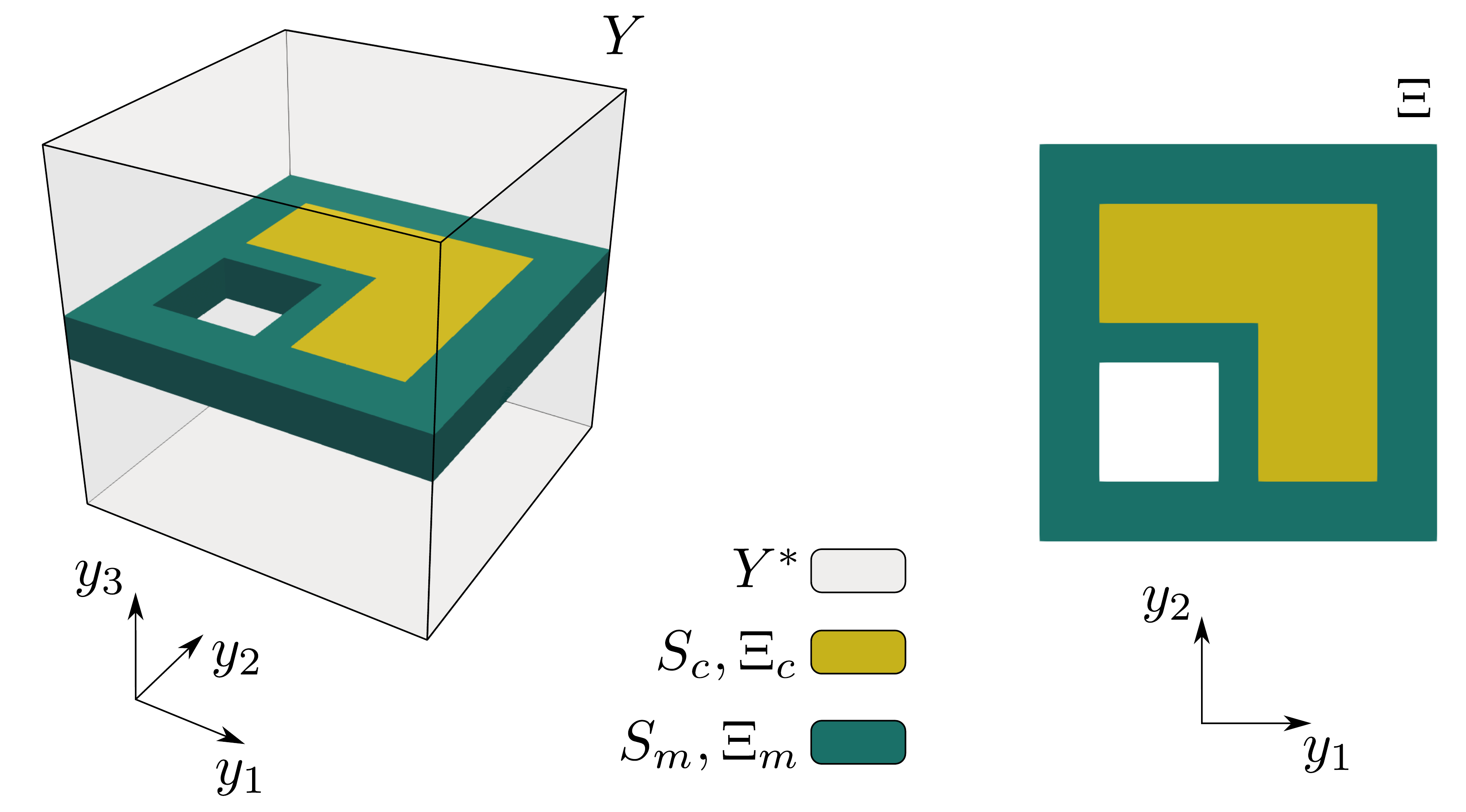}
    \caption{The computational domains $Y$ and $\Xi$
             involved in the calculations of the characteristic responses
             and the homogenized coefficients.}
    \label{fig:num-domain-mic}
\end{figure}

The characteristic responses and the homogenized coefficients are calculated
within the representative cells $Y$ and $\Xi$, see
Fig.~\ref{fig:num-domain-mic}. The 3D computational domain $Y$ consists of the
fluid domain $Y^\ast$, the elastic matrix $S_m$ and the elastic inclusion
$S_c$. In the similar way, the 2D plate domain $\Xi$ comprises the matrix
and inclusion parts $\Xi_m$ and $\Xi_c$. The acoustic fluid is characterized by
its density $\rho_0 = 1.55$\,kg\,m$^3$ and by the sound speed $c =
343$\,m\,s$^{-1}$. We consider the aluminium matrix with the density $\rho_m =
2700$\,kg\,m$^3$, Young modulus $E_m = 70$\,GPa and Poisson ratio $\nu_m =
0.34$. In order to obtain the contrast in the elasticity of the constituents,
we employ the rubber inclusion with material parameters $\rho_c =
1200$\,kg\,m$^3$, $E_c^\varepsilon = 0.1$\,GPa, thus $E_c = E_c^\varepsilon /
\varepsilon^2$, and $\nu_c = 0.48$. The scaling parameter $\varepsilon$
defining the size of the holes and elastic inclusions is chosen as $\varepsilon
= 0.01$, which means that the thickness of the plate is $h^\varepsilon = \bar h
\varepsilon$ for $\bar h = 0.12$.

Computational analysis of the resonant frequencies of subproblems \eq{eq-lp7} and
\eq{eq-lp3}--\eq{eq-lp6} results in the values presented in
Tab.~\ref{tab:num-crit_freqs_I} and Tab.~\ref{tab:num-crit_freqs_II}, where
only the first six lowest frequencies for each subproblem are displayed.

\begin{table}[ht!]
    \begin{tabular}{c||l|l|l|l|l|l}
        $\omega^{cf_{I}} = \sqrt{\lambda}$ [Hz]&
        246051 &
        263714 &
        274403 &
        301639 &
        326247 &
        351988 \\
    \end{tabular}
    \caption{Critical frequencies related to problem \eq{eq-lp7},
        where $\lambda$ are problem eigenvalues, see \eq{eq-lp8}.}
    \label{tab:num-crit_freqs_I}
\end{table}

\begin{table}[ht!]
    \begin{tabular}{c||l|l|l|l|l|l}
        $\omega^{cf_{II}} = \sqrt{1/\epsilon}$ [Hz]&
        36035 &
        47332 &
        54616 &
        67843 &
        68262 &
        77368
    \end{tabular}
    \caption{Critical frequencies related to problems \eq{eq-lp3}--\eq{eq-lp6},
        where $\epsilon$ are eigenvalues calculated according to \eq{eq-nm9}.}
    \label{tab:num-crit_freqs_II}
\end{table}

\begin{figure}[ht]
    \centering
    \includegraphics[width=0.48\linewidth]{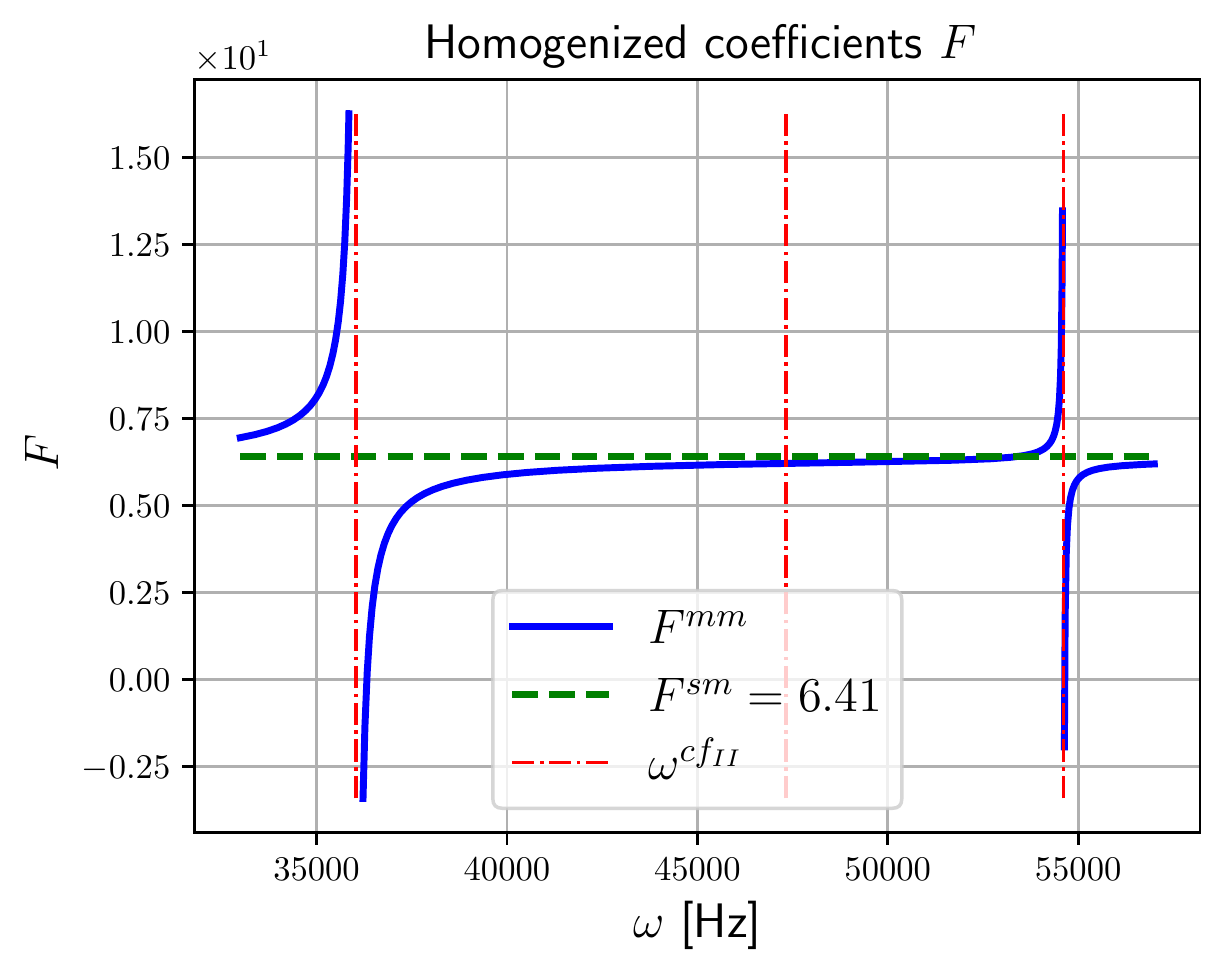}\hfil
    \includegraphics[width=0.48\linewidth]{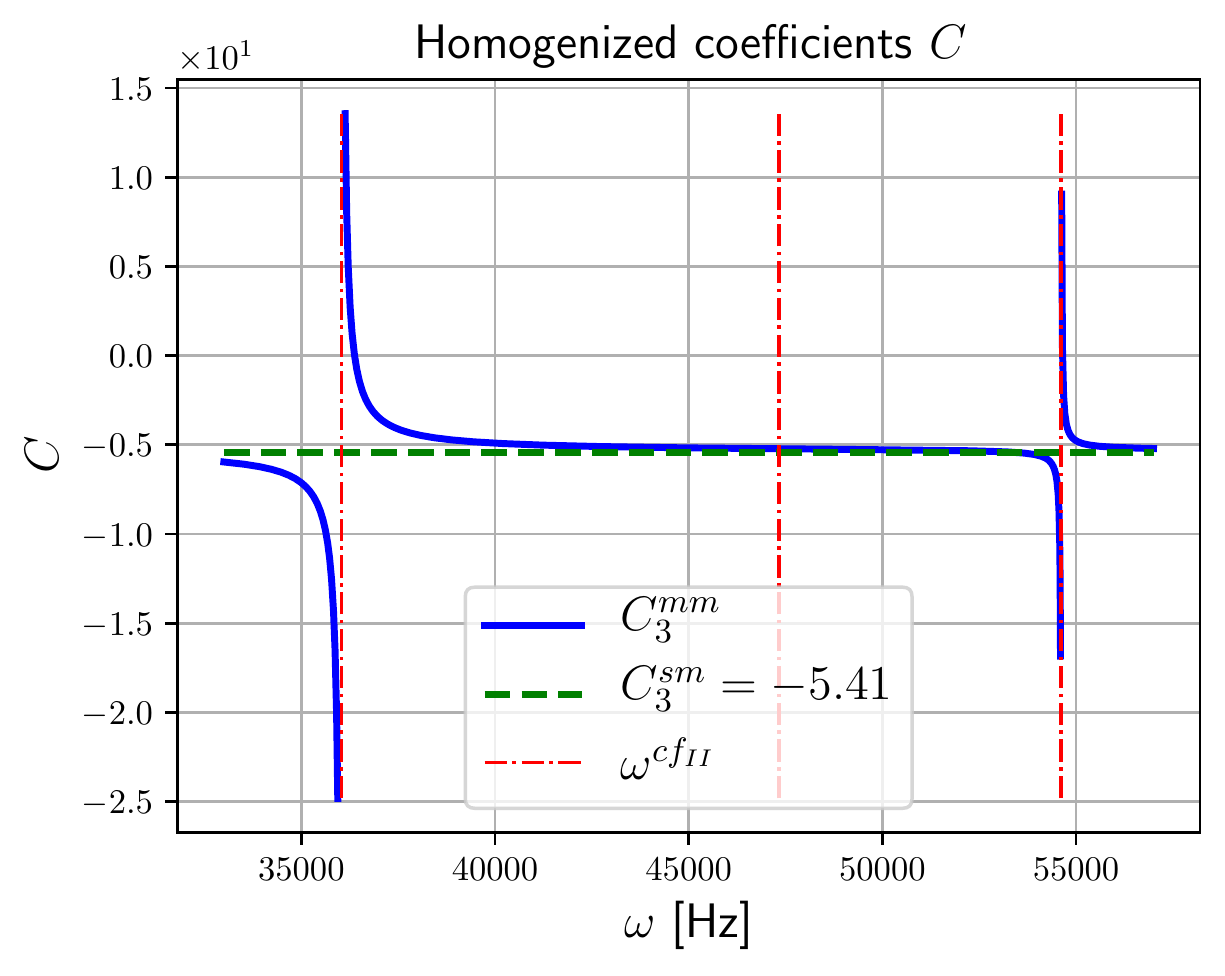}\\
    \includegraphics[width=0.48\linewidth]{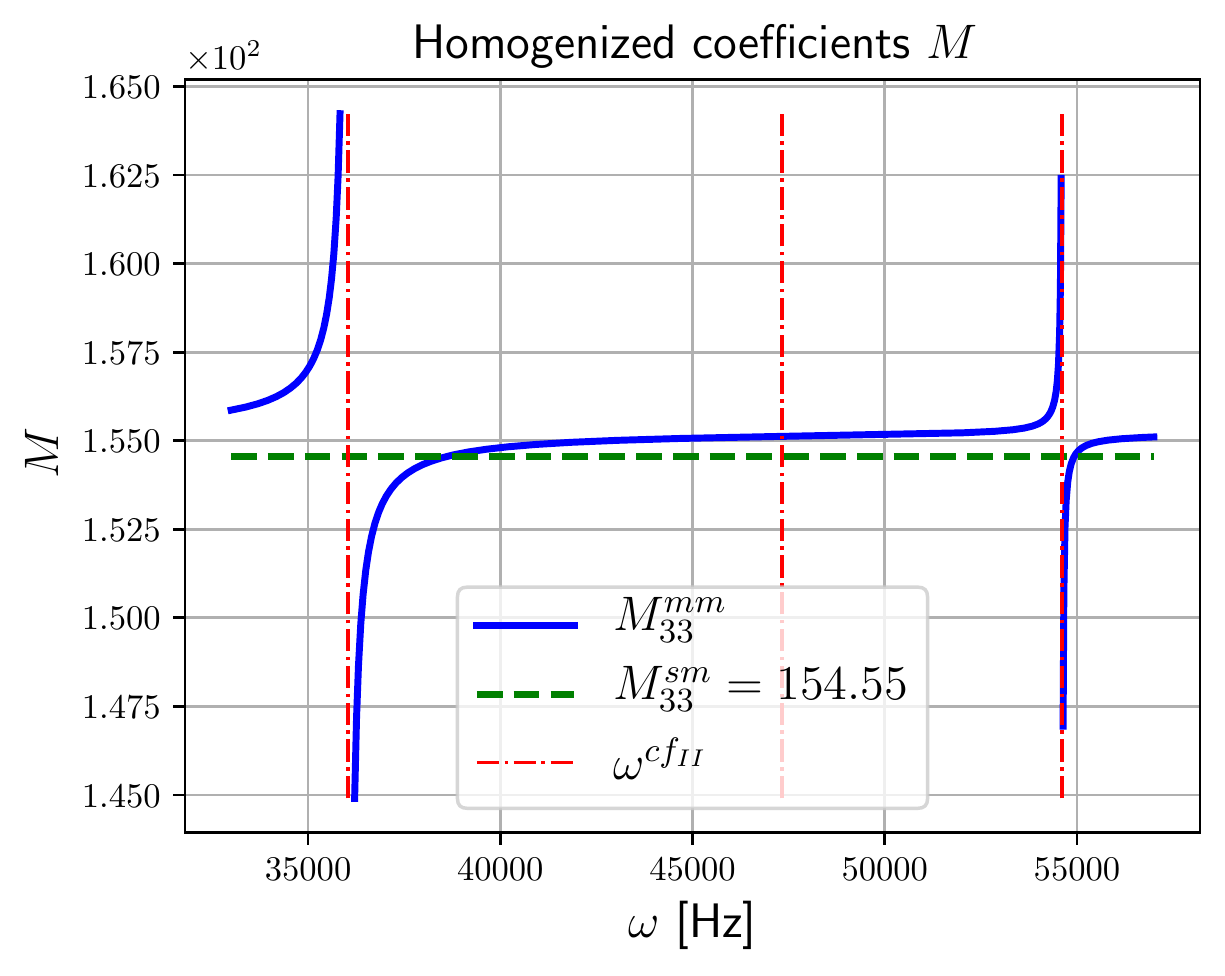}\hfil
    \includegraphics[width=0.48\linewidth]{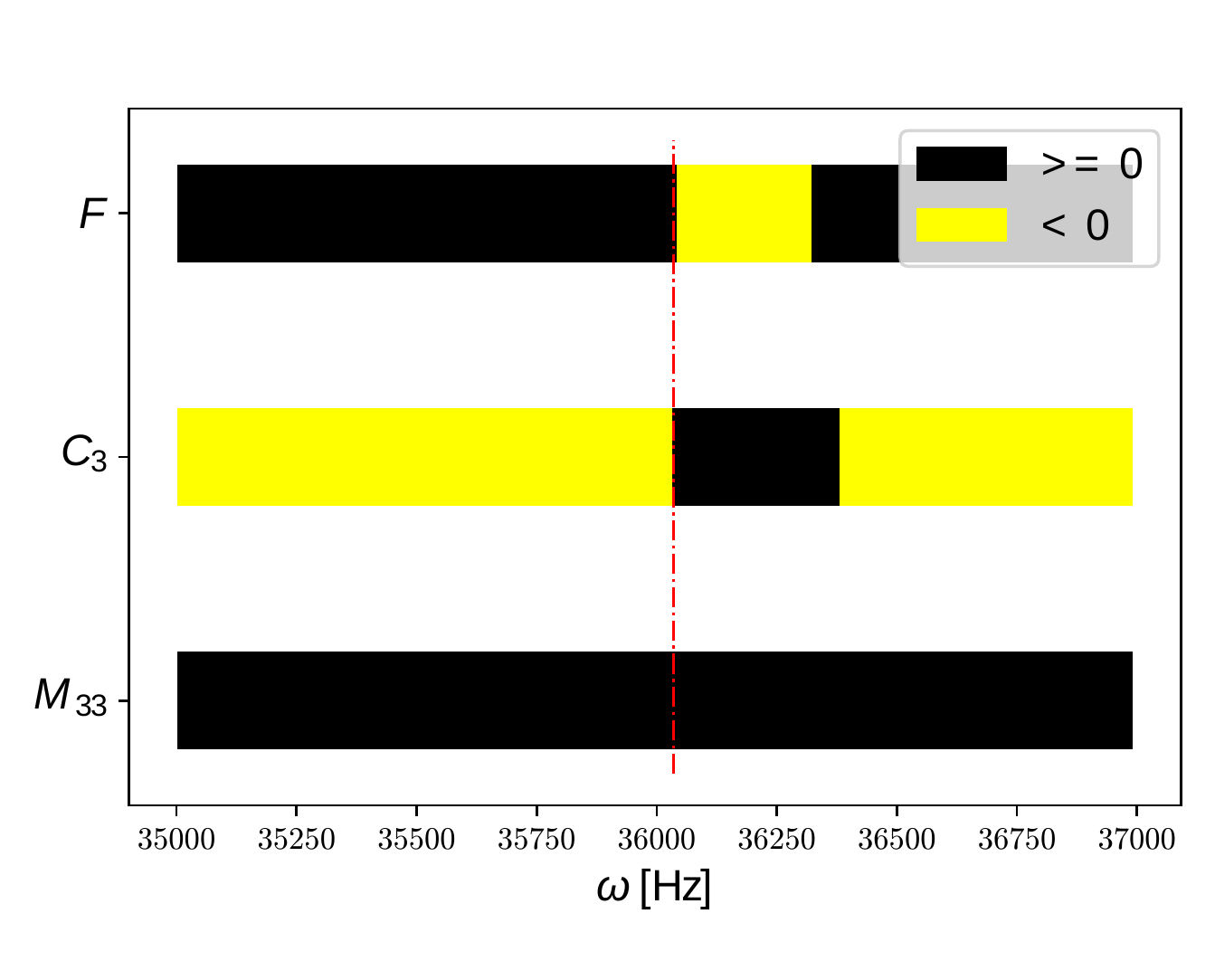}\\
    \caption{The frequency dependent homogenized coefficients $F$, $C_3$ and $M_{33}$ 
        calculated for the frequency range
         33\,KHz -- 57\,KHz.}
    \label{fig:num-coefs}
\end{figure}

\begin{figure}[ht]
    \centering
    \includegraphics[width=0.85\linewidth]{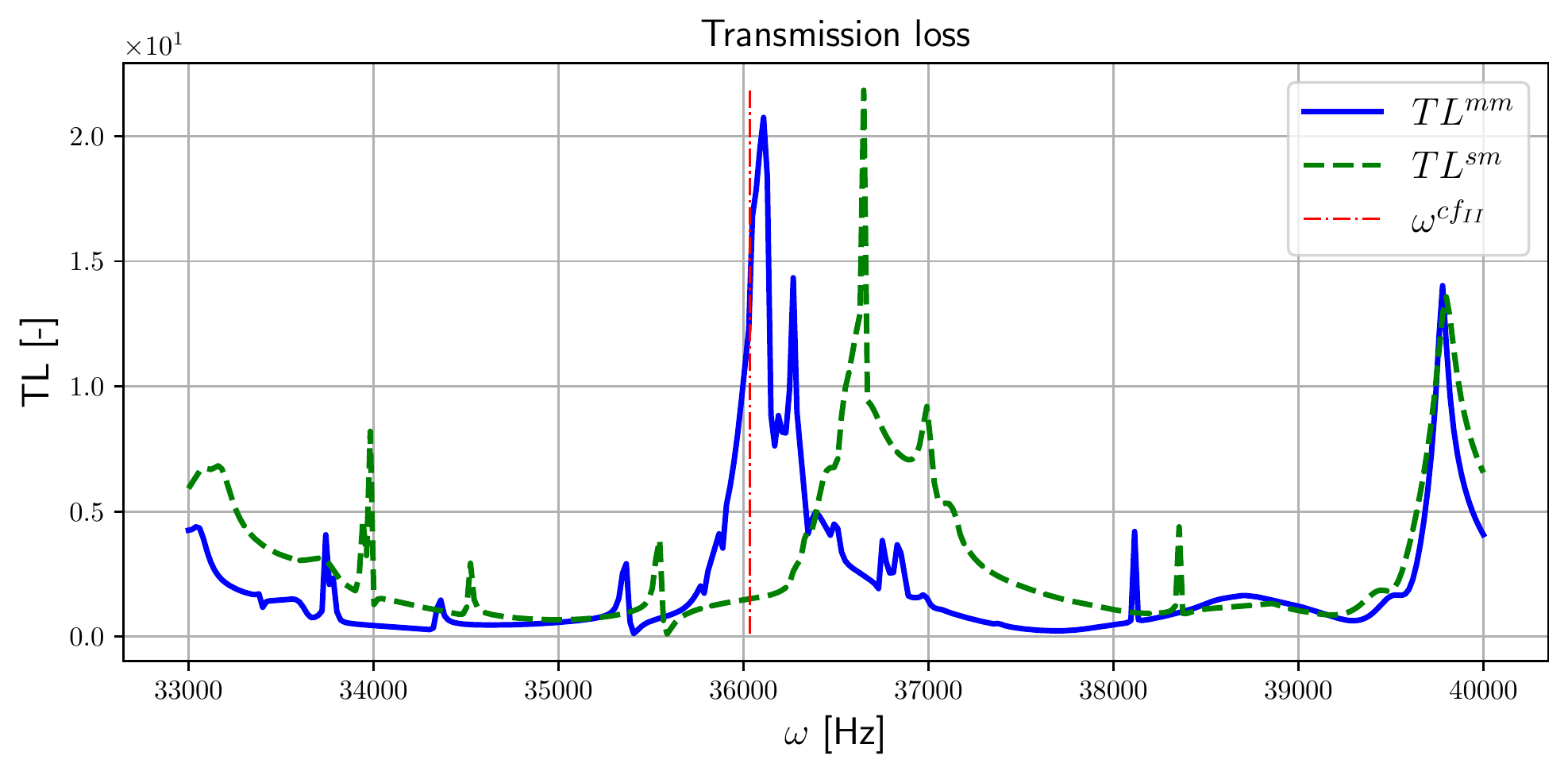}
    \caption{The transmission loss curves calculated by the metamaterial ($TL^{mm}$)
             and standard ($TL^{sm}$) models.}
    \label{fig:num-tl_curves}
\end{figure}

In the following part, we compare the responses of the proposed model, which
involves the frequency varying homogenized coefficients, with those obtained by
the vibroacoustic model with frequency independent coefficients published in
\cite{Rohan-Lukes-AMC2019}. According to the notation discussed in the
beginning of this section, the calculated values are labelled by superscripts
$^{mm}$ (metamaterial model) and $^{sm}$ (standard model). We assume the
frequency range 33\,KHz -- 57\,KHz, involving the first three critical
frequencies of problems \eq{eq-lp3}--\eq{eq-lp6}. In Fig.~\ref{fig:num-coefs}
we plot the frequency dependent homogenized coefficients $F$, $C_3$, $M_{33}$
and the constant values of these coefficients obtained by the standard model.
To compare a global (macroscopic) response of the models, we calculate the
transmission loss $\mbox{TL}$ employing the following expression:

$$ \mbox{TL} = 10\log_{10} \frac{\int_{\Gamma_{in}} \vert p \vert^2}{\int_{\Gamma_{out}}
    \vert p \vert ^2}. $$
The TL values for the metamaterial and the standard models around the resonant frequency $\omega^{cf_{II}}_1$
 are shown in Fig.~\ref{fig:num-tl_curves}.

\begin{figure}[ht]
    \centering
    %
    \includegraphics[width=0.48\linewidth]{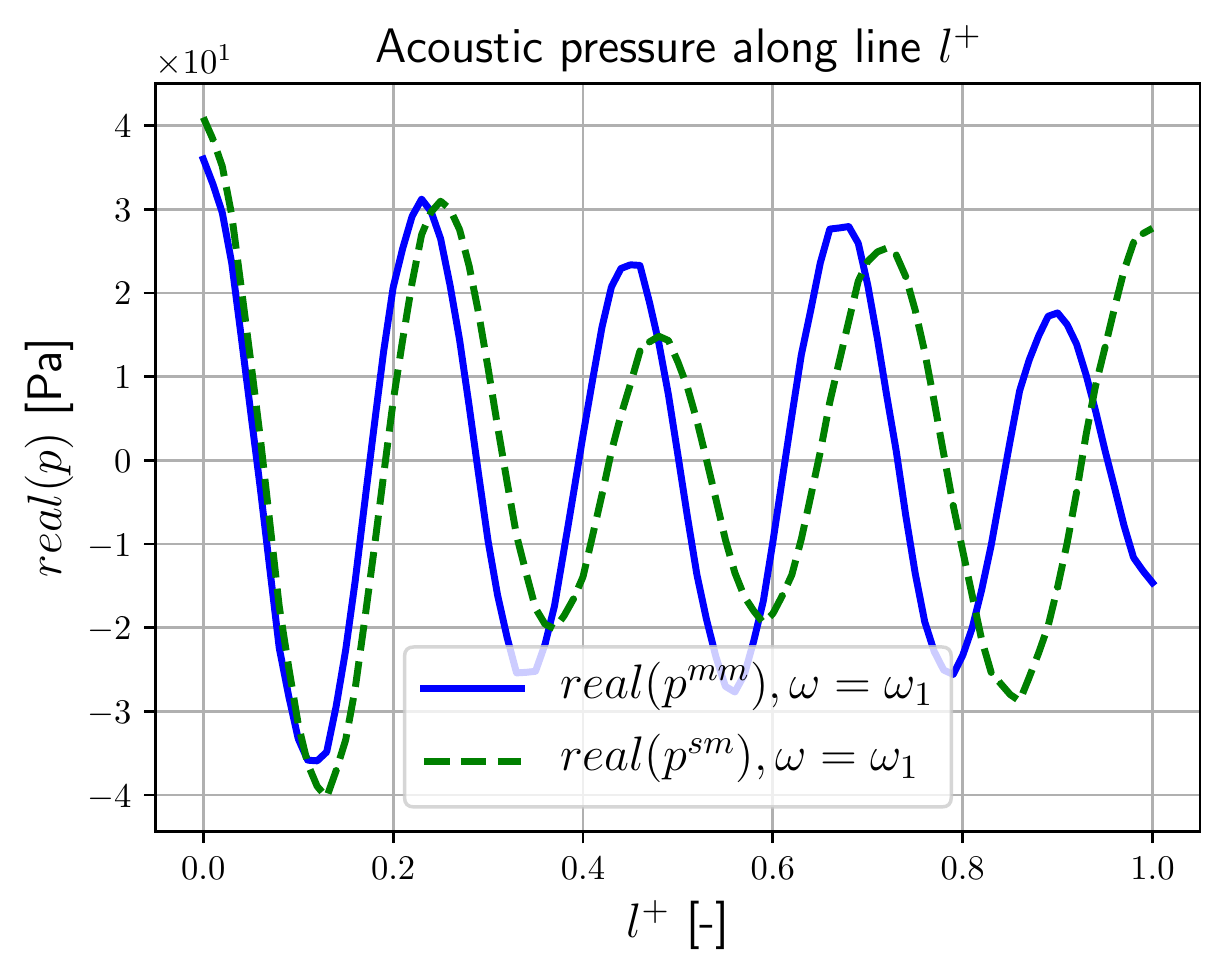}\hfil
    \includegraphics[width=0.48\linewidth]{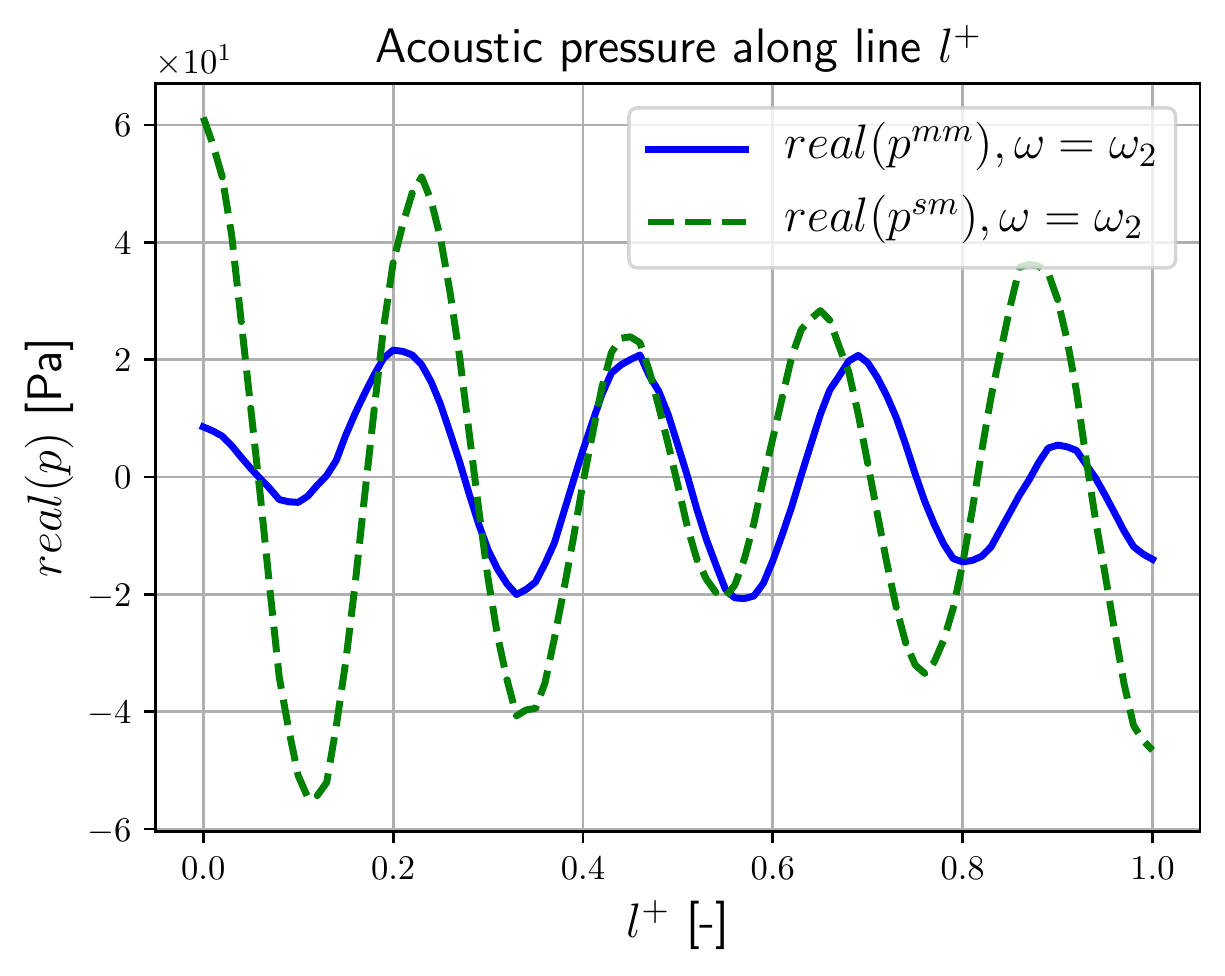}\\
    %
    \includegraphics[width=0.48\linewidth]{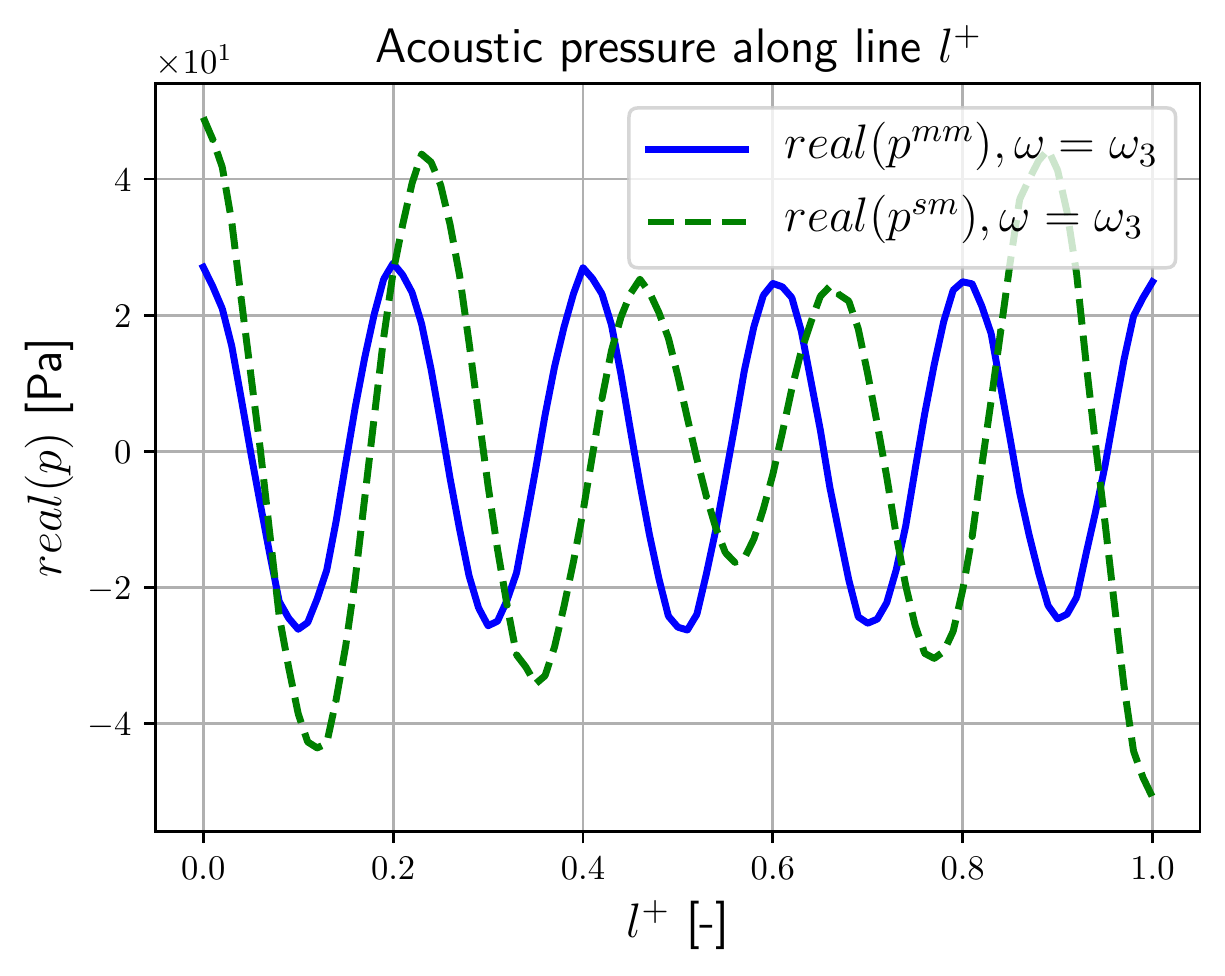}\hfil
    \includegraphics[width=0.48\linewidth]{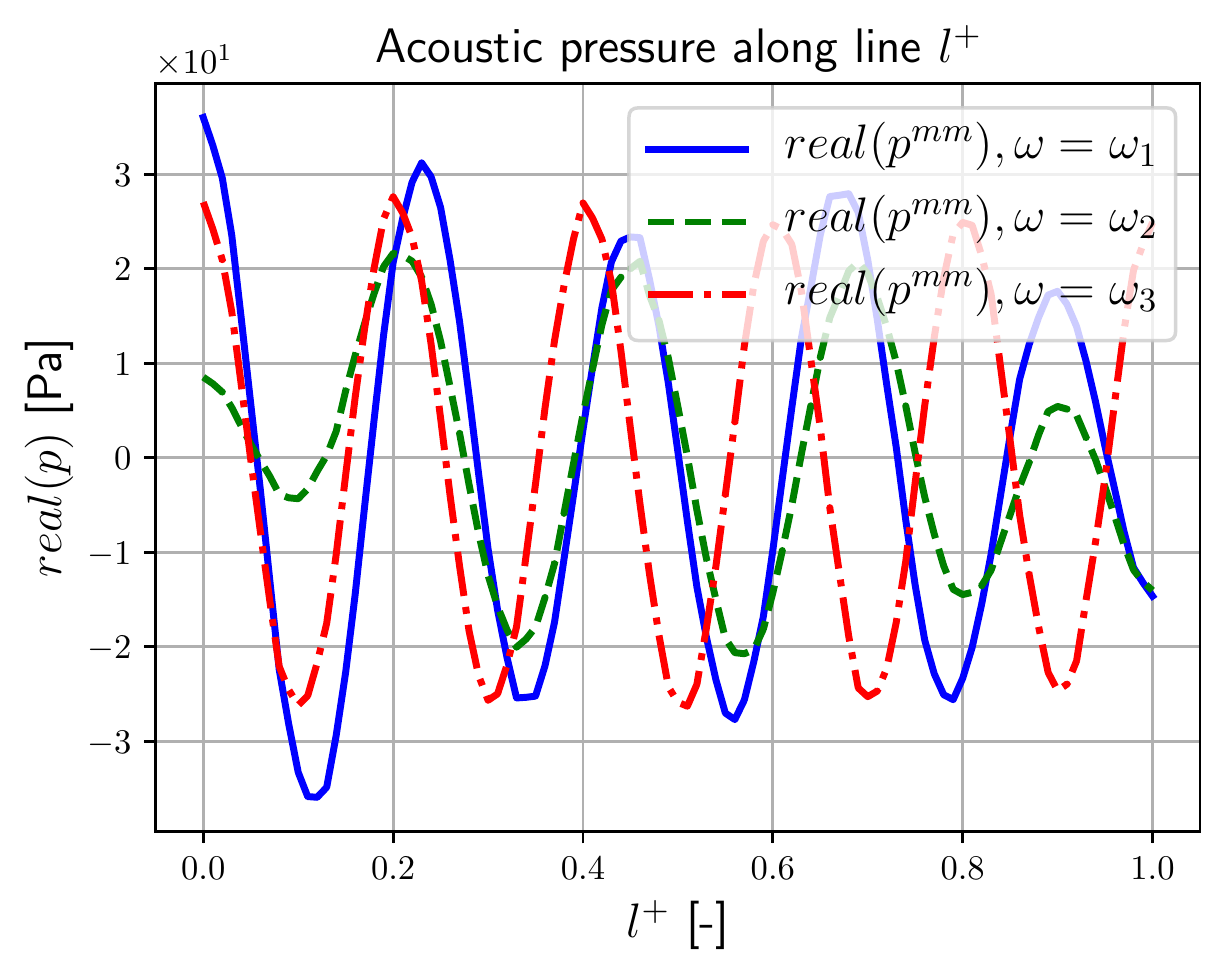}\\
    %
    %
    \caption{Comparison of the acoustic pressures calculated by the standard ($p^{sm}$)
        and the metamaterial ($p^{mm}$) models. The real parts of the pressure are plotted along line $l^+$
        for frequencies $\omega_1=33000$\,Hz, $\omega_2=35900$\,Hz and $\omega_3=36200$\,Hz.}
    \label{fig:num-lp_curves_p}
\end{figure}

\begin{figure}[ht]
    \centering
    \includegraphics[width=0.8\linewidth]{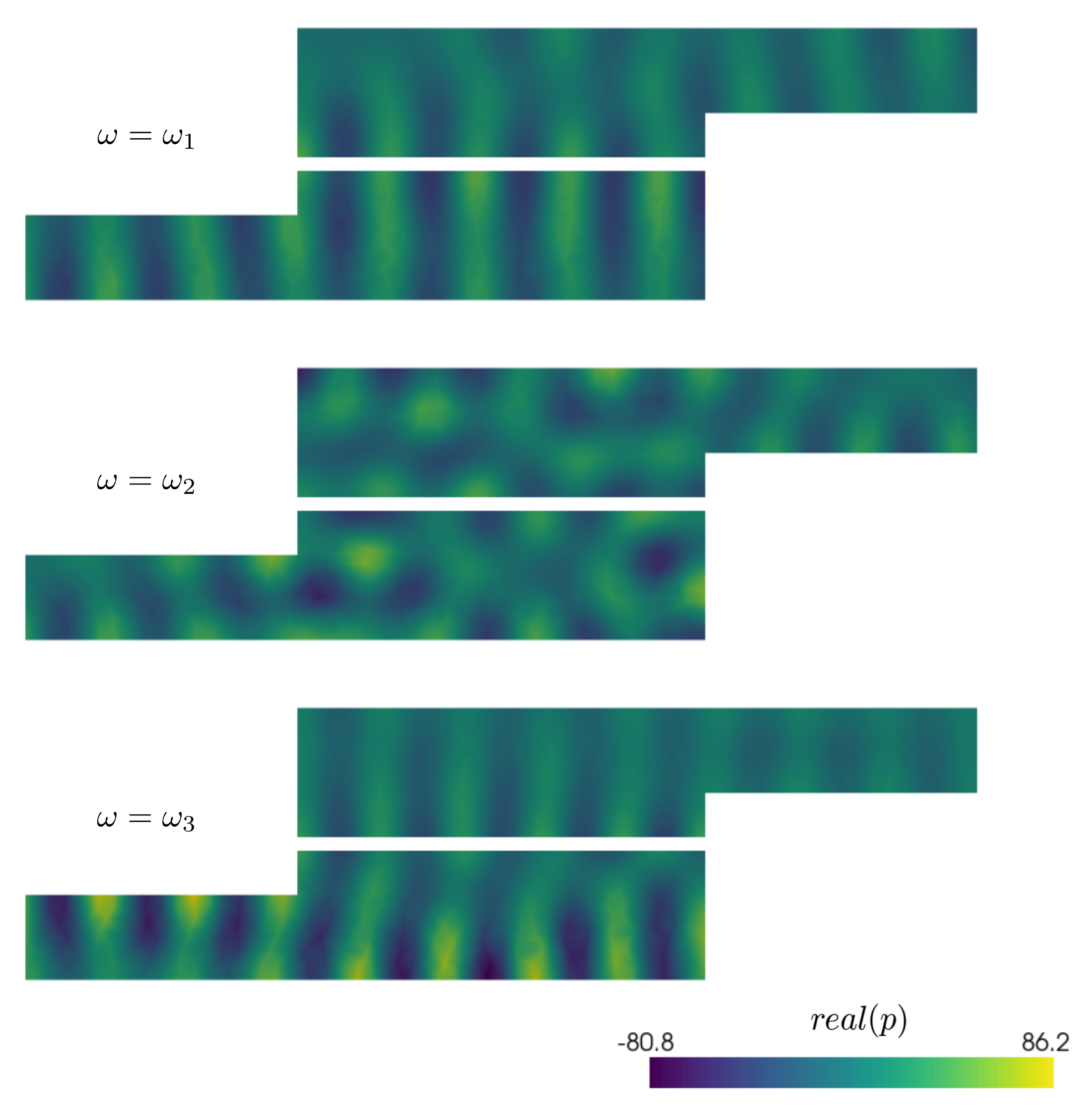}
    \caption{The distribution of the acoustic pressure (real part of $p^{mm}$) in the macroscopic domain $\hat\Omega^+ \cup \hat\Omega^-$.}
    \label{fig:num-distribution_p}
\end{figure}

\begin{figure}[ht]
    \centering
    \includegraphics[width=0.48\linewidth]{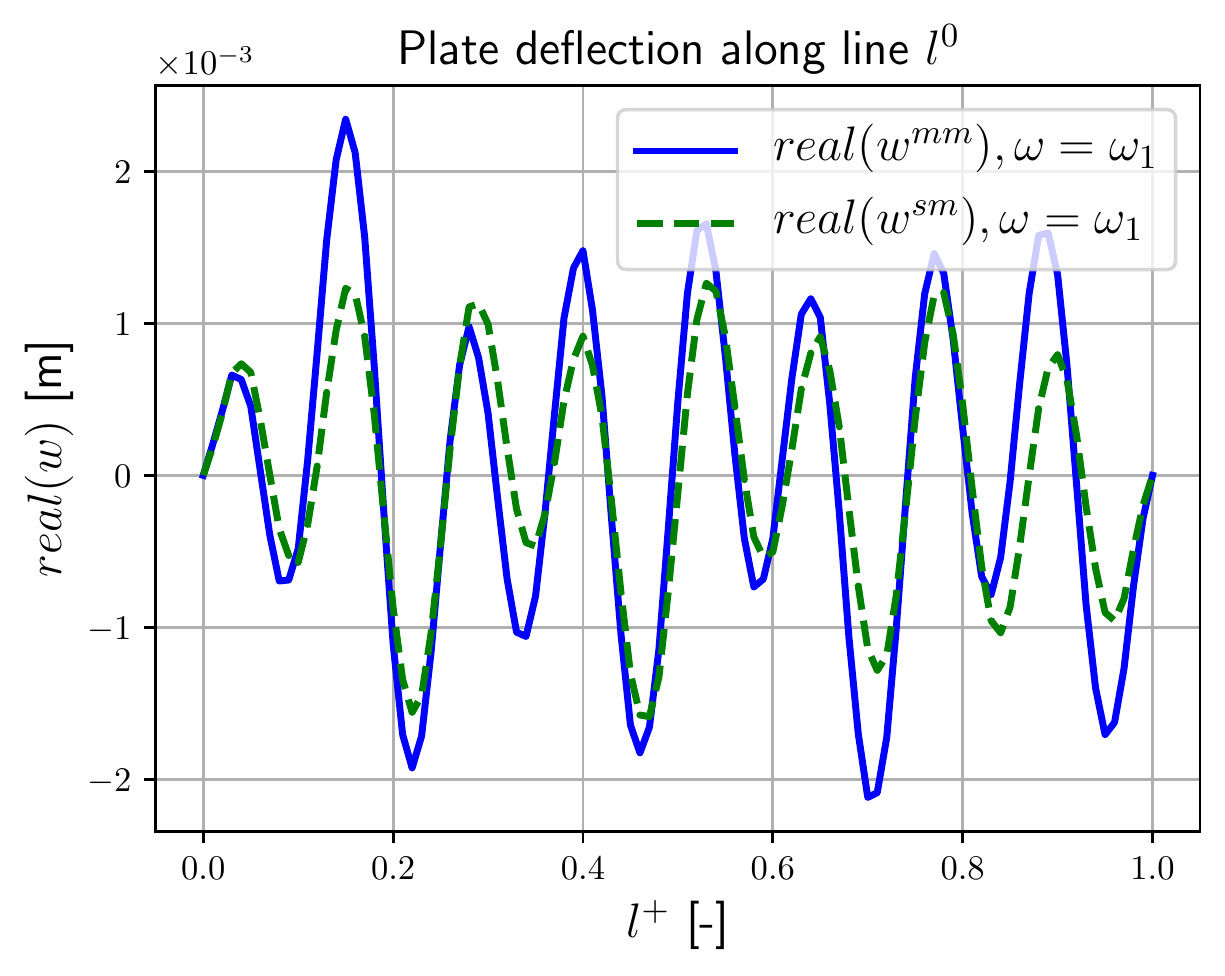}\hfil
    \includegraphics[width=0.48\linewidth]{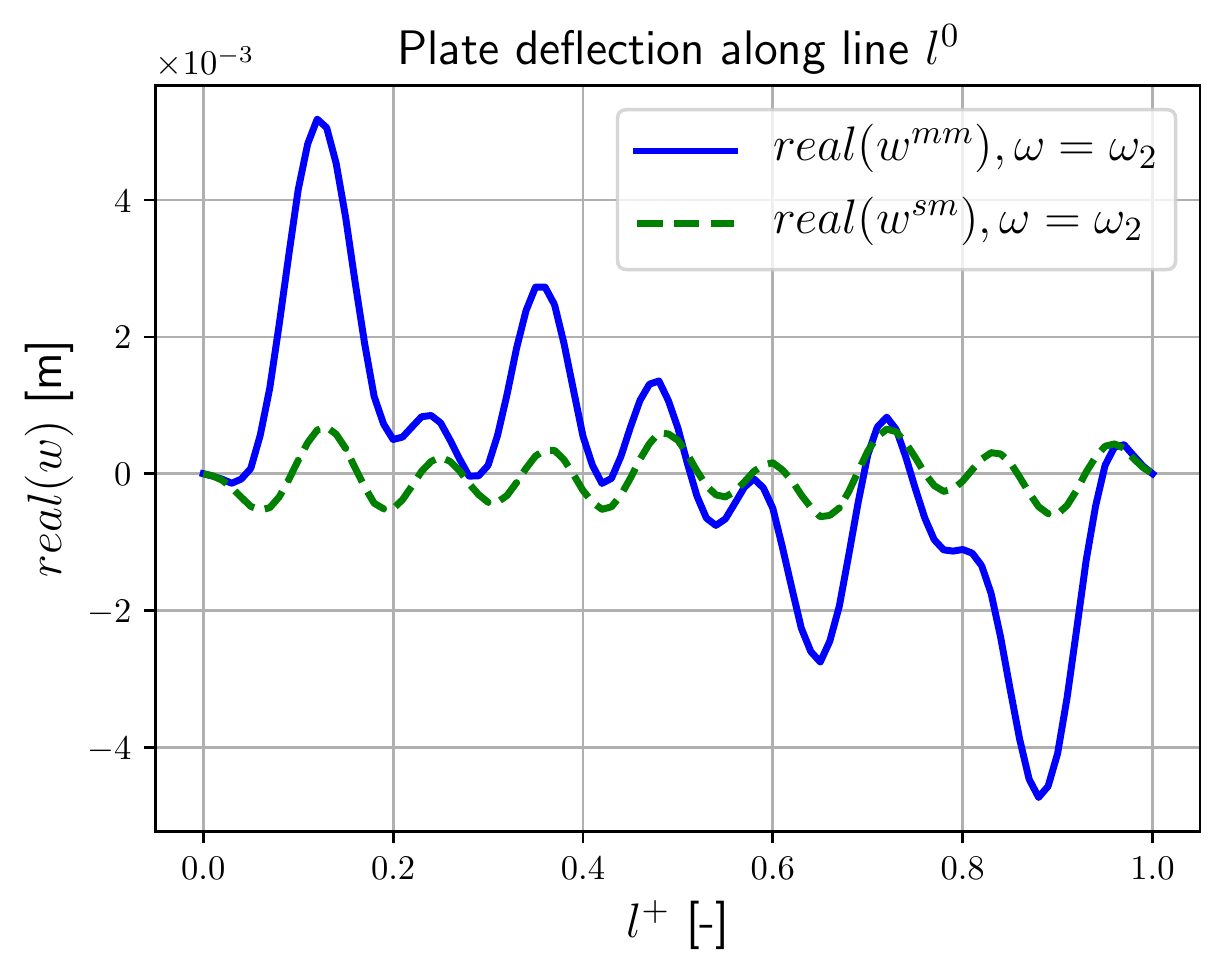}\\
    \includegraphics[width=0.48\linewidth]{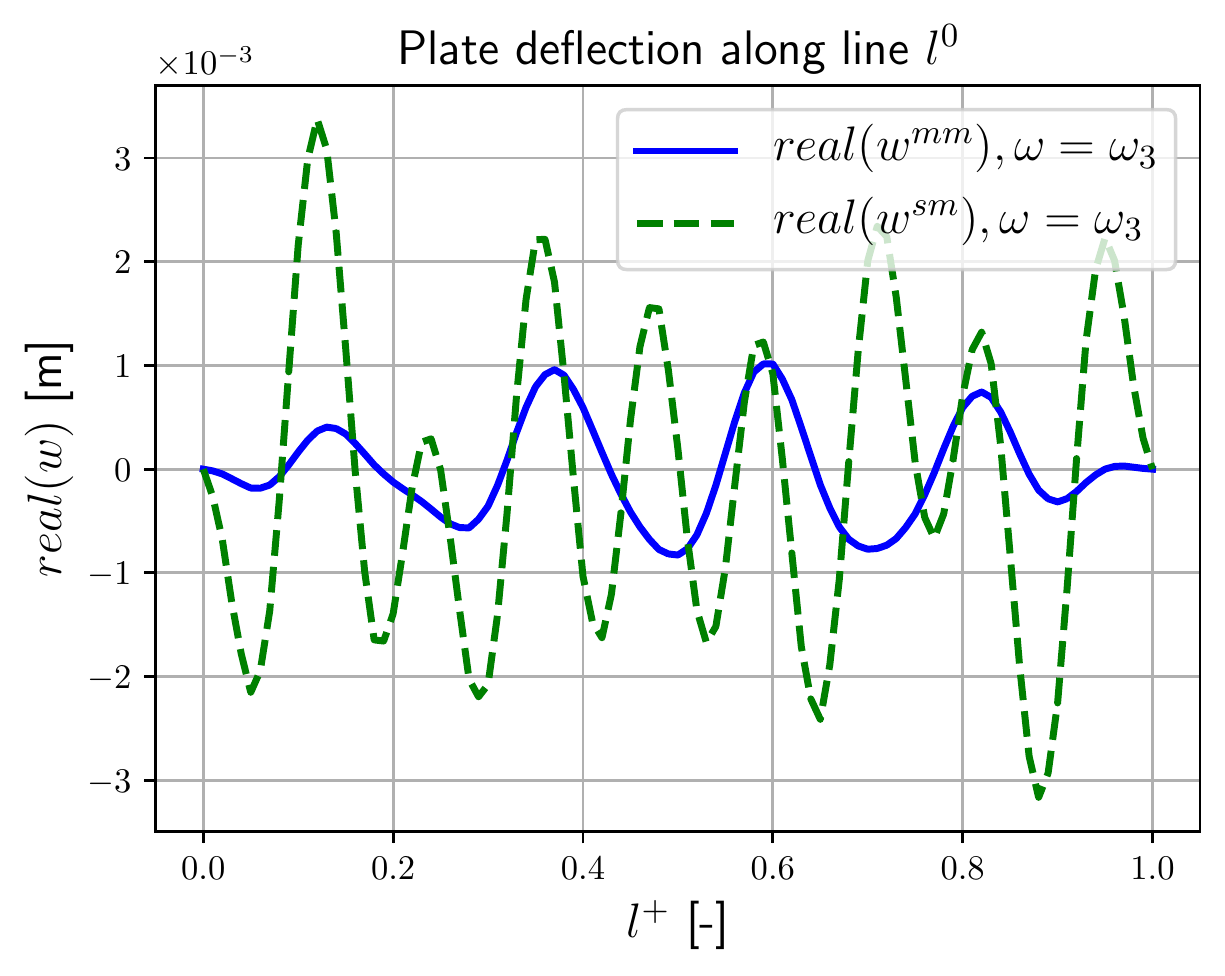}\hfil
    \includegraphics[width=0.48\linewidth]{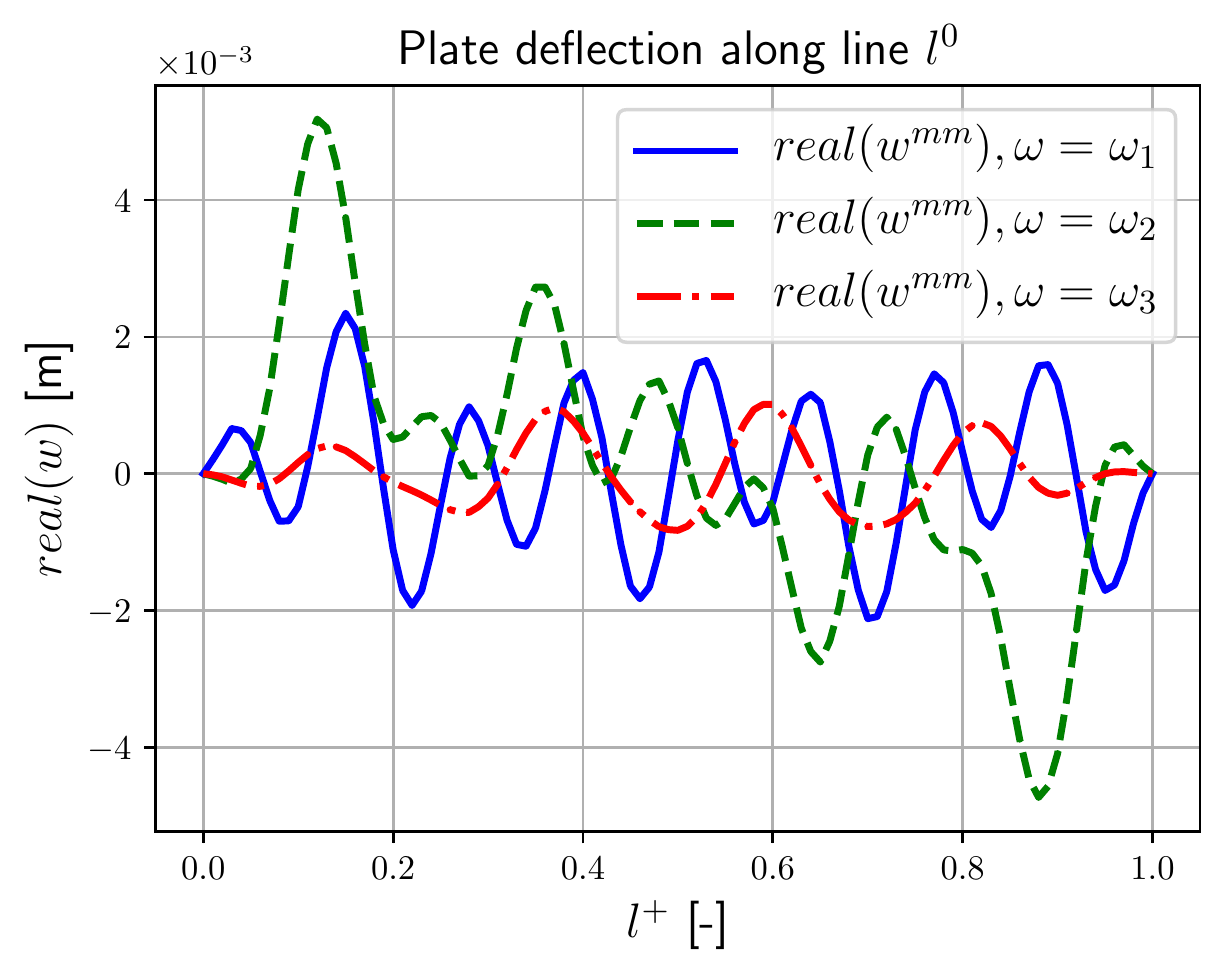}\\
    \caption{Comparison of the plate deflections calculated by the standard ($^{sm}$)
        and the metamaterial ($^{mm}$) models. The real parts of the plate deflection are plotted along line $l^0$
        for frequencies $\omega_1=33000$\,Hz, $\omega_2=35900$\,Hz and $\omega_3=36200$\,Hz.}
    \label{fig:num-lp_curves_w}
\end{figure}

\begin{figure}[ht]
    \centering
    \includegraphics[width=0.99\linewidth]{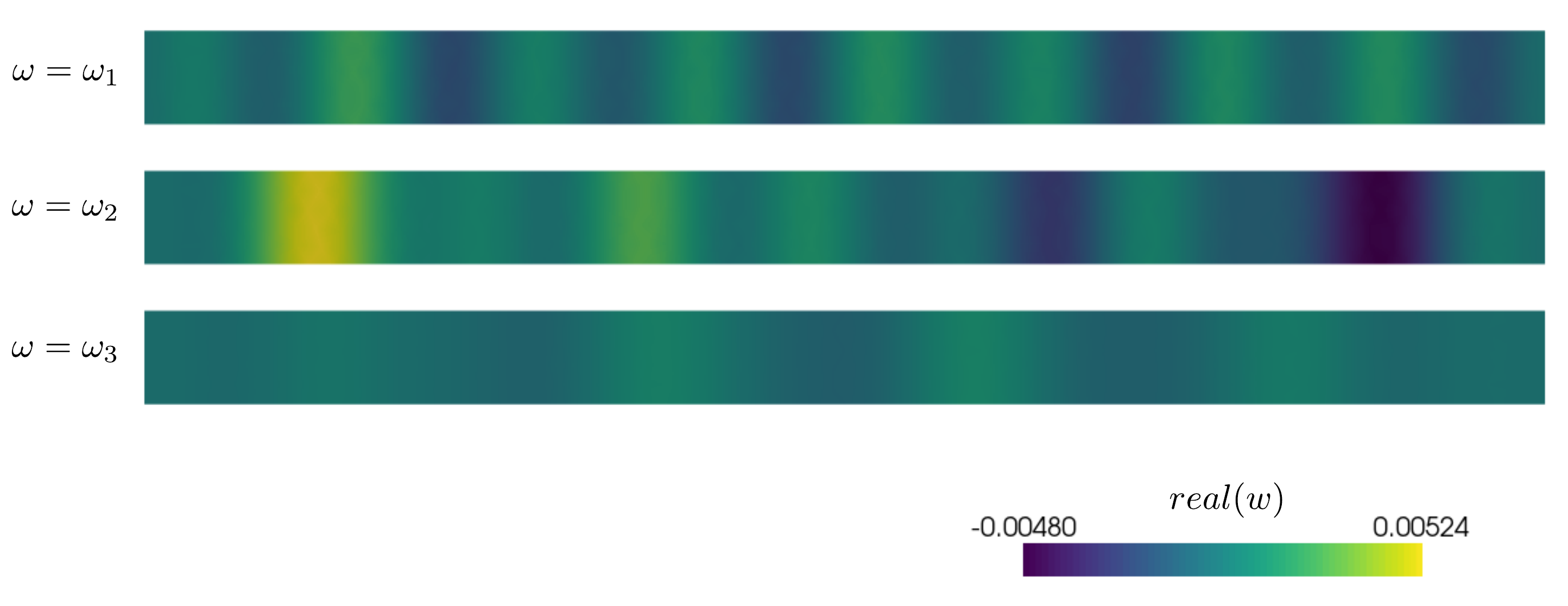}
    \caption{The distribution of the plate deflection (real part of $w^{mm}$) in the macroscopic domain $\Gamma_0$.}
    \label{fig:num-distribution_w}
\end{figure}

Another comparison is made in Fig.~\ref{fig:num-lp_curves_p}, where we compare
the results of both the models along the line probes $l^+$ and $l^0$ which are
parallel to $x_1$-axis and their positions are marked out in
Fig.~\ref{fig:num-domain-mac}. Further, we choose three frequencies
$\omega_1=33000$\,Hz, $\omega_2=35900$\,Hz and $\omega_3=36200$\,Hz; the
first one is far enough from the lowest critical frequency $\omega^{cf_{II}}_1
= 36035$\,Hz, see Tab.~\ref{tab:num-crit_freqs_II}, the second frequency is
slightly lower ($\omega^{cf_{II}}_1 - \omega_2 = 135$\,Hz) and the third one is
slightly higher ($\omega_3 - \omega^{cf_{II}}_1 = 165$\,Hz) than $\omega^{cf_{II}}_1$.
The right bottom subfigure of Fig.~\ref{fig:num-lp_curves_p} shows the acoustic
pressures along the probe line $l^+$ calculated by the metamaterial model for
$\omega_1$, $\omega_2$, $\omega_3$. The other subfigures compare the results
$p^{mm}$ obtained by the newly proposed model with the reference values
$p^{sm}$. The distributions of the local acoustic pressure for given
frequencies are depicted in Fig.~\ref{fig:num-distribution_p}.
As for the acoustic pressure field, the similar comparisons for the plate
deflection (transversal displacement) in $\Gamma_0$ are presented in
Fig.~\ref{fig:num-lp_curves_w} and Fig.~\ref{fig:num-distribution_w}.
For the out-of-resonance frequency the standard and the metamaterial models
provides the similar results as expected. Near resonance, the macroscopic
fields of the metamaterial model are affected by varying homogenized
coefficients so that they are either attenuated or amplified. 

For illustration of the microscopic characteristic responses, we show the
corrector functions $\pi^\beta$, $\xi$, $\hat\varsigma$, see the local problems
\eq{eq-lp3} and \eq{eq-lp4}, calculated in the fluid domain $Y^\ast$ and in the
elastic inclusion $\Xi_c$.
\begin{figure}[ht]
    \centering
    \includegraphics[width=0.95\linewidth]{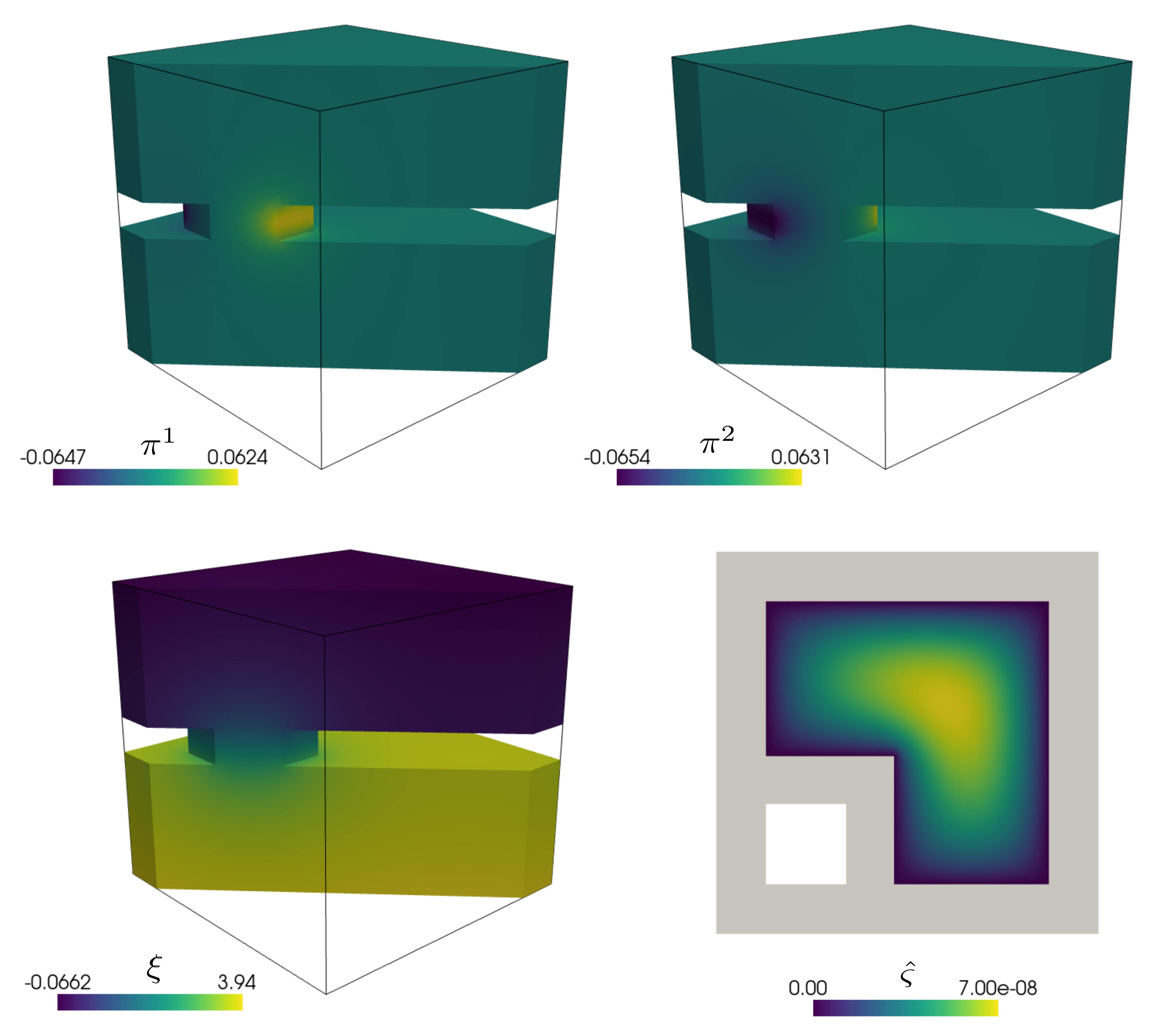}\hfill
    \caption{The characteristic responses of the local problems \eq{eq-lp3} and \eq{eq-lp4}:
        top -- Eq.\eq{eq-lp3}, correctors $\pi^\beta$ ($\hat w^\beta = 0$),
        bottom -- Eq.\eq{eq-lp4}, correctors $\xi$, $\hat\varsigma$.}
    \label{fig:num-mic-correctors}
\end{figure}

\subsection{Tuning of resonant frequencies}

\begin{figure}[ht]
    \centering
    \includegraphics[width=0.45\linewidth]{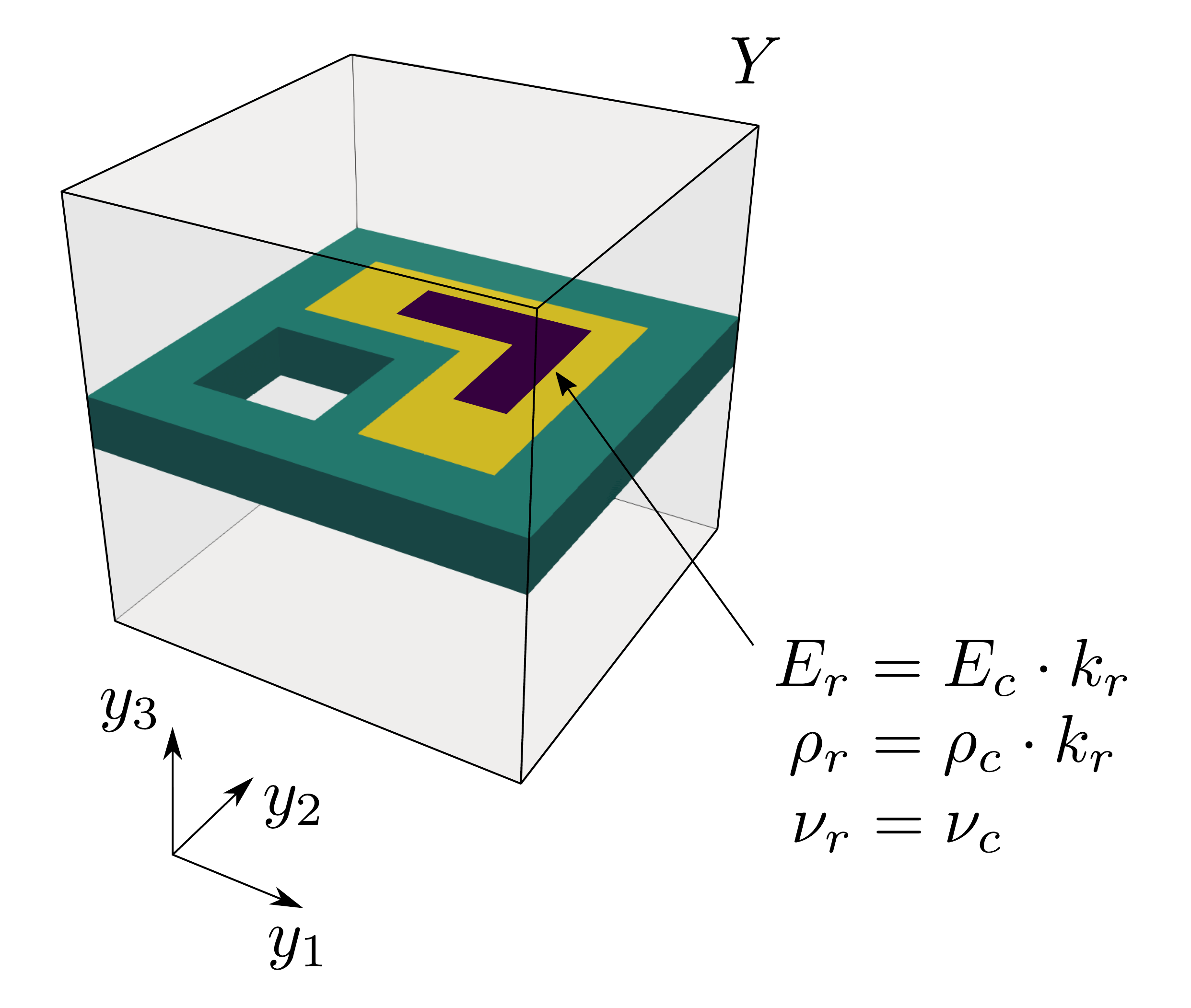}\hfill
    \includegraphics[width=0.54\linewidth]{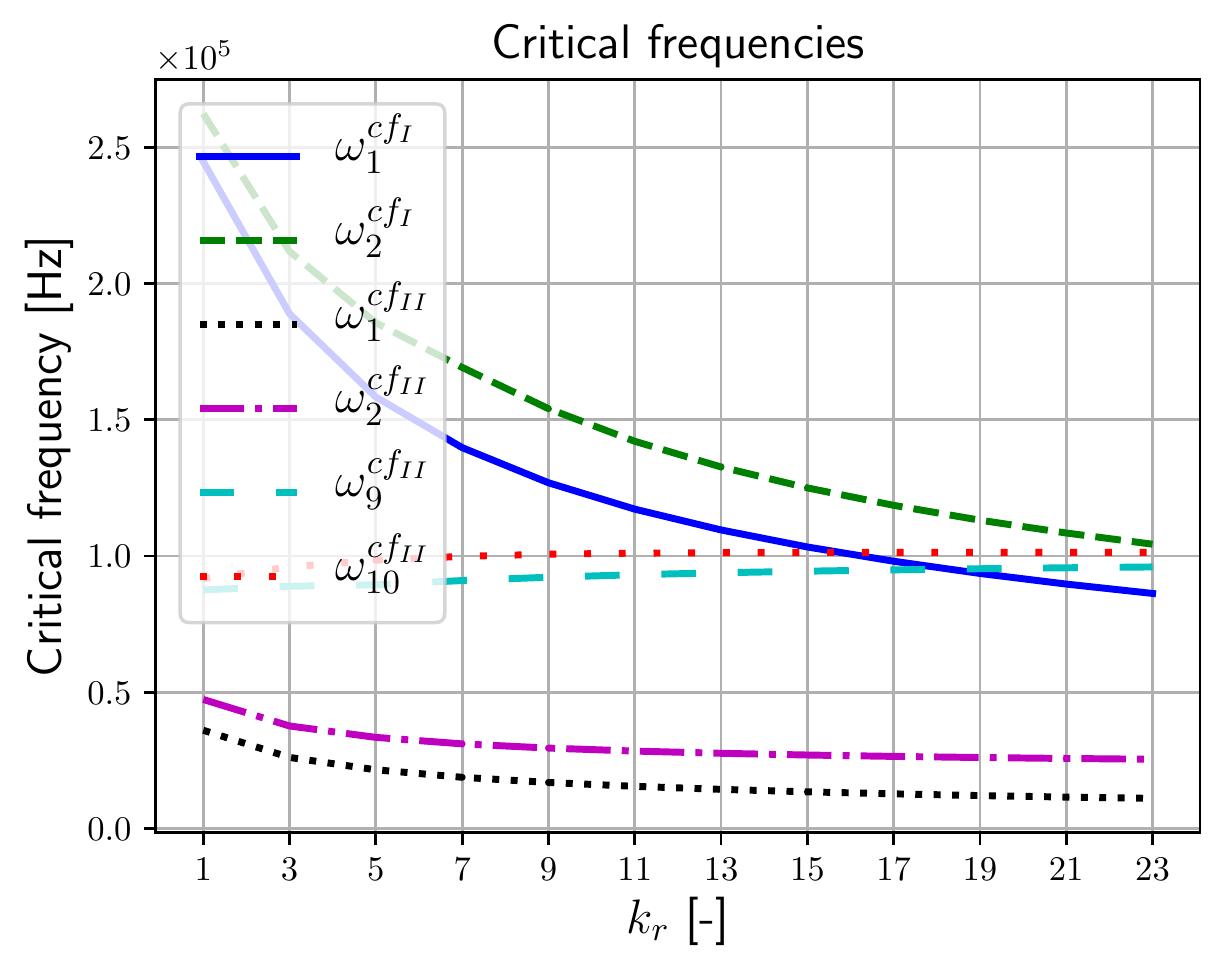}
    \caption{Left: the microscopic geometry with the resonator embedded in the elastic inclusion.
             Right: the dependence of the critical frequencies $\omega^{cf_{I}}$ and $\omega^{cf_{II}}$
             on the material properties of the resonator part which are parametrized by $k_r$.}
    \label{fig:num-domain-mic-resonator}
\end{figure}

In the case of the metamaterial model, the critical frequencies are dependent
on the geometrical arrangement of the perforated structure and on the material
properties of the constituents. To demonstrate the effect of varying material
parameters, we modify the soft inclusion (domain $S_c$) by an embedded part,
see the left subfigure of Fig.~\ref{fig:num-domain-mic-resonator}, the specific
weight and stiffness of which we parametrize as follows: $E_r = E_c \cdot k_r$
and $\rho_r = \rho_c \cdot k_r$. The dependence of the first few critical
frequencies $\omega^{cf_I}$ and $\omega^{cf_{II}}$ on $k_r$ is shown in
Fig.~\ref{fig:num-domain-mic-resonator} right. We observe the significant drop
of $\omega^{cf_I}_1$, $\omega^{cf_I}_2$ and a slight decrease of
$\omega^{cf_II}_1$, $\omega^{cf_II}_2$ with increasing $k_r$, while the higher
critical frequencies, e.g. $\omega^{cf_II}_9$, $\omega^{cf_II}_{10}$, increase.

\begin{figure}[ht]
    \centering
    \includegraphics[width=0.49\linewidth]{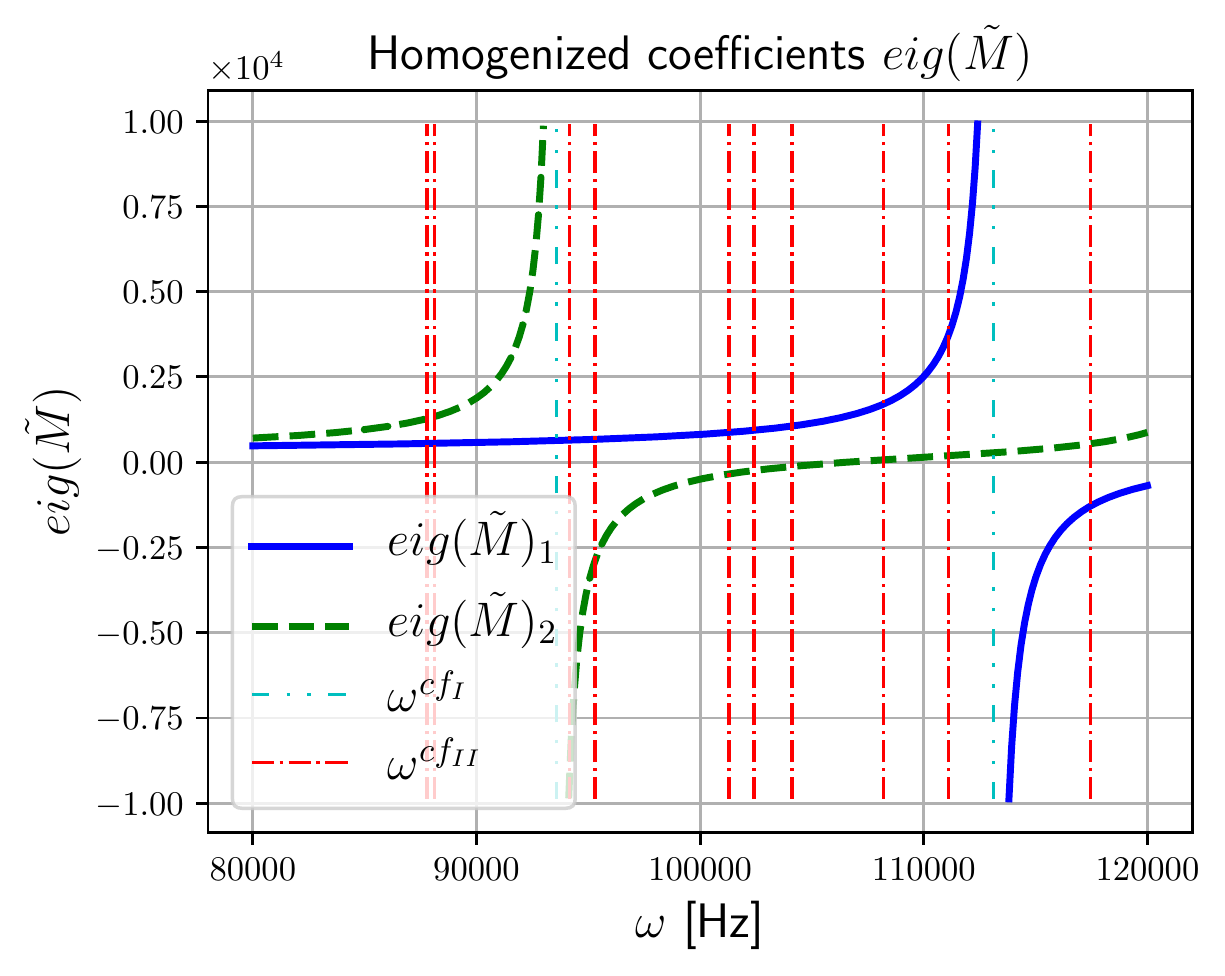}\hfill
    \includegraphics[width=0.49\linewidth]{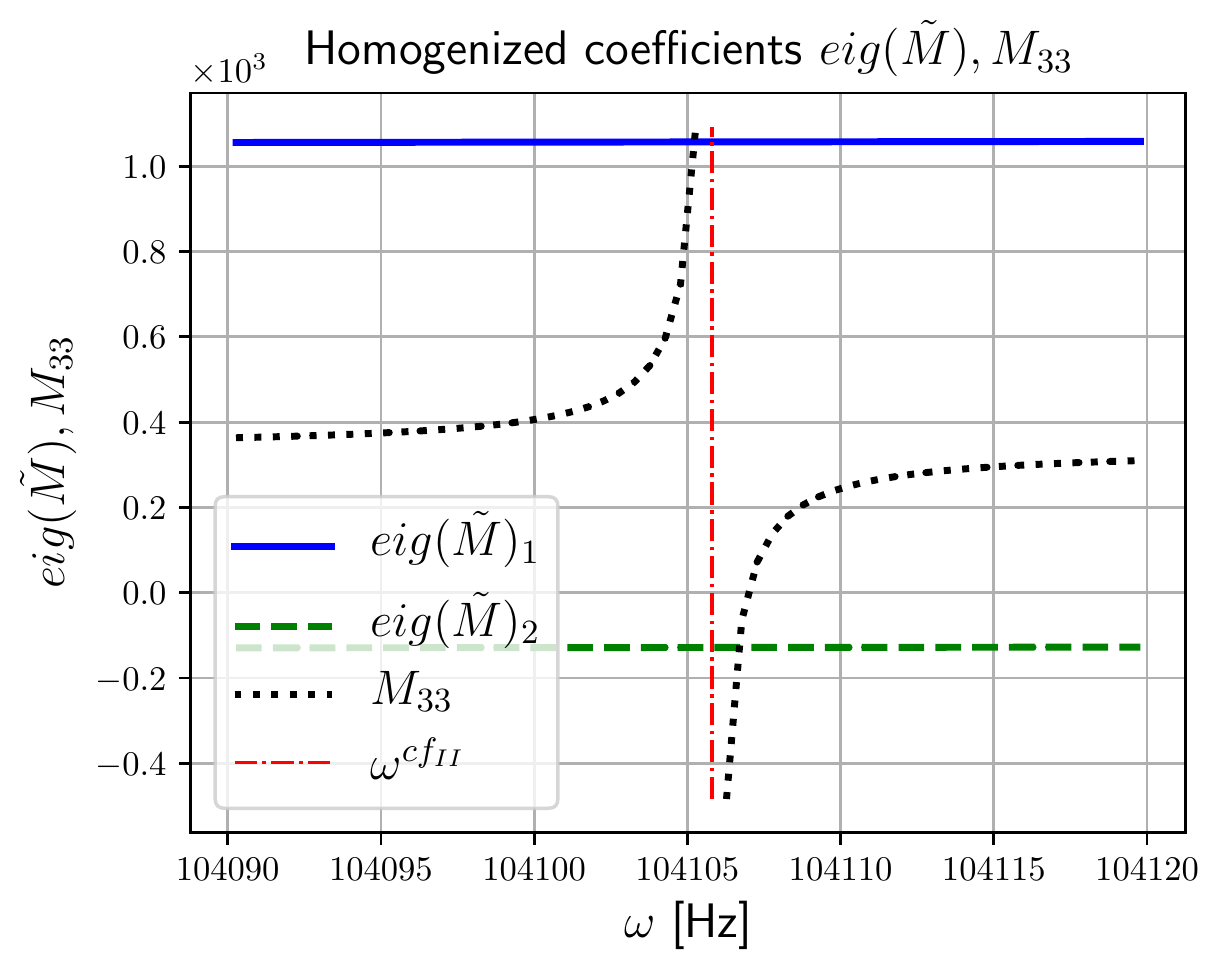}\\
    \caption{Left: the eigenvalues of homogenized coefficients $\tilde M$.
        Right: the eigenvalues of $\tilde M$ and coefficient $M_{33}$ around a resonant frequency.
    }
    \label{fig:num-resonator-coefM}
\end{figure}

Let $k_r = 19$, then we get the mingled critical frequencies $\omega^{cf_I}$
and $\omega^{cf_II}$ starting from $\omega = 93577$\,Hz, see
Fig.~\ref{fig:num-resonator-coefM} left. Among these higher frequencies we are
able to locate a range, where at least one eigenvalue of coefficient $\tilde M$
is negative and also $M_ {33}$ is less than zero as depicted in
Fig.~\ref{fig:num-resonator-coefM} right. This indicates a possible band gap
effects in a given frequency range.

\section{Conclusion}\label{sec-concl}

We derived vibro-acoustic transmission conditions describing the interaction
between the metamaterial plate and the surrounding acoustic fluid. The plate
with periodic distribution of soft inclusions has been treated as the large
contrast elastic medium with the scale-dependent stiffness of the resonators,
following the idea of \cite{rohan-miara-ZAMM2015}. Due to this, the
homogenization and the ``3D to 2D''dimensional reduction enable to replace the
3D problem imposed in the transmission layer with a complicated geometry by the
Dirichlet-to-Neumann operator given in an implicit form representing the
metasurface. Frequency dependent effective coefficients of the homogenized
model can be evaluated for thousands of frequencies in a time slot of seconds
due to a very efficient algorithm based on a spectral decomposition of the
representative periodic cell response. The numerical examples are rather
illustrative, although they show some effects of the metamaterial plate model
-- in a neighbourhood of the resonance frequencies associated with the spectral
decomposition of the fluid-structure interaction, the impedance and some other
coefficients can change their signs which affects the acoustic transmission and
the plate vibrations. Such a behaviour cannot be captured by the standard model
\cite{Rohan-Lukes-AMC2019}, derived without the scale-related contrast
property. Suitable design modifications (the size of holes, mechanical
properties of the resonators) enable that all coefficients representing the
metasurface mass density can become negative at least in narrow frequency
intervals, thus, yielding the band gaps in the wave transmission. This
motivates the further research towards optimized microstructures using the
shape \cite{isma2012,rohan-lukes-icovp2013}, or topology
\cite{noguchi2020topology} optimization up to now considered for the standard
perforated plates. Further perspectives involve effects of flows through the
perforated metamaterial plates and wave scattering in cascades of blades, \cf
\cite{Maierhofer-Peake2020}, or porous layered (sandwich) structures
\cite{Mikelic-Czochra-plate2015,Webster-Bociu-2020-multilayered-poroelasticity}
introducing naturally the attenuation.

\bigskip
\paragraph{Acknowledgment}

The research at the University of West Bohemia has been supported by the grant
project GACR 19-04956S of the Czech Scientific Foundation. and in a part due to
the European Regional Development Fund-Project ``Application of Modern
Technologies in Medicine and Industry'' {(No.~CZ.02.1.01/0.0/
0.0/17~048/0007280)} provided by the Czech Ministry of Education, Youth and
Sports.

\appendix

\section{Limit of the unfolded vibroacouctic problem}\label{apx-1}

Upon unfolding the fluid equation \eq{eq-wf6}, its \lhs yields the limit
\begin{equation}\label{eq-f1}
\begin{split}
& c^2 \int_{\Gamma_0} \intY_{Y^*} \Tuf{\gradpl p^\veps} \cdot \Tuf{\gradpl q^\veps}
+ \frac{c^2}{\veps^2} \int_{\Gamma_0} \intY_{Y^*}\pd_z \Tuf{p^\veps}
\pd_z \Tuf{q^\veps} - \om^2 \int_{\Gamma_0} \intY_{Y^*} \Tuf{p^\veps} \Tuf{q^\veps}\\
\rightarrow & c^2 \int_{\Gamma_0} \intY_{Y^*} (\gradplx p^0 + \gradply p^1)\cdot
(\gradplx q^0 + \gradply q^1) + c^2 \int_{\Gamma_0} \intY_{Y^*}\pd_z p^1 \pd_z q^1 - \om^2 \int_{\Gamma_0} \intY_{Y^*} p^0 q^0\;.
\end{split}
\end{equation}
The unfolded \rhs interaction terms in \eq{eq-wf6}, below divided by $\imu\om c^2$, 
\begin{equation}\label{eq-f4}
\begin{split}
& \frac{1}{\veps}\int_{\Gamma_0}\intY_{\Xi_S}(u_3^0+\veps u_3^1 + \chi_c \hat u_3) \veps \Jump{q^1}{\hh}\\
& +\frac{\hh}{\veps}\int_{\Gamma_0} \intY_{\pd\Xi_S} \int_{-1/2}^{1/2} \ol{\nb}\cdot \left(\ol{\ub}^0 + \veps \ol{\ub}^1 + \chi_c \hat{\ol{\ub}} -\veps\hh \zeta(\thetabf^0 + \veps\thetabf^1 + \chi_c \hat\thetabf)\right)(q^0 +\veps q^1)\;,
\end{split}
\end{equation}
converge, as follows:
\begin{equation}\label{eq-f4b}
\begin{split}
 & \int_{\Gamma_0}\left( u_3^0\intY_{\Xi_S} \Jump{q^1}{\hh}
+ \ol{\ub}^0\cdot\intY_{\pd\Xi_S}\ol{\nb}\hh\int_{-1/2}^{1/2}q^1 \dd\zeta + 
q^0 \hh \intY_{\pd\Xi_S}\ol{\nb}\cdot\ol{\ub}^1 + \intY_{\Xi_c}  \hat u_3 \Jump{q^1}{\hh}\right)
\end{split}
\end{equation}

In the plate equation \eq{eq-wp7}, the unfolded inertia terms at the \lhs  converge, as follows
\begin{equation}\label{eq-wp7-L1} 
\begin{split}
& \om^2 \int_{\Gamma_0}\intY_{\Xi_S}\Tuf{\rho^\veps}\Tuf{\ub^\veps}\cdot \Tuf{\vb^\veps}+ 
  \om^2 \frac{h^2}{12} \int_{\Gamma_0}\intY_{\Xi_S} \Tuf{\rho^\veps}\Tuf{\thetabf^\veps}\cdot \Tuf{\psibf^\veps}\\
  \rightarrow  & -\om^2 \int_{\Gamma_0} \rho_S \left(( \ub^0 + \chi_c \hat u_3)\cdot(\vb^0 + \chi_c \hat v_3) + 
\frac{h^2}{12} (\thetabf^0  + \chi_c\hat\thetabf)\cdot(\psibf^0 + \chi_c\hat\psibf)\right)\;,
\end{split}
\end{equation}
where $\rho_S(y') = \chi_m(y')\rho_m + \chi_c(y')\rho_c$ is the solid density. 
Then we obtain the limit of the unfolded elasticity terms with substituted heterogeneity anstaz \eq{eq-elast1},
\begin{equation}\label{eq-wp7-L2} 
\begin{split}
& - \int_{\Gamma_0}\intY_{\Xi_S} [\Tuf{\Eop^\veps} \gradplS \Tuf{\ol{\ub}^\veps} ]: \Tuf{\gradplS \ol{\vb}^\veps} 
- \frac{h^2}{12}\int_{\Gamma_0}\intY_{\Xi_S} [\Tuf{\Eop^\veps} \gradplS \Tuf{\thetabf^\veps}]: \Tuf{\gradplS \psibf^\veps} \\
& -\int_{\Gamma_0}\intY_{\Xi_S} [\Tuf{\Sb^\veps} (\Tuf{\gradpl u_3^\veps} - (\Tuf{\thetabf^\veps})]\cdot
(\Tuf{\gradpl v_3^\veps} - \Tuf{\psibf^\veps}) \\
 \rightarrow
& +  \int_{\Gamma_0}\left(\intY_{\Xi_m}[\Eop_m(\gradplxS\ol{\ub}^0 + \gradplyS\ol{\ub}^1)]:(\gradplxS\ol{\vb}^0 + \gradplyS\ol{\vb}^1) +\intY_{\Xi_c} [\Eop_c \gradplyS\hat{\ol{\ub}}:\hat{\ol{\vb}}\right)
  \\
& + \int_{\Gamma_0} \left(\intY_{\Xi_m}[\Sb_m(\gradpl_x u_3^0 + \gradpl_y u_3^1 - \thetabf^0)]\cdot
(\gradpl_x v_3^0 + \gradpl_y v_3^1 - \psibf^0)
+ \intY_{\Xi_c}[\Sb_c \gradpl_y \hat u_3] \cdot\gradpl_y \hat v_3\right)\\
& + \frac{h^2}{12}\int_{\Gamma_0} \left(\intY_{\Xi_m}[\Eop_m(\gradplxS \thetabf^0 + \gradplyS \thetabf^1)]:(\gradplxS \psibf^0 + \gradplyS \psibf^1) + \intY_{\Xi_c}[\Eop_c \gradplyS \hat\thetabf]:\gradplyS \hat\psibf \right)\\
=: & \int_{\Gamma_0}\left(\Pcal_m((\ub^0,\ub^1,\thetabf^0,\thetabf^1),(\vb^0,\vb^1,\psibf^0,\psibf^1)) + \Pcal_c((\hat\ub,\hat\thetabf),(\hat\vb,\hat\psibf))\right)\;.
\end{split}
\end{equation}

The unfolded \rhs term in \eq{eq-wp7}
\begin{equation}\label{eq-p2}
\begin{split}
& \frac{\imu\om\rho_0}{\veps\hh}\int_{\Gamma_0}\intY_{\Xi_S}(v_3^0+\veps v_3^1 + \chi_c \hat v_3) \veps \Jump{p^1}{\hh}\\
& +\frac{\imu\om\rho_0}{\veps}\int_{\Gamma_0} \intY_{\pd\Xi_S} \int_{-1/2}^{1/2} \ol{\nb}\cdot \left(\ol{\vb}^0 + \veps \ol{\vb}^1 +  \chi_c \hat{\ol{v}}-\veps\hh \zeta(\psibf^0 + \veps\psibf^1 + \chi_c \hat\psibf )\right)(p^0 +\veps p^1)\;,
\end{split}
\end{equation}
converges to the limit expression
\begin{equation}\label{eq-p2b}
\begin{split}
  & \frac{\imu\om\rho_0}{\hh}\int_{\Gamma_0}\left(v_3^0\intY_{\Xi_S}\Jump{p^1}{\hh}
  +\intY_{\Xi_c}\hat v_3 \Jump{p^1}{\hh}\right) \\
  & + \imu\om\rho_0 \int_{\Gamma_0}\left(\ol{\vb}^0 \cdot \int_{-1/2}^{1/2}\intY_{\pd\Xi_S}\ol{\nb} p^1 + p^0 \intY_{\pd\Xi_S}\ol{\vb}^1\cdot \ol{\nb}\right).
\end{split}
\end{equation}

\section{Properties of the homogenized coefficients}\label{apx-2}
\subsection{Proof of the homogenized coefficients symmetry relationship}
We prove the symmetry relationships claimed in Proposition~\ref{thm-3}, assertion (i).
The following symmetry holds $M_{3\alpha} =  M_{\alpha 3}$,
\begin{equation}\label{eq-M5}
\begin{split}
   M_{3\alpha} & = \frac{\rho_0}{\hh}\ippYc{\TT(\hat \vpi^3,\eta^3)}{(-\hat \vpi^\alpha,\eta^\alpha)}=
  \frac{\rho_0}{\hh}\ippYc{\TT(\hat \vpi^\alpha,\eta^\alpha)}{(-\hat \vpi^3,\eta^3)} \\
  & = - \rho_0\intY_{\pd\Xi_S} n_\alpha  \int_{-1/2}^{1/2}\eta^3 =  M_{\alpha 3}\;.
\end{split}
\end{equation}
The symmetry $M_{\alpha\beta} =  M_{\beta\alpha}$, $\alpha,\beta$ is straightforward,
\begin{equation}\label{eq-M5a}
  \begin{split}
 M_{\alpha\beta} = \ippYc{\TT(\hat \vpi^\alpha,\eta^\alpha)}{(\hat \vpi^\beta,\eta^\beta)} =   M_{\beta\alpha}\;.
\end{split}
\end{equation}

Further, $C_\alpha = 0$ by some symmetry agrguments.
Symmetry of $D_{3\alpha} = D_{\alpha 3}^*$ and {$D_{\alpha\beta}^* = D_{\beta\alpha}$} is obtained using the autonomous local problems. Due to \eq{eq-f11}, we get
\begin{equation}\label{eq-M7}
\begin{split}
  D_{\alpha\beta}^* & =  \intY_{Y^*} \pd_\alpha^y \eta^\beta = \ippYc{\TT(\hat w^\alpha,\pi^\alpha)}{(\hat \vpi^\beta,-\eta^\beta)} \\
  & = \ippYc{\TT(\hat \vpi^\beta,\eta^\beta)}{(\hat w^\alpha,-\pi^\alpha)} = \hh \intY_{\pd \Xi_S} n_\beta \int_{-1/2}^{1/2} \pi^\alpha \dd\zeta =  D_{\beta\alpha}\;. \\
  D_{\alpha 3}^* & =  \intY_{Y^*} \pd_\alpha^y \eta^3
= \ippYc{\TT(\hat w^\alpha,\pi^\alpha)}{(\hat \vpi^3,-\eta^3)} 
= \ippYc{\TT(\hat \vpi^3,\eta^3)}{(\hat w^\alpha,-\pi^\alpha)}  \\
& = -\imu\om\frac{\hh}{\rho^0} \intY_{\Xi_c} \rho \hat w^\alpha +
\ipXiS{\Jump{\pi^\alpha}{\hh}}{1} =  D_{3\alpha}\;.
\end{split}
\end{equation}

 Symmetries declared in \eq{eq-M10}$_{2,3}$ can be proved easily using the local problems \eq{eq-lp3}-\eq{eq-lp6}. Coefficient $F$ can be expressed using \eq{eq-lp4} (with $q = \xi$), as follows:
\begin{equation}\label{eq-M10a}
\begin{split}
  F & = - \intY_{I_y^+} \xi + \intY_{I_y^-} \xi =  \ippYc{\TT(\hat \vsigma,\xi)}{(\hat \vsigma,\xi)} \\
  & = \frac{\hh}{\rho_0} \left[ \bhYc{\hat \vsigma}{\hat \vsigma} - \om^2 \ipXic{\rho \hat \vsigma}{\hat \vsigma}\right] + \intY_{Y^*}\nabla_y \xi \cdot \nabla_y \xi\;,
\end{split}
\end{equation}
if we chose the test function $\hat v = \hat \vsigma$ in \eq{eq-lp4}.

  The symmetry can be proved, ${C'}_k = C_k = 0$, $k= 1,\dots,3$. Firstly,
\begin{equation}\label{eq-M11}
\begin{split}
  {C'}_\alpha  &  = \intY_{I_y^+} \eta^\alpha - \intY_{I_y^-}\eta^\alpha = \ippYc{\TT(\hat \vsigma,\xi)}{(\hat \vpi^\alpha,-\eta^\alpha)}\\
  & = \ippYc{\TT(\hat \vpi^\alpha,\eta^\alpha)}{(\hat \vsigma,-\xi)} = \hh \intY_{\pd \Xi_S} n_\alpha \int_{-1/2}^{1/2} \xi(\cdot,\zeta)\dd\zeta = C_\alpha\;.
\end{split}
\end{equation}
In analogy (note $\intY_{\Xi_S} \Jump{\xi}{\hh} = 0$ due to the special shape on the holes, but ${C'}_3 \not = 0$ in general.)
\begin{equation}\label{eq-M12}
\begin{split}
  {C'}_3  &  = \intY_{I_y^+} \eta^3 - \intY_{I_y^-}\eta^3 = \ippYc{\TT(\hat \vsigma,\xi)}{(\hat \vpi^3,-\eta^3)}\\
  & = \ippYc{\TT(\hat \vpi^3,\eta^3)}{(\hat \vsigma,-\xi)} = -\imu\om\frac{\hh}{\rho_0} \intY_{\Xi_c} \rho \hat \vsigma + \ipXiS{\Jump{\hat \xi}{\hh}}{1} = C_3\;.
\end{split}
\end{equation}


\section{Homogenized coefficients -- computation using the discretized micro-problem}\label{apx-3}

Here we derive formulae for computing the homogenized coefficients reported in
Section~\ref{sec-FE-HC}. We disregard the assumption of the $z$-symmetric cells
$Y$, such that the cancellations declared in Proposition~\ref{thm-3} (iii) may
not apply, in general.

\bibliographystyle{elsarticle-num}
\bibliography{bib-acoustics2}

\end{document}